\documentclass[useAMS,usenatbib]{mn2e}
\usepackage{graphicx}
\usepackage{epsfig}
\usepackage{amssymb}
\usepackage{lscape}
\usepackage{ulem}
\usepackage{txfonts}

\title[ANN emission line calibrations] {Artificial neural network based
calibrations for the prediction of galactic [NII] $\lambda$ 6584
and H$\alpha$ line luminosities.}

\author[Teimoorinia \& Ellison] {Hossein Teimoorinia$^1$ \&
Sara L. Ellison$^1$\\
$^1$ Department of Physics \& Astronomy, University
of Victoria, Finnerty Road, Victoria, British Columbia, V8P 1A1,
Canada.}

\def\LaTeX{L\kern-.36em\raise.3ex\hbox{a}\kern-.15em
    T\kern-.1667em\lower.7ex\hbox{E}\kern-.125emX}

\begin{document}

\label{firstpage}

\maketitle

\begin{abstract}

The artificial neural network (ANN) is a well-established mathematical
technique for data prediction, based on the identification of
correlations and pattern recognition in input training sets.
We present the application of ANNs  to predict the emission line luminosities
of H$\alpha$ and [NII] $\lambda$6584 in galaxies. These important spectral
diagnostics are used for metallicities,  active galactic nuclei (AGN)
classification and star formation rates, yet are shifted into the
infra-red for galaxies above $z \sim 0.5$, or may not be covered in
spectra with limited wavelength coverage.  The ANN is trained with a
large sample of emission line galaxies selected from the Sloan
Digital Sky Survey using
various combinations of emission lines and stellar mass.  The ANN
is tested for galaxies dominated by both star formation and AGN; in
both cases the H$\alpha$ and [NII] $\lambda$6584 line luminosities
can be predicted
with a scatter $\sigma <$ 0.1 dex.  We also show that the performance
of the ANN does not depend significantly
on the covering fraction, mass or metallicity
of the data.  Polynomial functions are derived that
allow easy application of the ANN predictions to determine H$\alpha$ and
[NII] $\lambda$6584 line luminosities.    An ANN calibration
for the Balmer decrement (H$\alpha$/H$\beta$) based on line equivalent
widths and colours is also presented.  The effectiveness of the ANN
calibration is demonstrated with an independent dataset (the Galaxy
Mass and Assembly Survey).  We demonstrate the application of our
line luminosities to the determination of gas-phase metallicities and
AGN classification.  The ANN technique
yields a significant improvement in the measurement of metallicities
that require [NII] and H$\alpha$ when compared with the function based conversions
of Kewley \& Ellison.  The AGN classification is successful for 86 per cent
of SDSS galaxies.

\end{abstract}

\begin{keywords}
Astronomical databases: catalogues- Methods: data analysis- Methods :statistical- galaxies: ISM
\end{keywords}

\section{Introduction}

The physical properties of stars and galaxies are encoded in their
spectral properties.  Transcribing observational properties such as
colour and spectral line characteristics, into quantities such as
stellar mass, velocity dispersion, dust, temperature, density and
chemical abundances is a cornerstone of astronomical techniques.
However, access to key diagnostic features is often limited by either
atmospheric transmission, instrumental capability (e.g. limited
wavelength coverage) or detection thresholds.  To circumvent these
imposed limitations, it is common practice to determine indirect
measures of the underlying physical quantities via calibrations with
more readily measured quantities.  An excellent example is the
determination of the gas phase metallicity in galaxies.  Ideally, the
metallicity is determined through the measurement of the electron
temperature.  This requires detection of emission lines with very
different excitation potentials, such as [OIII] $\lambda$ 4363 and
[OIII] $\lambda \lambda$ 4959, 5007, which in turn permits a solution of
the electron temperature, density and oxygen abundance
(e.g. \citealt{OF-06}).  However, the weak strength of the
[OIII] $\lambda$ 4363 line means that this `direct' technique is only
easily applied to relatively metal-poor galaxies.  A large number of
indirect metallicity calibrations, which use stronger emission lines,
have therefore been proposed (see \citealt{KE08} for a review),
based either on predictions from theoretical models, or calibration
from observational data.

Whereas the strong line metallicity calibrations are motivated by the
detection threshold of a weak spectral feature, other calibrations are
driven by wavelength coverage.  For example, one of the most common
techniques for deriving star formation rate (SFR) in local galaxies is via
the luminosity of H$\alpha$ (e.g. \citealt{Kennicutt-94}).  At $z>0.5$
this line shifts out of the optical window, motivating calibration
against bluer lines such as [OII] $\lambda$3727 (e.g. \citealt{Gallagher-89,
Kennicutt-92,Kewley-02,Kennicutt-12}). At a similar redshift, the
[NII]$\lambda$6584 line is also lost from the optical
window.  [NII] $\lambda$ 6584 is used in a number of diagnostics,
including some metallicity calibrations (e.g. \citealt{DTT-02,PP-04}) and the very widely used active
galactic nucleus (AGN) classification technique of \citet{BPT-81}, or the `BPT' technique.  Alternative AGN
classification schemes that can be applied to $z>0.5$ galaxies use a
combination of bluer emission lines, the 4000 \AA\ break (D$_{\rm n
4000}$), colour and mass  (e.g., \citealt{Stasinska-06,Juneau-11,TBT-11,Yan-11}). For example, in the absence of near-infrared spectroscopy,
\cite{Juneau-11} show that combining [OIII]$\lambda$ 5007/H$\beta$ and
stellar mass successfully distinguishes between star formation and AGN
emission.

Although the application of calibrated metallicities, SFRs and AGN
classification is straight-forward and has proven a tremendously
useful tool in the study of galactic properties, each diagnostic is
designed for a single purpose prediction.  In this paper, we
investigate the use of artificial neural networks (ANNs) for
predicting the raw line luminosities of H$\alpha$ and [NII]$\lambda$6584,
quantities which can then be applied in the user's choice of calibration.
Moreover, the line luminosities themselves contain information
about physical conditions within the galaxy such as ionization
parameter and dust extinction \citep{Groves-12,Brinchmann-08,Kewley-13a}, which can be determined from the ANN
predicted luminosities.

The artificial neural network is a non-linear method that `learns'
from training sets of data to predict the properties of other samples.
There has been an increasing interest in using ANNs in astronomy;
recent applications include star-galaxy discrimination and galaxy
classification \citep{Lahav-96,Andreon-00,Cortiglioni-01,Ball-04,Teimoorinia-12}. Essentially any
problem that is based on pattern recognition or correlations can be
tackled with ANNs.  The potential downside is that the ANN machinery
is complex and requires significant overhead to construct and train.
This is in contrast with many of the calibration approaches described
above that provide simple analytic formulae to predict physical
quantities based on linear correlations.

In this paper, we aim to combine the power of the artificial neural
network with the user-friendliness of an analytic solution in order to
provide ANN-calibrated equations that predict the luminosities of H$\alpha$
and [NII]$\lambda$ 6584.  The paper is presented as follows.  In
Section \ref{ann_sec} we describe the basics of regression techniques,
including their mathematical foundations and the specific cases of
the ANN and basis function regression.  Section \ref{balmer_sec} presents
a simple case study of how the ANN can be applied to spectral data.
The Balmer decrement (H$\alpha$/H$\beta$) of star-forming galaxies
in the Sloan Digital Sky Survey (SDSS) is predicted and compared
to the recently published linear calibration of \cite{Groves-12}.
Readers who are solely interested in the prediction of H$\alpha$
and [NII]$\lambda$6584 luminosities may wish to skip directly to Sections
\ref{sf_sec} and \ref{mixed_sec} where we present the ANN calibrations
for star-forming galaxies and emission line galaxies of unknown classification
respectively.  In Section \ref{fx_sec} we describe the widely applicable
polynomial and matrix-based forms of our ANN outputs.  The robustness
of the calibrations in tested in Section \ref{test_sec}, and the
application to gas-phase metallicities and AGN classification are demonstrated
in Section \ref{app_sec}.

Throughout the paper we assume a cosmology with $\Omega_{\rm tot}$,
$\Omega_{\rm M}$, $\Omega_{\rm \Lambda}$= 1.0, 0.3, 0.7  and H$_0$ =
70 km s$^{-1}$ Mpc$^{-1}$.  Line fluxes (corrected for Galactic extinction and underlying
stellar continuum),
stellar masses, equivalent widths
and other galactic parameters are
taken from the MPA/JHU catalogs (unless otherwise noted), with errors scaled
accordingly\footnote{The MPA/JHU catalogs, details on Galactic extinction
corrections and error scalings can
be found at http://www.mpa-garching.mpg.de/SDSS/DR7/raw\_data.html}.

\section{Regression methods}\label{ann_sec}

The modelling of data is a mathematical problem that is fundamental
in the scientific method.
Linear models are not always sufficient to describe physical
phenomena, so that non-linear representations may be required to
find more complicated relations between different parameters in a
given data set.  In particular, when there is insufficient information on the nature
of the data under study, or the function describing the connection
between the parameters of the data is not known \textit{a priori}, non-linear
methods are particularly powerful,

In general, regression methods deal with a function in the following
form:

 \begin{equation}
\textbf{Tar}=f(\textbf{A}_i,\textbf{X}_i).
\label{eq-reg}
 \end{equation}

 For a defined function with known input vectors \textbf{X$_i$} the
aim is to find parameters \textbf{A$_i$} in order to retrieve the
target vector \textbf{Tar}.  When the analytic form of $f$ is not
explicitly known, then linear or non-linear representations of the
function must be derived in order to identify the optimal form.  In
this regard, non-linear functions may be preferable, since they can
solve more complicated problems.  Selecting a special form for the
function (and optimizing the values of \textbf{A$_i$}) is referred to as
a regression method. Amongst regression methods,
artificial neural networks and basis function regression (BFR)
are two of the most commonly used.

\subsection{Artificial Neural Networks}

ANNs are widely used in a large variety of applications in various
disciplines. A neural network is, in essence, an attempt to simulate
the brain. It is a mathematical model that tries to simulate the
functionalities of biological neural networks. Its basic building
block is a simple mathematical model. A neural network is a computing
system made up of a number  highly
interconnected processing elements that are called neurons. An ANN
contains three layers: input, hidden and output, where
each layer has a different
number of  neurons. The input layer and output layers are connected to the
input and target data, respectively. The hidden layer connects these two layers
in a more complicated way. In this way, all possible combinations of the
input and the target data can be considered. Choosing the number of neurons
depends on the nature of the problem under study. Using this model, each
input datum is weighted  and then  added up with a bias, \textbf{{B}}, to
make a desired target, \textbf{{Tar}}, by an activation function, $f$.
A non-linear activation function introduces non-linearity
into the network, which  usually is selected as a sigmoid  function
(e.g., see Vanzella et al 2004). In vector form we have the following
relationship:

 \begin{equation}
\textbf{Tar}=f(\textbf{W*X$_{tr}$+B})
\label{weight-tr}
 \end{equation}

A network can contain several layers with different $f$, \textbf{W} and \textbf{B}. The output
of each layer can be considered as the input of the next layer. In this way, each element of the
input vector \textbf{X$_{tr}$}
(the input training data set) is connected to each neuron input through
the weight matrices \textbf{W}.  A network can be trained  using  supervised
or unsupervised methods (\citealt{Hagan-94,Kohonen-97}). In a
supervised mode, for example, \cite{Vanzella-04} use the magnitudes of
galaxies using the HDF-S and SDSS data  as  input data, \textbf{X$_{tr}$},
to train a network  using the  redshift information of the galaxies as
target values, \textbf{Tar}.

Briefly, after collecting the data (i.e., \textbf{Tar} and
\textbf{X$_{tr}$}) a network is created and then
configured. Configuration is the step in which the network is arranged
to be compatible with the problem under study (e.g., a fitting or a
classification problem). Initialization of the ANN parameters (weights
and biases) and training the network (finding suitable weights and
biases) are the next steps. The latter can be done using one of
several algorithms. For example, in a gradient descent algorithm, and
in an iterative way, the aim may be to improve (or upgrade) weights
and biases in the direction where a performance function (e.g., a mean
square error function) decreases most rapidly (i.e., in the direction
of negative gradient).  If the $W_{k}$ are the current weights and
biases and $G$ is the current gradient then the algorithm may be
written as:

 \begin{equation}
 W_{k+1}=W_{k}-\alpha_{k} G_{k},
 \label{algorithm}
 \end{equation}

where $\alpha$ is a learning rate that is an adjustable parameter.
$\alpha$ controls the size of the weight and bias changes during the
learning process. We use the Levenberg-Marquardt  optimization method
(Marquardt et al. 1963) to update weight and bias values. Finally, after
the validation of a network, it will be ready to use (or test) for new data.
In the validation step, the trained network (i.e., the network with defined
weights and biases) is generally tested  on an independent  dataset,
\textbf{X$_{va}$}, set to verify the network's performance is robust.

 \begin{equation}
\textbf{Out$_{va}$}=f(\textbf{W*X$_{va}$+B}).
\label{weight-va}
 \end{equation}

If the output of a network, \textbf{Out$_{va}$}, is close to the
actual values then this means that the network is well trained.
Generally, this also shows that there has been a good correlation
between \textbf{X$_{tr}$} and \textbf{Tar}. In some cases, we wish to
parameterize the mapping from D-dimensional inputs to the desired output. In
other words, we have multi-dimensional inputs and a known output.
The question of which input data-set (from $n$ available input
vectors) make this correlation the strongest is an important issue.
Specifically, in this work, we will consider which combinations of
strong emission lines, and stellar mass, can be best combined to give
a reliable prediction of the H$\alpha$ and [NII] line luminosities.

In equation \ref{weight-tr}, the input data generally contain different
pieces of physical information:

\begin{equation}
\bold{X_{tr}=X_{tr}(X_{m1}, X_{m2},..., X_{mn}}).
\end{equation}

In other words, \textbf{X$_{tr}$} can be a matrix containing $n$
(column) vectors. For example, it can comprise the information of
colour (as vector $\bold{X_{m1}})$ as well as information of
equivalent width (EW) of a certain emission line ($\bold{X_{m2}}$) of
a data set of $m$ galaxies.  Again, to give a concrete example, Groves
et al. (2012) show that the Balmer decrement ($H{\alpha}/H{\beta}$)
can be measured from the emission line EWs and colours. Therefore, in a
neural network approach, $\bold{X_1}$ and $\bold{X_2}$ can be
considered as EWs and colours, respectively.

\subsection{Basis Function Regression}

 In regression problems, models have the same basic form of
$Y=f(C_i,X_i)$ in which Y and X are target and independent input data,
respectively. In principle, $f$ can have any form. In any selected
form, coefficients $C_i$ should be found according to a minimization
algorithm. In a basis function regression we can select a polynomial
function in which $f(C,X)$ can, for example, be a second or third
order polynomial function.  In order to provide a user-friendly method
for the application of the ANN predictions, we will present the
polynomial functions that best encapsulate the networks'
predictions. To this end, we use a second order polynomial function
and a Levenberg-Marquardt algorithm to fit the data. If \textbf{X} is an
N-dimensional column vector, then we can write a quadratic function in
a general form as:

  \begin{equation}
f(C,X)=\sum\limits_{i,j=1}^N C_{ij}X_iX_j+\sum\limits_{i=1}^N C^{'}_iX_i+C^{''}.
\label{eq-basis-function}
 \end{equation}

Equivalently, in matrix form we have: $f(C,X)=X^{T} C X + X^{T} C^{'}
+ C^{''}$.  N is the number of physical parameters used as input data
and C$_{ij}$ is a symmetric matrix. For example, for a certain run
which utilizes three physical parameters such as X$_1$=mass,
X$_2$=OIII and X$_3=$H$\beta$, N=3 and there is a total of 10
independent coefficients to be determined. By this equation, we
consider different connections between the physical parameters in a
more complicated form.  We will give an example of using this equation
in the next section.

\section{A case study: An ANN calibration of the Balmer decrement}\label{balmer_sec}

In order to demonstrate the application of the ANN to spectral data, and
to test its performance against the analytic calibrations, we apply the
ANN technique to the prediction of the Balmer decrement
(H$\alpha$/H$\beta$).   Since the ANN predicts the line luminosities
that we would expect to measure in the observed spectrum, the
predicted H$\alpha$ luminosity can be combined with the observed H$\beta$
luminosity to determine the amount of dust extinction (see below).  The
determination of extinction is
a critical component of spectral analysis since, once determined,
a correction can be applied to other line fluxes.

The performance of the ANN can be compared
directly to the analytic prediction of the Balmer decrement
presented by \cite{Groves-12}. A major component of the work presented
here is the translation of the ANN solutions into polynomial equations
that can be simply applied by any user.  We therefore also present
and test a second order polynomial basis function that is based on the
ANN, i.e., Eq. \ref{eq-basis-function},
to find the associated coefficients that can be used to predict
H$\alpha$/H$\beta$ for a given set of input.

Under the assumption of either an optically
thin or thick medium, and the temperature of the region, it is
possible to predict the intrinsic ratio of the Balmer emission line
series (e.g. \citealt{DS-03,OF-06}). The observed ratio of fluxes can
therefore be combined with predictions
from extinction laws (e.g. \citealt{CCM-89,Pei-92,Calzetti-94}) to
derive the attenuation through a particular region.
As described in detail by \cite{Groves-12}, the accurate measurement
of H$\alpha$ and H$\beta$ line fluxes requires sufficiently high S/N
and resolution to be able to account for underlying absorption
in the stellar continuum.  Equivalent widths, however, are relatively
easy to measure, but do not account for the required absorption corrections.
For this reason, some studies (e.g. \citealt{KP-03}) have
proposed the use of EWs instead of fluxes, with a fixed correction
for underlying stellar absorption.  A more sophisticated approach
is proposed by \cite{Groves-12}.
Based on a calibration of H$\alpha$ and H$\beta$ EWs, continuum-corrected
fluxes and colours, \cite{Groves-12} provide a calibration between
EWs and the Balmer decrement that statistically accounts for the stellar
absorption.

Based on a sample of emission line (star-forming and AGN dominated)
galaxies in the SDSS, the analysis of Groves et al. (2012) determines
the Balmer decrement
by first quantifying an absorption correction factor to the H$\alpha$
and H$\beta$ EWs, and then uses the rest-frame fibre $g-r$ colour as
a proxy for
the stellar continuum.  Groves et al. (2012) hence determine the
following prediction of the Balmer decrement (where EWs are in \AA):

\begin{equation}
\rm{log}(H{\alpha}/H{\beta})
=\rm{log}[\frac{EW(H{\alpha})+A_{1}}{EW(H{\beta})+A_{2}}]+[A_3(g - r)+A_4].
\label{formula-Groves}
\end{equation}

Generally, the absorption EW is not the same for all the Balmer lines,
i.e., A$_{1}$ $\neq$ A$_{2}$. \cite{Groves-12} use a constant Balmer
absorption correction for both EW(H$\alpha$) and EW(H$\beta$). In this
case a correction factor, which depends also on the EW range of the sample,
is derived.  The correction factor is determined from minimizing the offset
between the SDSS galaxies'
EW Balmer decrements and the measured H$\alpha$/H$\beta$ (with an
error-based weighting scheme, as described in \cite{Groves-12}.

Here, we select a subset of the full data that have $10<\rm{EW(H}\alpha)<30$
\AA, for which A$_1$= A$_2$ = 3.5 \AA\ (see also Fig. 4 of Groves et
al. 2012).  The values of A$_3$ and A$_4$ are determined from the
best-fit between continuum
fluxes measured at  the H$\alpha$ and H$\beta$ wavelengths and fibre
rest-frame $g-r$ colour, and are found by Groves et al. (2012)
to be 0.39 and $-0.26$, respectively.

\begin{figure}
\centering
\includegraphics[width=8cm]{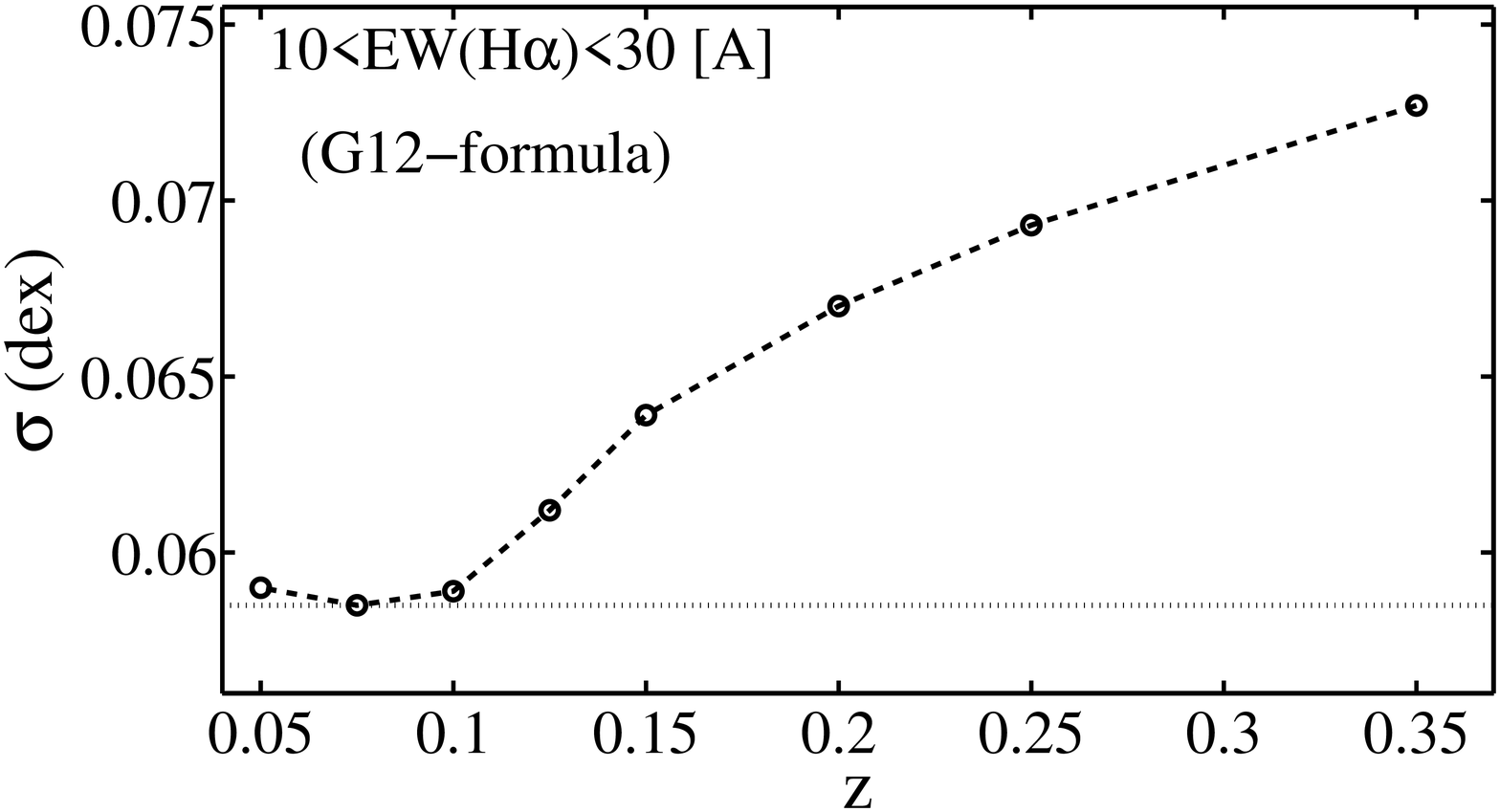}
\includegraphics[width=8cm]{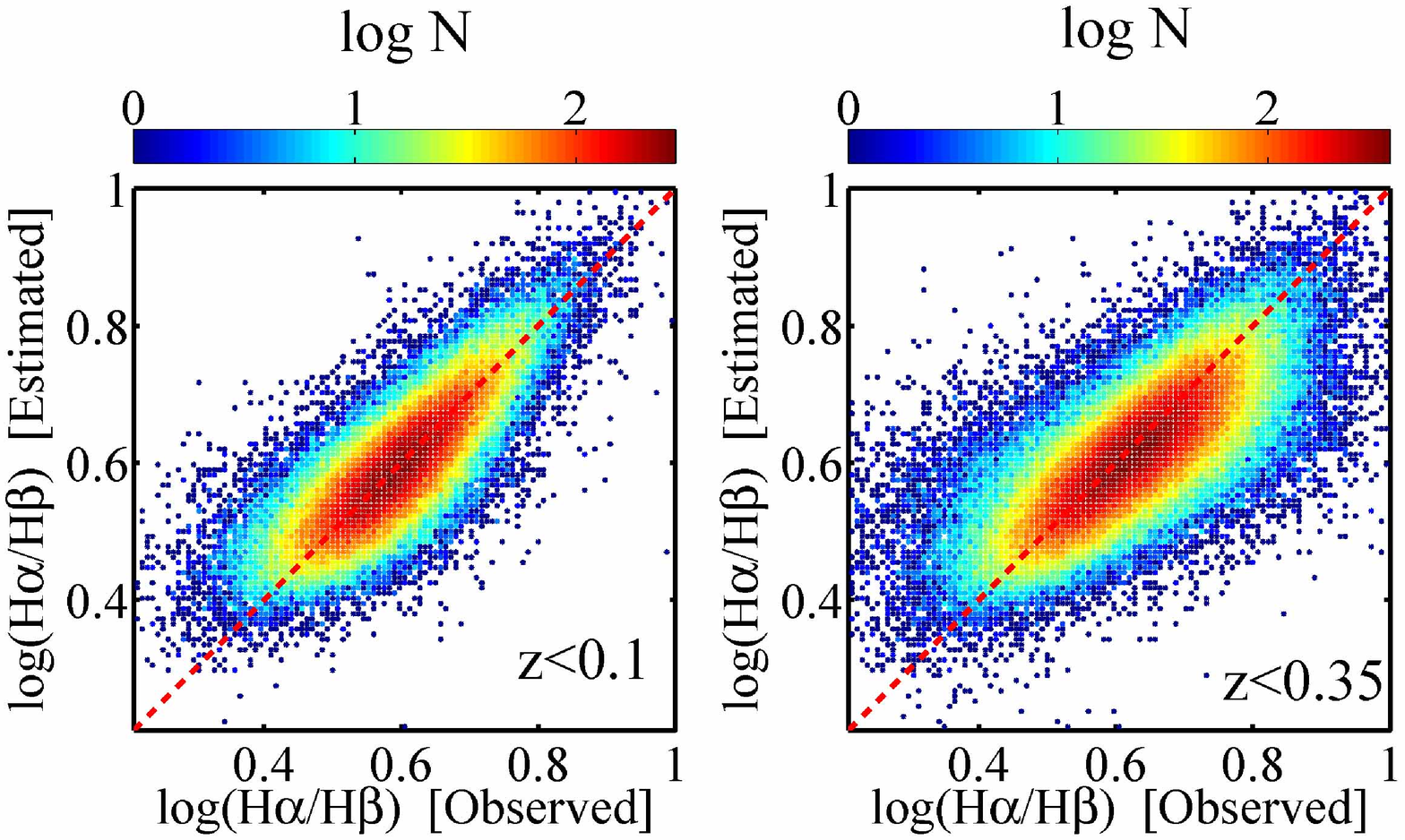}
\caption{The top panel: the standard deviation   between the
predicted log(H$\alpha$/H$\beta$) and observed ratio versus redshift.
The horizontal dashed line shows scatter obtained by  Equation
\ref{formula-Groves} from Groves et al. (2012).
The lower panels: the estimated log(H$\alpha$/H$\beta$) obtained from
Equation \ref{formula-Groves} versus the value presented in the MPA/JHU
catalogs for two cases,  $z<0.1$ (left) and $z<0.35$ (right).
The colours and associated
colour bars indicate the number of galaxies in each coloured point.}
\label{groves}
\end{figure}

It is useful to begin our analysis with a test of the
Balmer decrement prediction from Eq. \ref{formula-Groves} based
on EWs extracted from the MPA/JHU catalogs, observed Petrosian
magnitudes obtained from the SDSS and our own corrections to
rest-frame quantities.  Using Eq. \ref{formula-Groves}, in Fig. \ref{groves}
we plot the standard deviation between the predicted ratio of
log(H$\alpha$/H$\beta$)
and the observed ratio as a function of redshift.  The top panel of
Fig. \ref{groves} shows that the scatter agrees with that quoted by
\cite{Groves-12} for $z < 0.1$ and this EW range: $\sigma = 0.058$ dex. However, for higher redshifts, the scatter in the comparison between the
observed and predicted Balmer decrements increases steadily.
This effect is highlighted in the lower two panels of Fig.
\ref{groves} which compares the observed and predicted ratio of
log(H$\alpha$/H$\beta$) in two redshift ranges,   $z<0.1$ (left)
and $z<0.35$ (right).   These panels include all emission
line galaxies with a S/N$>$3 (based on the criterion applied by Groves
et al. 2012) in each of the H$\alpha$ and H$\beta$
lines.  The colours and associated colour bar indicate the number
of galaxies in each coloured point.  We
speculate that the decline in performance of Eq. \ref{formula-Groves}
for our dataset might be caused by a difference in calculation
of rest-frame colours.  Regardless of the source of this redshift dependence,
in order to fairly compare the performance
of the ANN method and that of Groves et al. (2012), we simply restrict
ourselves to the redshift range $z<0.1$.

\subsection{The ANN and basis function approaches}

We will now compare the performance of the ANN and also the basis
function (polynomial) method to the same application,
namely the prediction of the Balmer decrement based on the observed
equivalent widths of H$\alpha$ and H$\beta$ and rest-frame $g-r$ colour.
The artificial neural network is calibrated by sets of training data
that are defined by the vector \textbf{X$_{tr}$}, see equation
\ref{weight-tr}.  For example, to replicate the input data of the
\cite{Groves-12} analysis we define X$_1$=log(EW(H$\alpha))$,
X$_2$=log(EW(H$\beta$)), X$_3$=$g$, and X$_4$=$r$, as four individual components
of \textbf{X$_{tr}$}. These four components (N=4) can be also used in
Equation \ref{eq-basis-function} to find the 15 independent coefficients of this equation.
At the start of the training procedure, the network
must be initialized and the choice of the initialization parameters
may affect the outcome of the trained network.  We therefore
repeat the training procedure 20 times
with different initializations and average the results for the 10 best
trained networks in terms of performance.

We begin with the same sample used by \cite{Groves-12} described
above: all emission line galaxies in the SDSS DR7 in the MPA/JHU
catalogs with H$\alpha$ and H$\beta$ S/N$>$3, which yields a sample of
$\sim$ 95,000 galaxies. Of the $\sim$ 95,000 galaxies in the sample
we randomly select half of them for training the network and the rest
are used as a validation set. The validation step compares the
performance of the ANN on target data (i.e. data not used in the
training step) to ensure that accuracy is preserved in its
application.  A demonstration of the training and validation
sets is shown in Section \ref{test_size_sec}.

\begin{figure}
\centering
\includegraphics[width=7.4cm,height=3.4cm,angle=0]{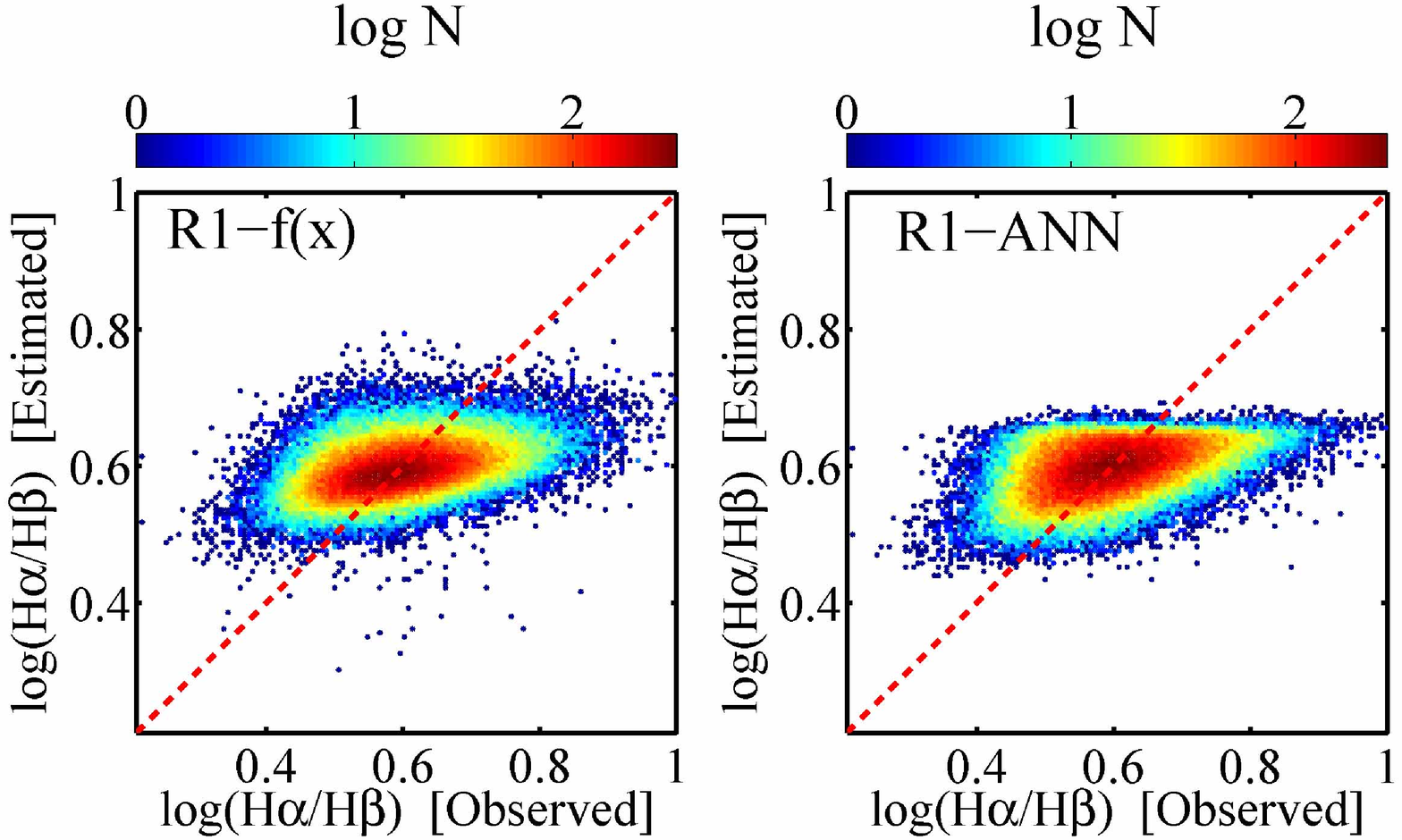}
\includegraphics[width=7.4cm,height=3.4cm,angle=0]{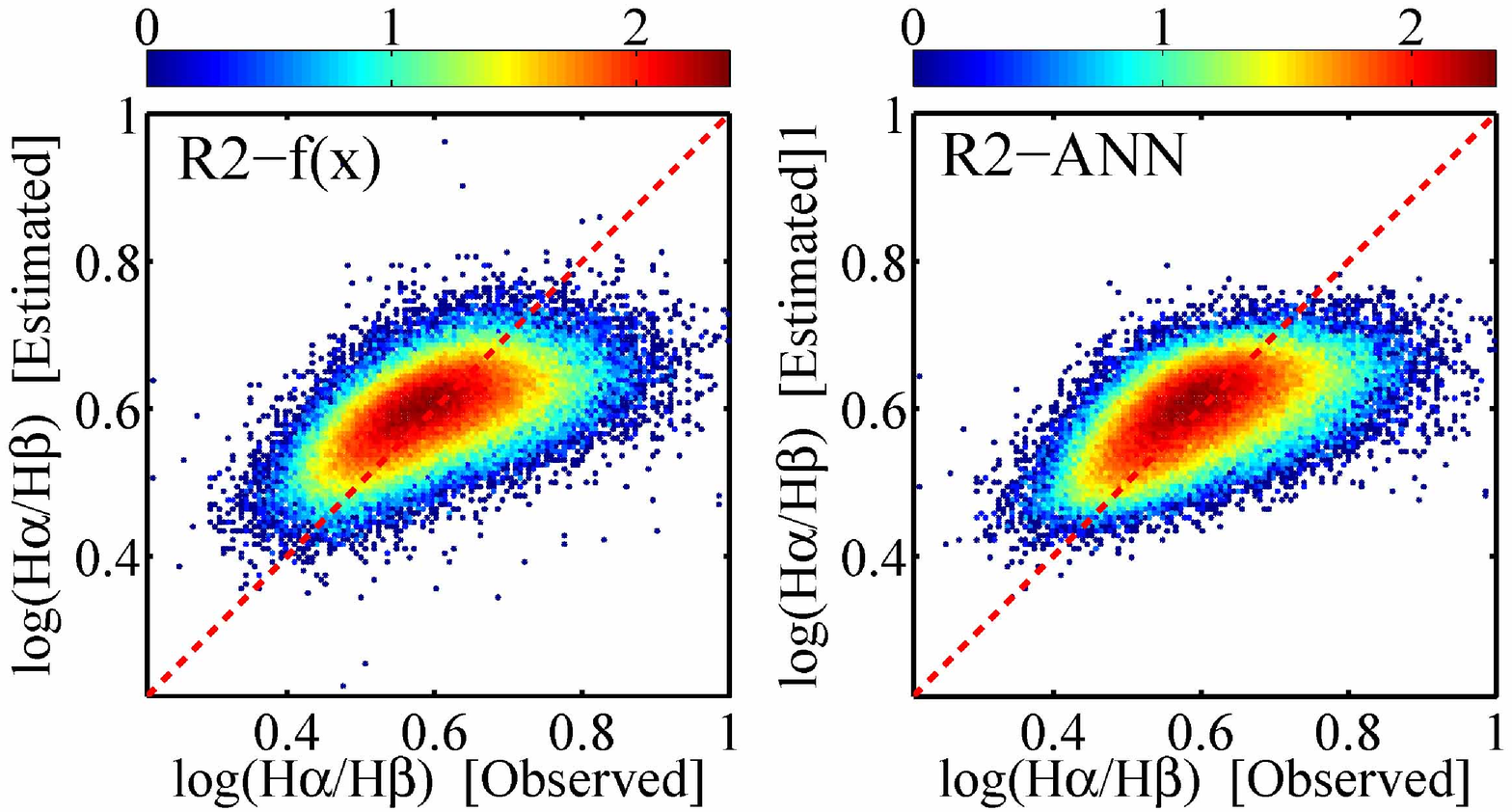}
\includegraphics[width=7.4cm,height=3.4cm,angle=0]{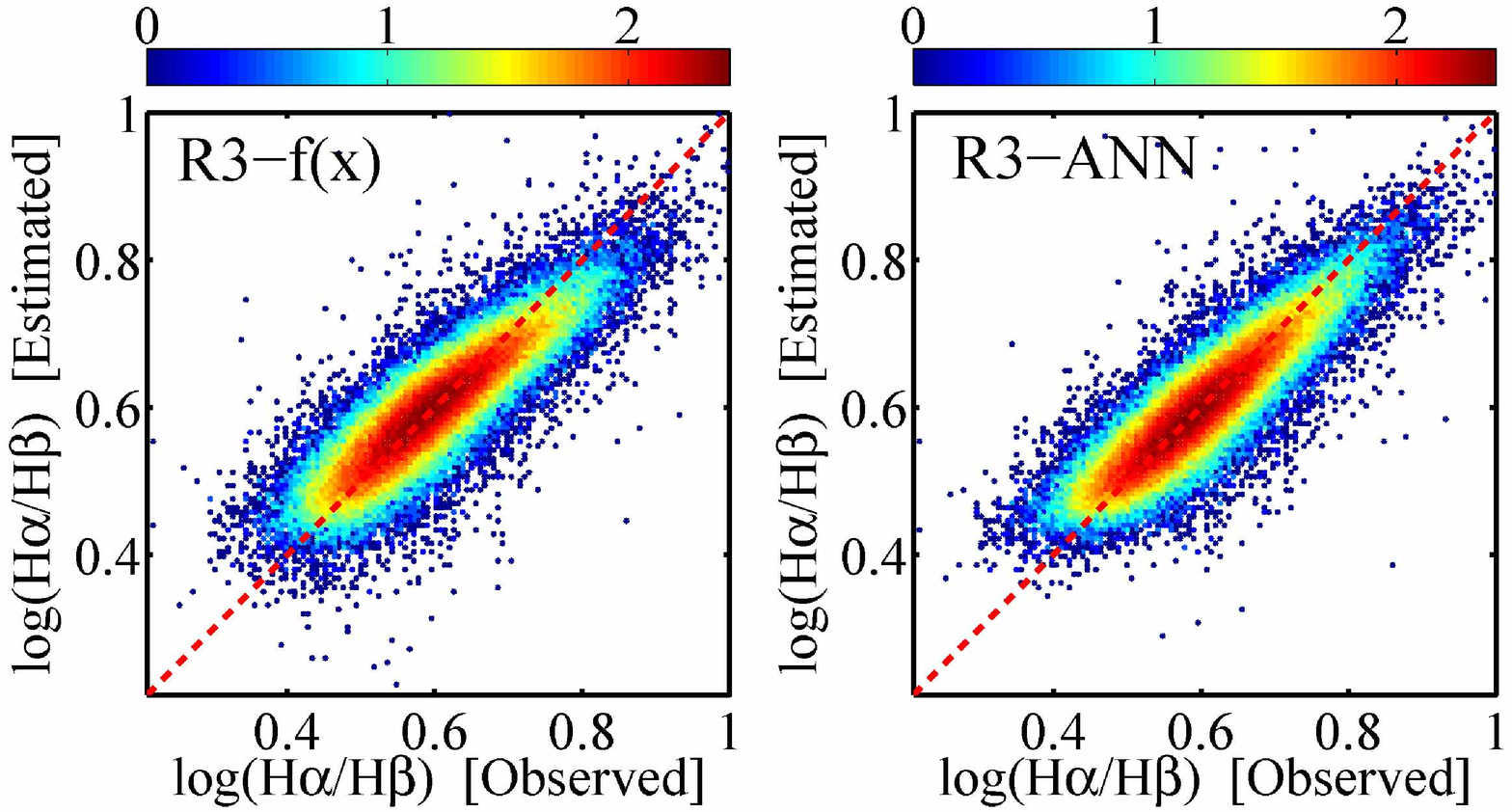}
\includegraphics[width=7.4cm,height=3.4cm,angle=0]{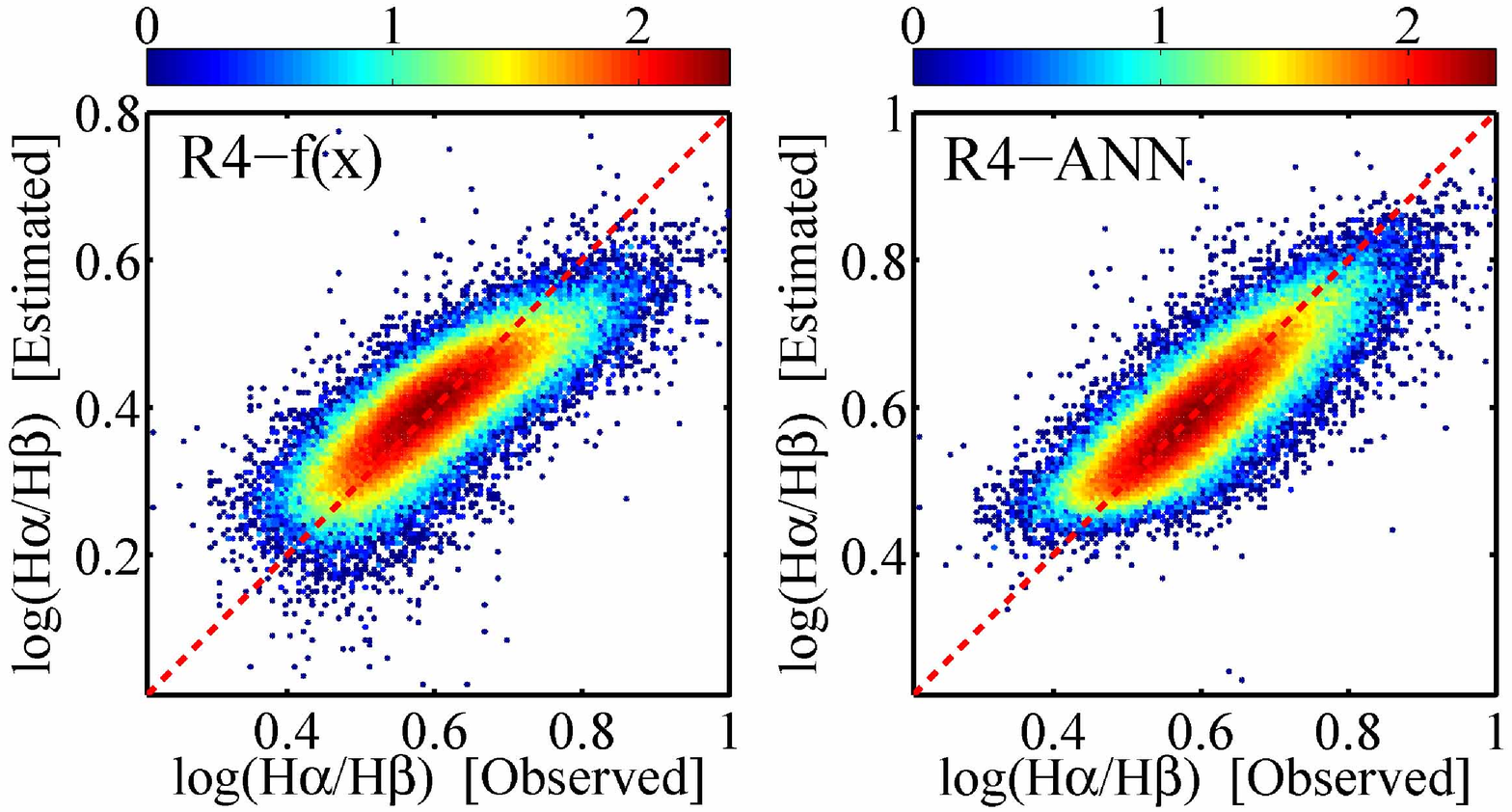}
\includegraphics[width=7.4cm,height=3.4cm,angle=0]{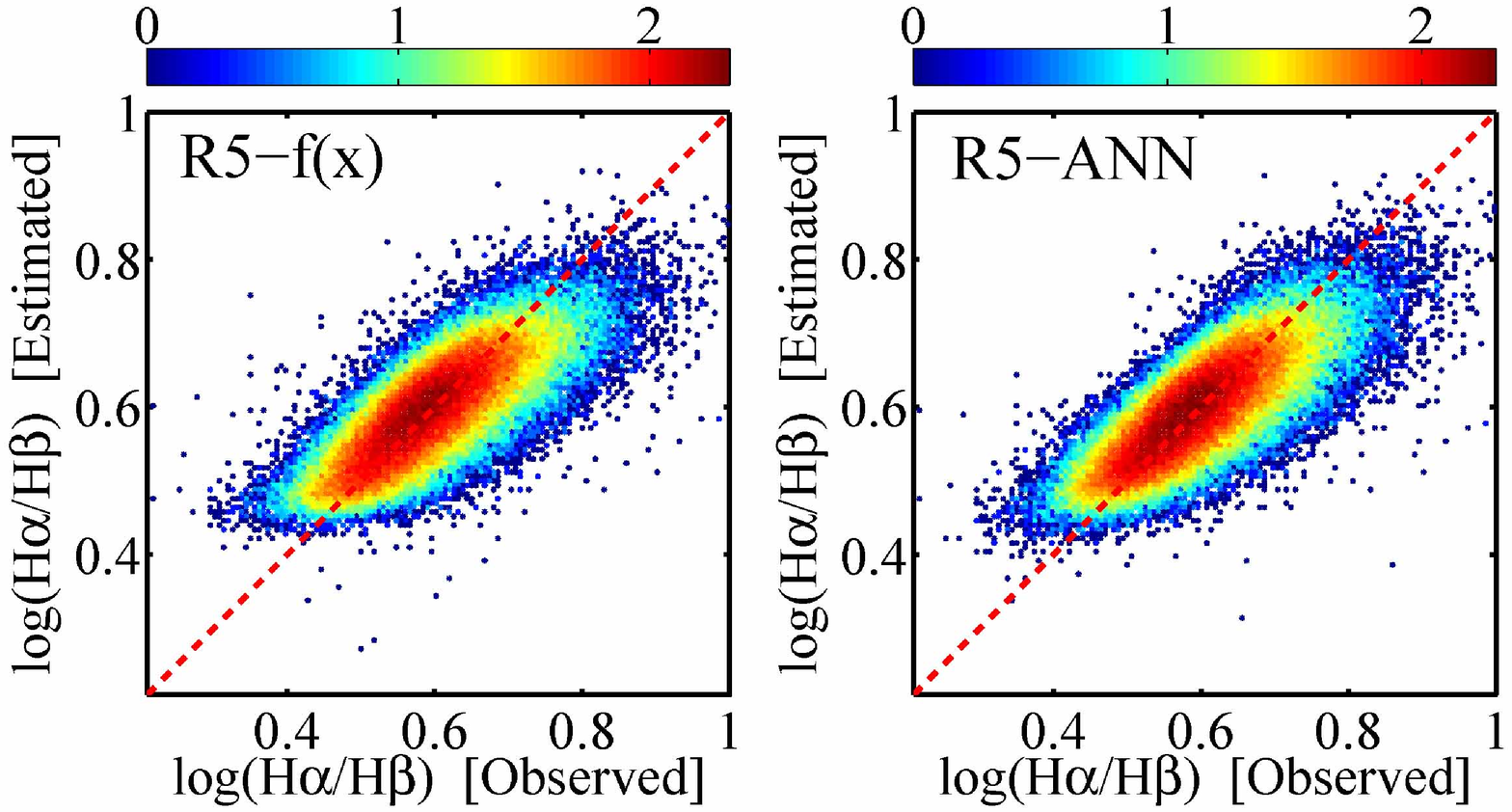}
\includegraphics[width=7.4cm,height=3.4cm,angle=0]{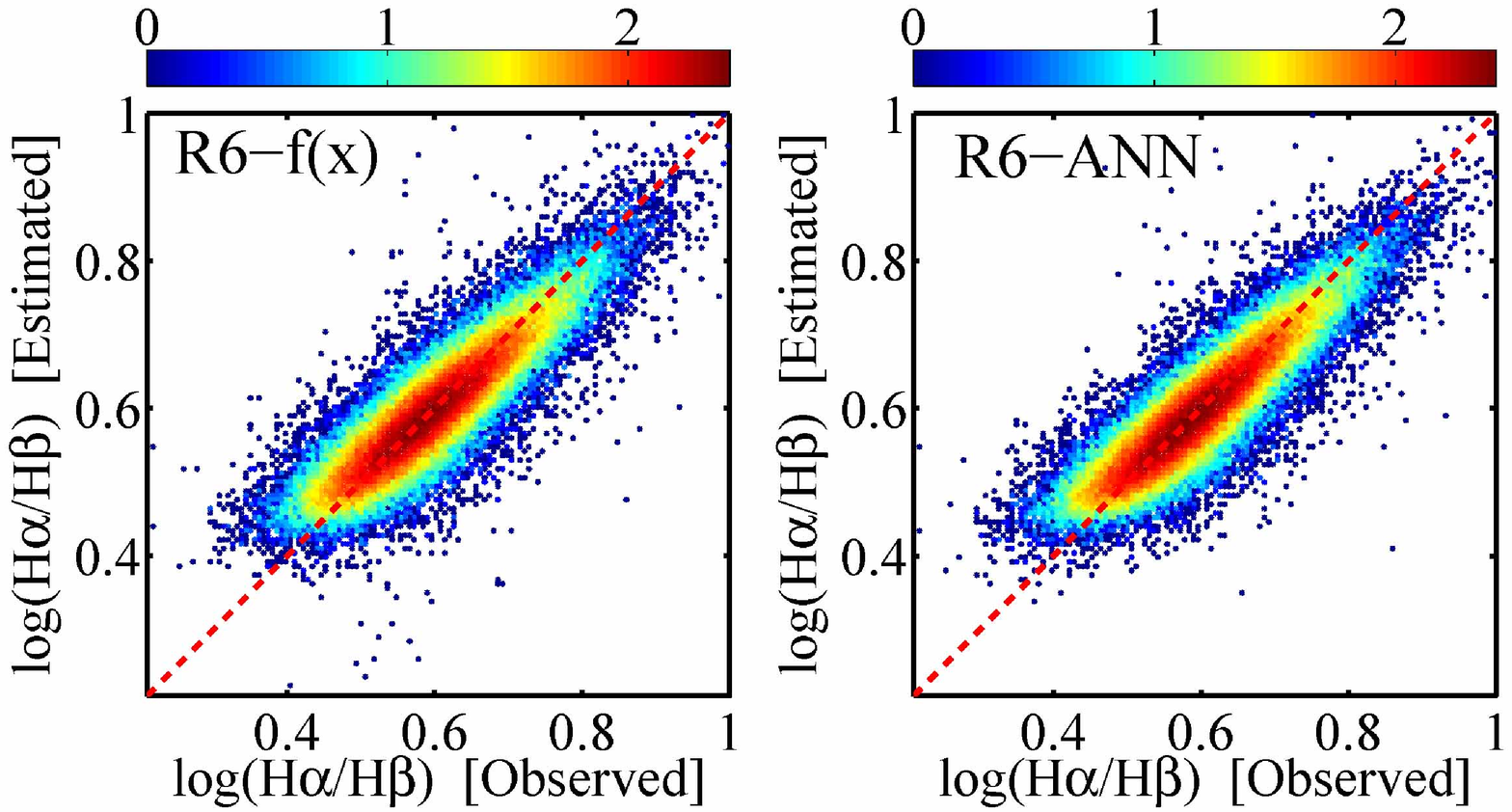}
\caption{Each panel shows the value of log(H$\alpha$/H$\beta$) estimated
from the basis function (left panel) and ANN (right panel) versus the measured value in the MPA/JHU catalog.
The figure comprises the six different runs listed in
Table \ref{balmer_runs}.  Run 3 uses the same input data as the
calibration of Groves et al. (2012), using  X$_1$=log(EW(H$\alpha))$ and
X$_2$=log(EW(H$\beta$)), $g$, and $r$, i.e. the panel labelled `R3-ANN' is
directly comparable with Fig. \ref{groves}.
The colours and associated colour bars indicate the number of galaxies in
each data point.   }
\label{ann_balmer}
\end{figure}

The ANN is set up with six different runs, which represent six different
sets of training data.  Although the default comparison with \cite{Groves-12} uses EW(H$\alpha)$, EW(H$\beta$), $g$ and $r$, it is
instructive to investigate other parameter combinations.  The data used
in the input runs are listed in Table \ref{balmer_runs}, where run
R3 uses the same set of input data as Groves et al. (2012).  Fig.
\ref{ann_balmer} compares the performance of the ANN and
basis function methods, with each row representing a different
combination of input parameters, summarised by the runs in
Table \ref{balmer_runs}.  The ANN
predicted values are compared to those measured in the MPA/JHU
catalogs in the right panels of Figure \ref{ann_balmer}. The left panels
show the results obtained from the basis function approach,
denoted by f(x).

\begin{table}
\begin{center}
\caption{Balmer decrement training runs; all magnitudes are extinction
corrected, absolute, rest-frame
 and
calculated within the SDSS fibre. EWs are in \AA.}
\begin{tabular}{c|c}
\hline
Runs & Input vectors (X)  \\
\hline
R1  & log(EW(H$\alpha$)), log(EW(H$\beta$)) \\
R2  &  log(EW(H$\alpha$)), log(EW(H$\beta$)),$r$ \\
R3  &  log(EW(H$\alpha$)), log(EW(H$\beta$)),$g, r$  \\
R4  &  log(EW(H$\alpha$)), log(EW(H$\beta$)),$i, z$  \\
R5  &  log(EW(H$\alpha$)), log(EW(H$\beta$)),log(M$_{\rm fib}/M_\odot)$ \\
R6  & log(EW(H$\alpha$)), log(EW(H$\beta$)),$u, g, r, i, z$ \\
\hline
\end{tabular}
\label{balmer_runs}
\end{center}
\end{table}

\begin{figure}
\centering
\includegraphics[width=8cm,height=4.4cm,angle=0]{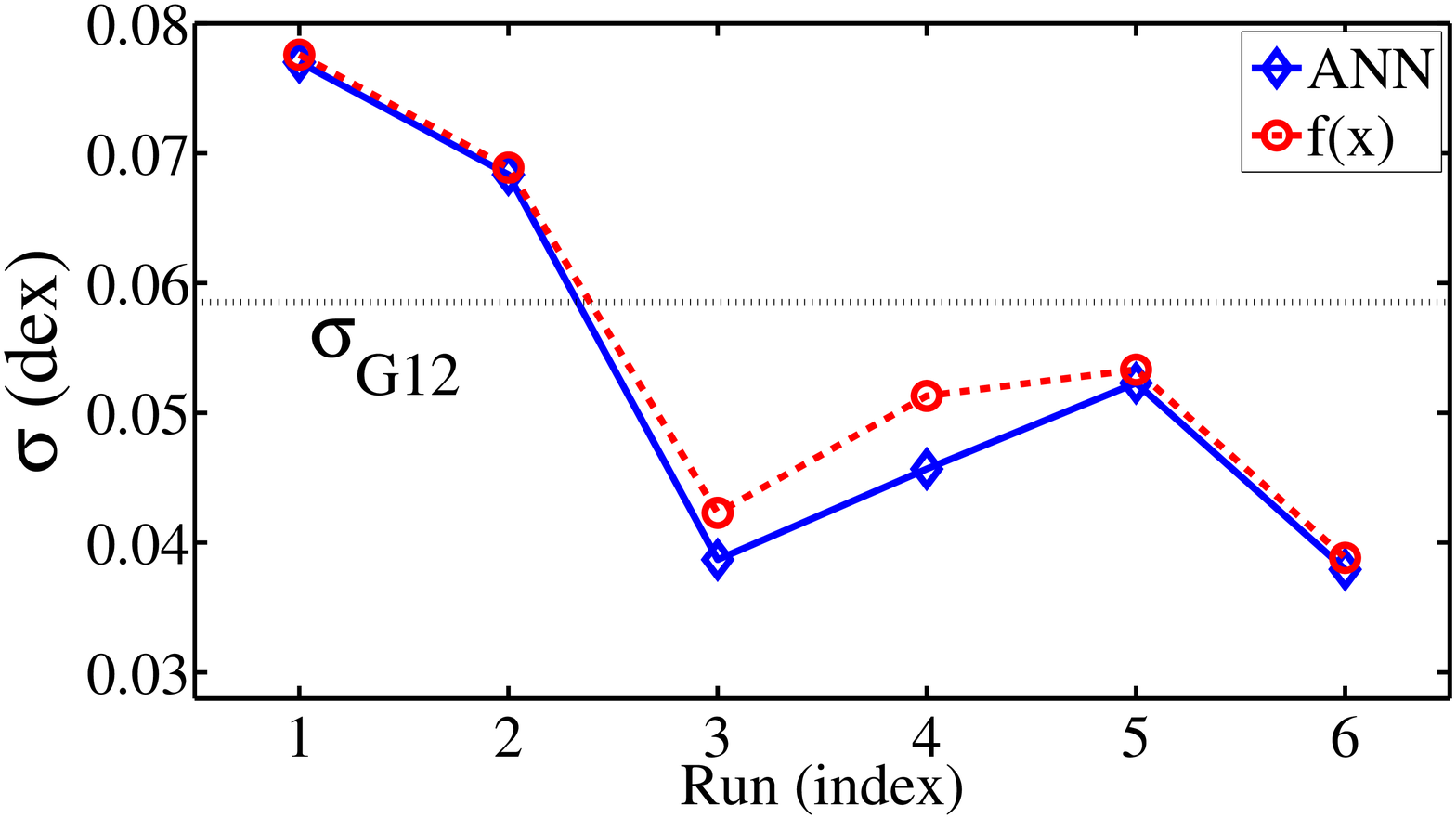}
\caption{The standard deviation of the six different runs listed in Table
\ref{balmer_runs} and shown in Fig. \ref{ann_balmer} for ANN (blue solid line)
and the basis function (red dashed line).
For comparison, the standard deviation of the Groves et al. (2012) calibration is shown
as a horizontal dotted line.  Runs whose standard deviations are below
this line (R3 -- R6) show a smaller scatter between the ANN predictions
and measured values than Groves et al. (2012).}
\label{balmer_sd}
\end{figure}

The top panel of Figure \ref{ann_balmer} (run R1) shows that an ANN with
input data \textbf{X$_1$}=log(EW(H$\alpha$)) and
\textbf{X$_2$}=log(EW(H$\beta$)) alone does not contain sufficient
information to accurately reproduce the target value,
H$\alpha$/H$\beta$.  The standard deviation for this run is relatively
large, almost 0.08 dex.  Adding information about the brightness of the
galaxy, in the case of run R2, the $r$-band  absolute magnitude,
does not significantly improve the situation.  Run R3 has the same set
of input data as the calibration of Groves et al. (2012):
EW(H$\alpha$), EW(H$\beta$), $g$ and $r$ (absolute magnitudes) and now the third panel from top
of Figure \ref{ann_balmer} shows a good correlation between the ANN
prediction and measured Balmer decrement.  As described by \cite{Groves-12},
this is expected as the colour provides information on the stellar
continuum.  The standard deviation for run R3 using the ANN
is 0.038 dex, compared
to 0.058 dex for the \cite{Groves-12} predictions from equation
\ref{formula-Groves} (Figure \ref{groves}).  Although this improvement
is incremental, it does demonstrate the power and applicability of the
ANN approach. The ANN also has the advantage of avoiding an explicit
fit between continuum measured at the H${\alpha}$ and H${\beta}$
wavelengths and colour; the coefficients A$_{3}$ and
A$_{4}$ are produced naturally as part of the ANN output.
Run R4 substitutes $i$ and $z$-band magnitudes for the
$g$ and $r$ used in R3; the correlation is still good ($\sigma$ = 0.046 dex),
but not as good as in R3.  Using the fibre stellar mass (also
from the MPA/JHU catalog) instead of colour information also performs
more poorly than R3 with a $\sigma =0.052$ dex.   The run with the
least scatter between the
predicted and observed values is R6, which uses all 5 SDSS photometric
bands.  Run R6 performs marginally better
than R3 with a standard deviation of 0.037 dex.

Figure \ref{balmer_sd} summarises
the performance of each of the six runs; for each run
the standard deviation between the predicted and measured Balmer
decrement is plotted.   The value of the \cite{Groves-12} calibration is
shown for reference as a horizontal dotted line. The blue solid
 and red dashed lines show the result obtained from ANN
and f(x), respectively.   Figure \ref{balmer_sd} shows that runs R3 -- R6
all yield lower scatter than the calibration of \cite{Groves-12}.
Figure \ref{balmer_sd} also shows that the scatter for each ANN run is
slightly smaller than the result obtained by the basis function, although
the difference is always less than 0.01 dex.   The accuracy of predicted Balmer decrements obtained from
both ANN and f(x) approaches can clearly match (or exceed) that
of the formula presented by Groves et al. (2012).

The coefficients derived for runs R3 and R5 (the latter of
which is useful in the
absence of good optical photometry) are given below.  All EWs
are in \AA.  Due to the
large number of coefficients, a matrix representation of
Eq. \ref{eq-basis-function} is most convenient.
These coefficients therefore permit a straightforward analytic method
for calculating the Balmer decrement. For run R3, with X as:

\[ X= \left( \begin{array}{c}
\rm{log~EW(H\alpha)}\\
\rm{log~EW(H\beta)}\\
\rm{g}\\
\rm{r}\\
\end{array} \right)\]

and where $f(C,X)=X^{T} C X + X^{T} C^{'} + C^{''}$, C and C$^{'}$
are given by the following matrices:
\[ C= \left( \begin{array}{rrrr}
0.32005&-0.10277&0.14572&-0.14146\\
-0.10277&-0.03811&-0.08066&0.07727\\
0.14572&-0.08066&-0.03282&0.03883\\
-0.14146&0.07727&0.03883&-0.04356\\
\end{array} \right)\]
\[ C^{'}= \left( \begin{array}{r}
-0.38036\\
0.11262\\
0.32623\\
-0.30305\\
\end{array} \right)\]
and $C^{''}$=0.51103.\\

The scatter found using this relation to predict the Balmer decrement is $\sigma$=0.042 dex.\\

For run R5 with X as:
\[ X= \left( \begin{array}{c}
\rm{log~EW(H\alpha)}\\
\rm{log~EW(H\beta)}\\
\rm{log~(M_{fib}/M_\odot)}\\
\end{array} \right)\]
The coefficients are:

\[ C= \left( \begin{array}{rrr}
0.09724&-0.05493&0.04130\\
-0.05493&-0.03587&-0.01420\\
0.04130&-0.01420&0.02065\\
\end{array} \right)\]
\[ C^{'}= \left( \begin{array}{r}
-0.74362\\
0.29786\\
-0.39319\\
\end{array} \right)\]
$C^{''}= $2.28694, yielding a scatter of $\sigma=$0.053 dex.


The above formats are particularly well suited to matrix oriented software
(such as IDL and MATLAB), although the coefficients $C_{ij}$, $C^{'}_{i}$
and $C^{''}$ also can be directly used by considering the summation
form of Eq. \ref{eq-basis-function}.   For presentation purposes,
we will present coefficients in the matrix format for the remainder
of the paper.

\section{The prediction of [NII] and H$\alpha$ luminosities in star-forming galaxies}\label{sf_sec}

After the demonstration  in the previous section of our ANN technique,
we now move to the main objective of the current work: a prediction
of the H$\alpha$ and [NII]$\lambda 6584$ luminosities\footnote{The ANN
was also tested for the prediction of line fluxes, as well as luminosities.
However, the range in redshift in the SDSS was sufficient to increase the
scatter in the predicted fluxes, so luminosities are adopted for this work.}.
The methodology is the same
as in the previous section.  First a series of training runs are established,
each one with a different set of parameters to be used in the network.  The
training sets include half of the available number of galaxies.  The
other half is used in the verification phase, where we confirm that
the ANN predicts the line luminosities with the expected scatter. Finally,
in Section \ref{fx_sec} and Appendix \ref{coeff_sec}, we provide
analytic calibrations and matrix coefficients that
represent the ANN output to be used by the general community.

We begin by considering only star forming galaxies in this section,
but in the next section we go on to apply our technique to the general
sample of emission line galaxies, regardless of their classification.
The [NII]/H$\alpha$ ratios differ markedly between the star-forming
and AGN populations due to the more significant contribution of
forbidden lines in the presence of a hard
ionizaing background.  The different line flux ratios in star-forming
regions compared with AGN forms the basis for many of the well-known
AGN BPT classification schemes.  When only star-forming galaxies are
considered, there is a relatively tight relationship between the
[OIII]/H$\beta$ ratio and that of [NII]/H$\alpha$, which is largely
driven by a trend in metallicity (e.g. Dopita et al. 2006; Stasinska
et al. 2006).  This tight sequence, as we will see, lends itself to
excellent predictions of [NII] and H$\alpha$ luminosities in star-forming
galaxies.

In this section we select all galaxies from the SDSS that are
classified as star-forming according to the calibration of Stasinska
et al. (2006) and have available total and fibre stellar masses in the
MPA/JHU catalog.  Line fluxes have been corrected for Galactic extinction,
but not internal extinction, since this is essentially calibrated as
part of the ANN process. Before the imposition of any S/N requirements,
these basic criteria yield
120,444 galaxies.  Twelve  separate training runs are established,
containing between 1 and 8 parameters in different combinations, see Table
\ref{sf_runs}.  Each of the 11 runs between R2 and R12 requires at least
one emission line so the ANN is tested separately for 3 different
S/N criteria: S/N$>3$, S/N$>5$ and S/N$>8$.   R1 uses only fibre
stellar mass so does not need a S/N requirement.  For a given S/N
requirement, only the lines required as input in that run are required
to pass the criterion.  For example, in R2 only [OIII]$\lambda$ 5007
is required to pass the S/N criterion, and the same is true in run R5.
However, in run R8, both [OIII]$\lambda$ 5007 and H$\beta$ are required
to pass the criterion.  Table \ref{run_sizes} lists the number of galaxies
in each sample, according to the run number and S/N threshold.  Recall
that half of these samples will be used in the ANN training process.

\begin{table}
\begin{center}
\caption{The ANN run parameter combinations.}
\begin{tabular}{c|c}
\hline
Runs & Input vectors log(X)  \\
\hline
R1  &  M$_{\rm{fib}}/M_\odot$ \\
R2  &  OIII$\lambda$ 5007\\
R3  &  OII$\lambda$3727 \\
R4  &  H$\beta$ \\
R5  &  M$_{\rm{fib}}/M_\odot$ + OIII$\lambda$5007 \\
R6  &  M$_{\rm{fib}}/M_\odot$ + OII$\lambda$3727 \\
R7  &  M$_{\rm{fib}}/M_\odot$ + H$\beta$ \\
R8  &  M$_{\rm{fib}}/M_\odot$ + OIII$\lambda$5007 +  H$\beta$ \\
R9  &  M$_{\rm{fib}}/M_\odot$ + OII$\lambda$3727 + H$\beta$ \\
R10  & M$_{\rm{fib}}/M_\odot$ + OIII$\lambda$5007+ OII$\lambda$3727 + H$\beta$ \\
R11  &  R10+  OII$\lambda$3729, OIII$\lambda$4959  \\
R12  &  R11+  SII$\lambda\lambda 6717, 6731$ \\
\hline
\end{tabular}
\label{sf_runs}
\end{center}
\end{table}

\begin{table}
\begin{center}
\caption{The  number of galaxies in each run with different S/N thresholds
for star-forming galaxies.}
\begin{tabular}{c|c|c|c}
\hline\hline
Runs & N$_{gal}$ (S/N$>3$)& N$_{gal}$ (S/N$>5$)& N$_{gal}$ (S/N$>8$)\\
\hline
R1  & 120444 & 120444&120444\\
R2  & 93365 &69062&48249\\
R3  & 79309  &50262&26660\\
R4  & 118383 & 110186&87838\\
R5  & 93365 & 69062&48249\\
R6  & 79309 &50262&26660\\
R7  & 118383 &110186& 87838\\
R8  & 92970 &68149&46169\\
R9  & 79268 &50246&26650\\
R10 &76329  &48989&26309\\
R11 &47368  & 29143&16481\\
R12 &46910  & 28841&16270\\
\hline
\end{tabular}
\label{run_sizes}
\end{center}
\end{table}

The different training runs R1 -- R12 are chosen to both provide insight
into the most relevant parameters required for an accurate prediction of
the [NII] and H$\alpha$ luminosities, but also with consideration of the most
likely practical choices.  For example, we investigate the impact of
S/N on the ANN, and whether performance is improved when both members of
a given doublet (such as [OII] $\lambda \lambda$ 3727, 3729) are used.
Preference is given to relatively strong emission lines with blue
rest wavelengths, under the assumption that these will be the most readily
detectable in optical spectra.  We do, however, include [SII] $\lambda
\lambda$ 6716, 6731 in the last of the runs.  Although this doublet
is redwards of H$\alpha$ and [NII] $\lambda$ 6584, run R12 could
potentially be useful in the case of one of the lines being contaminated,
for example by telluric features.  Mass is also included as an input
parameter in some of the runs, due to its correlation with many
other galaxy properties (such as SFR and metallicity) that in turn
correlate with emission line strengths (or ratios).

\begin{figure}
\centering
\includegraphics[width=7.4cm,height=3.4cm,angle=0]{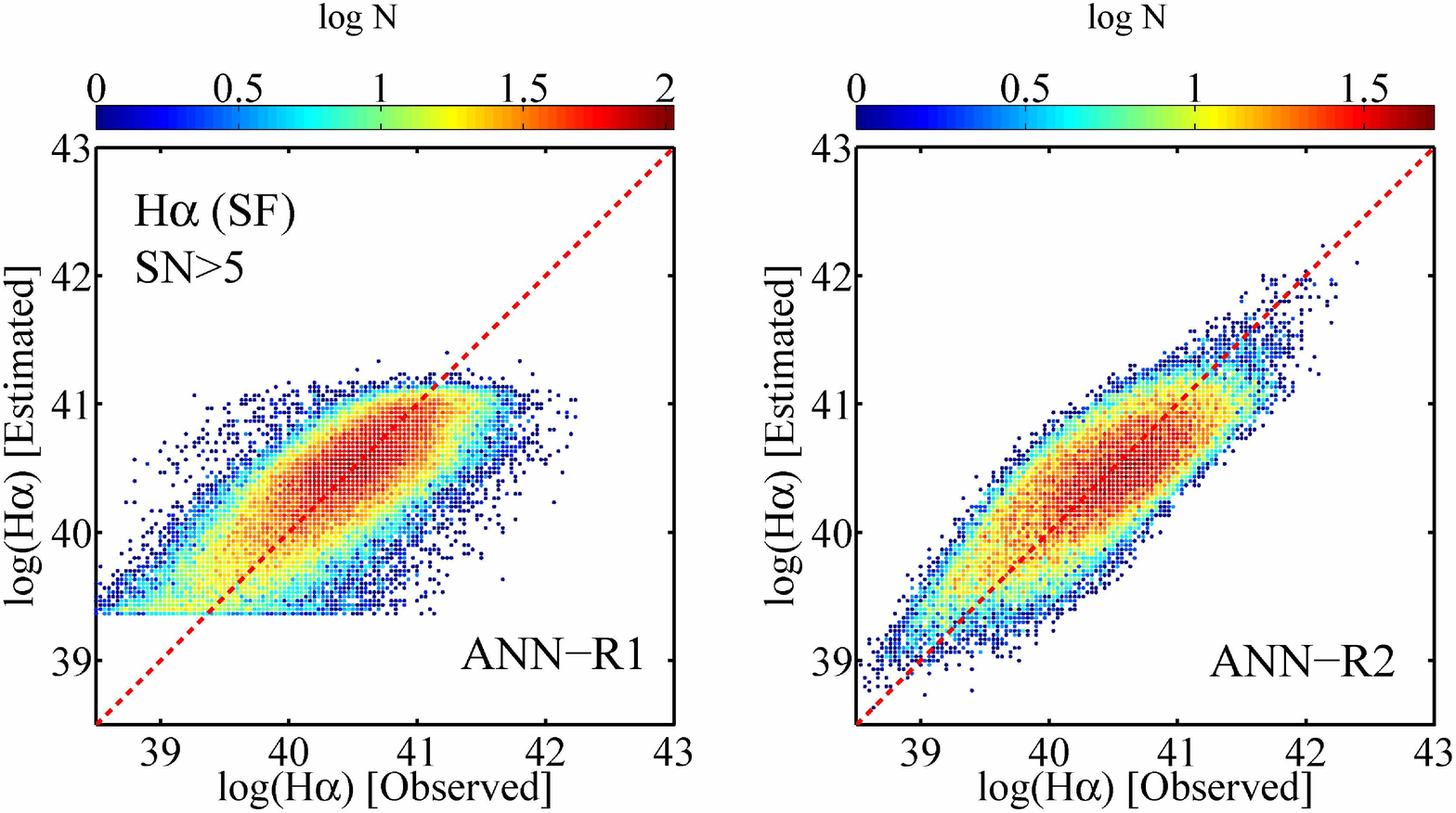}
\includegraphics[width=7.4cm,height=3.4cm,angle=0]{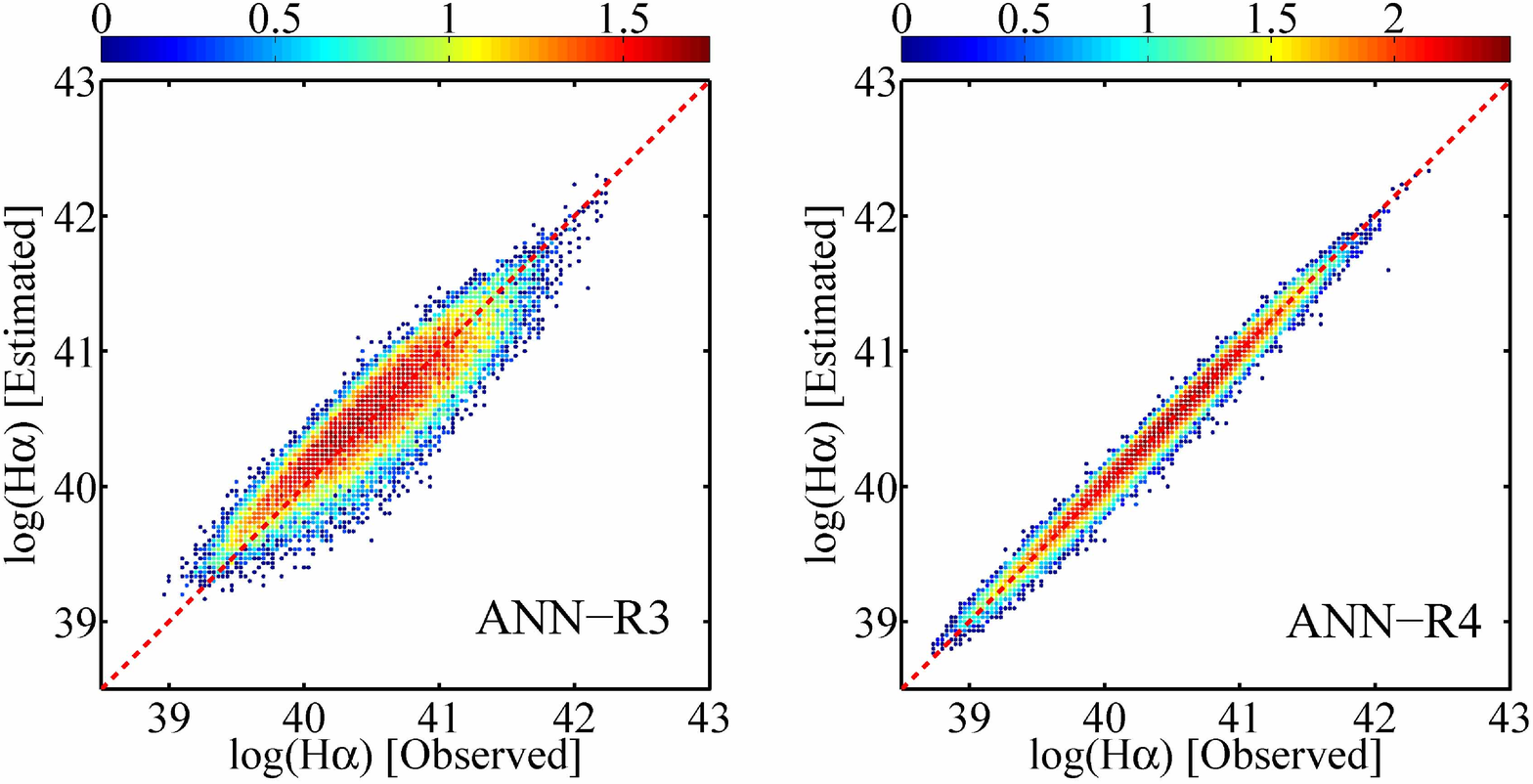}
\includegraphics[width=7.4cm,height=3.4cm,angle=0]{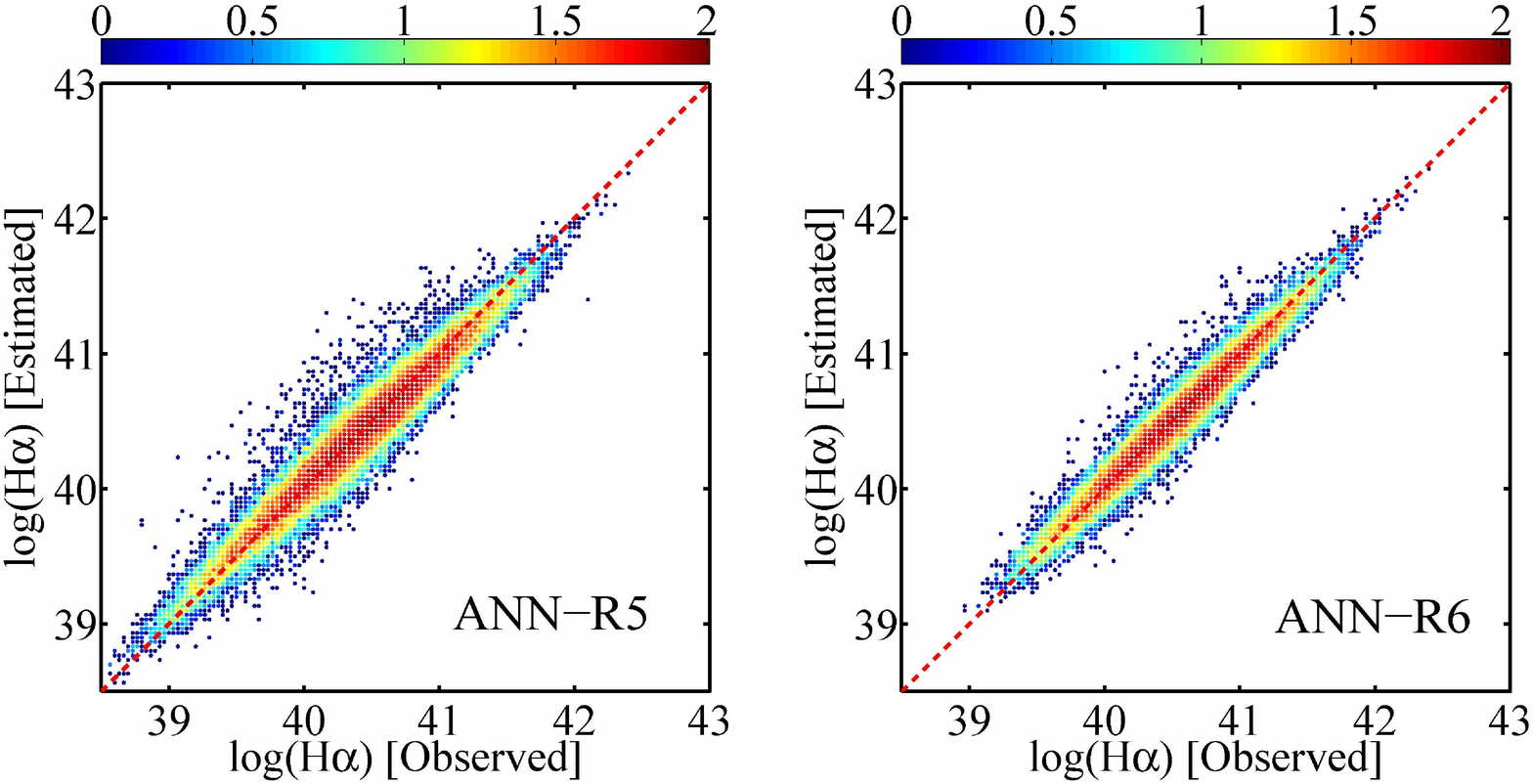}
\includegraphics[width=7.4cm,height=3.4cm,angle=0]{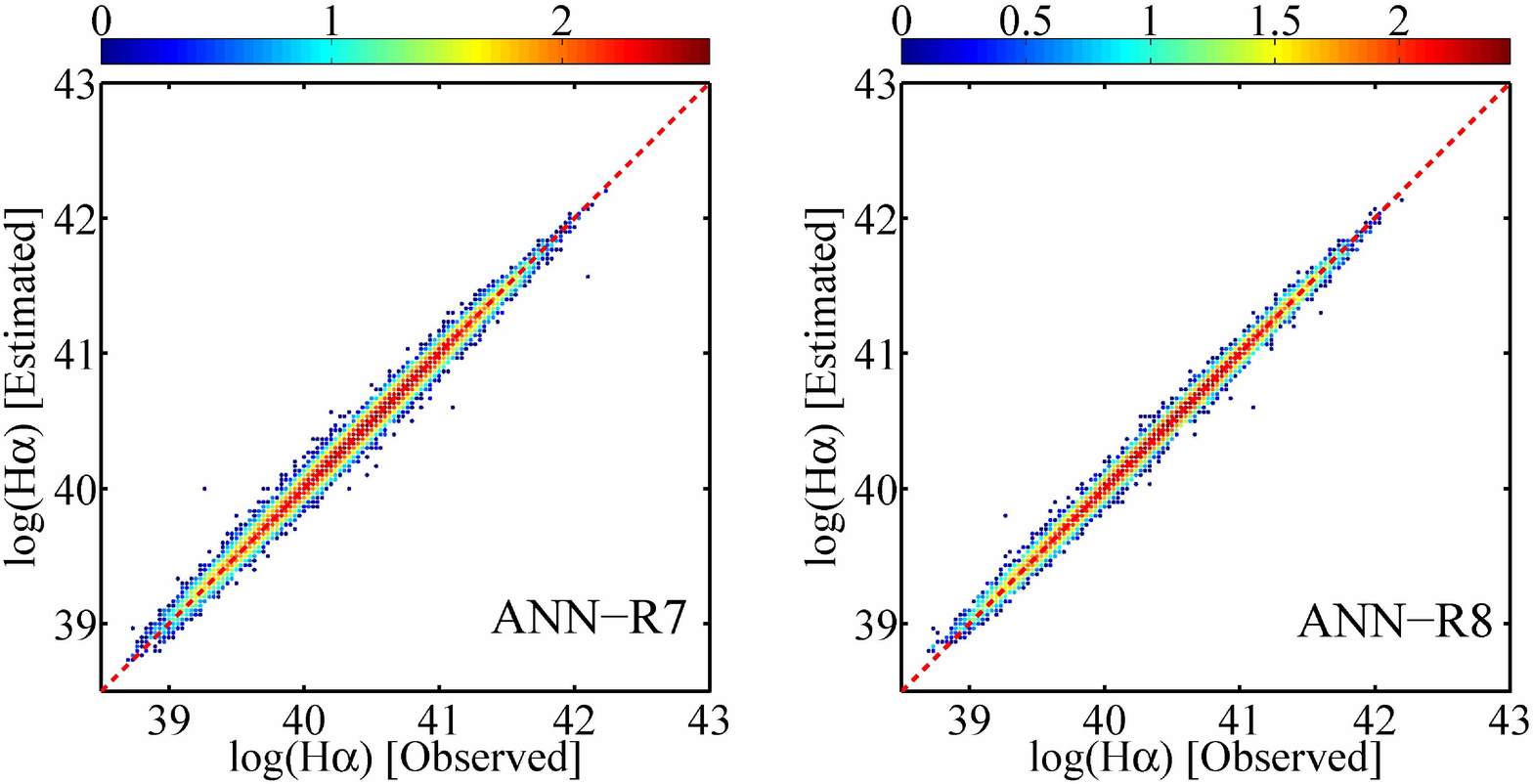}
\includegraphics[width=7.4cm,height=3.4cm,angle=0]{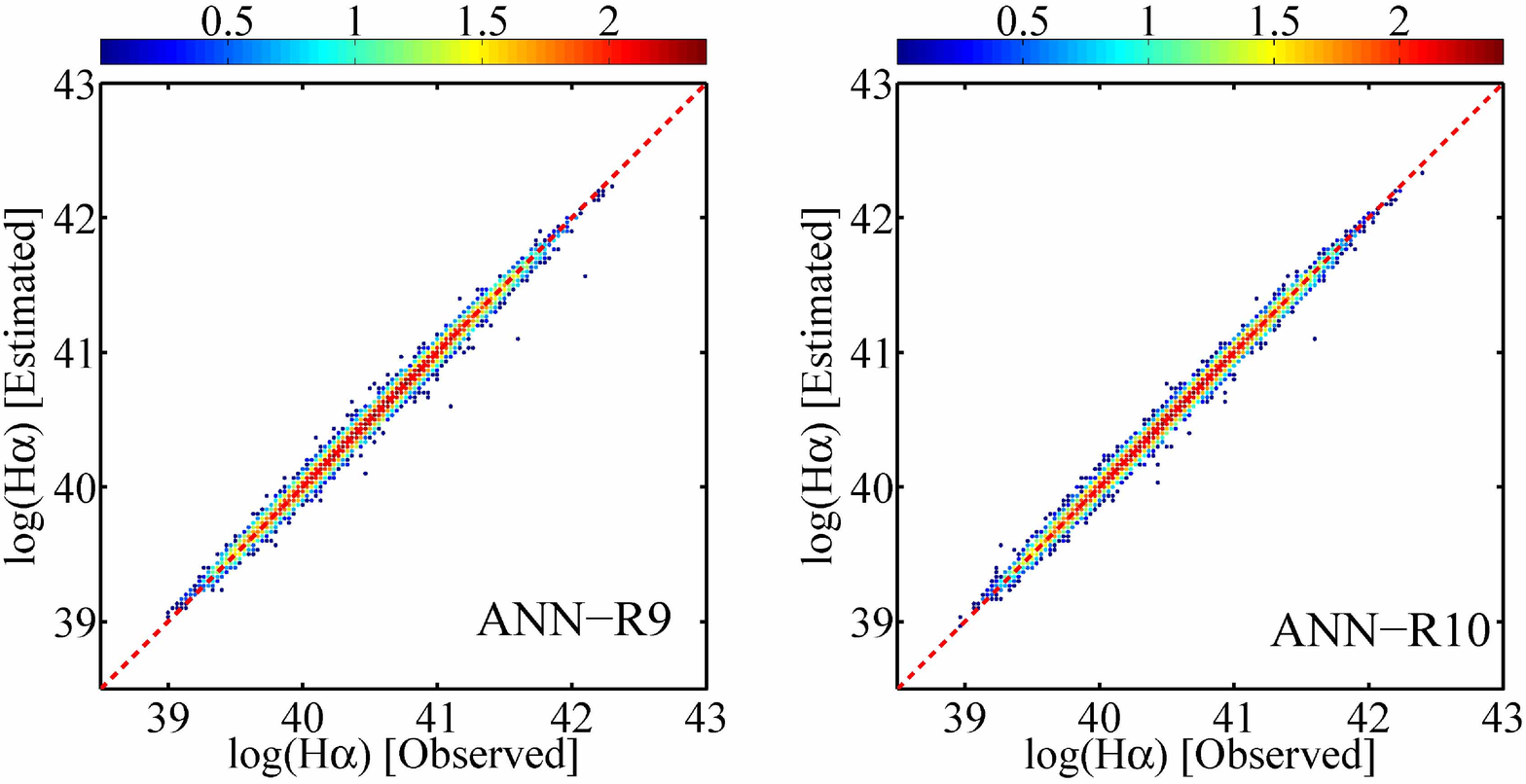}
\includegraphics[width=7.4cm,height=3.4cm,angle=0]{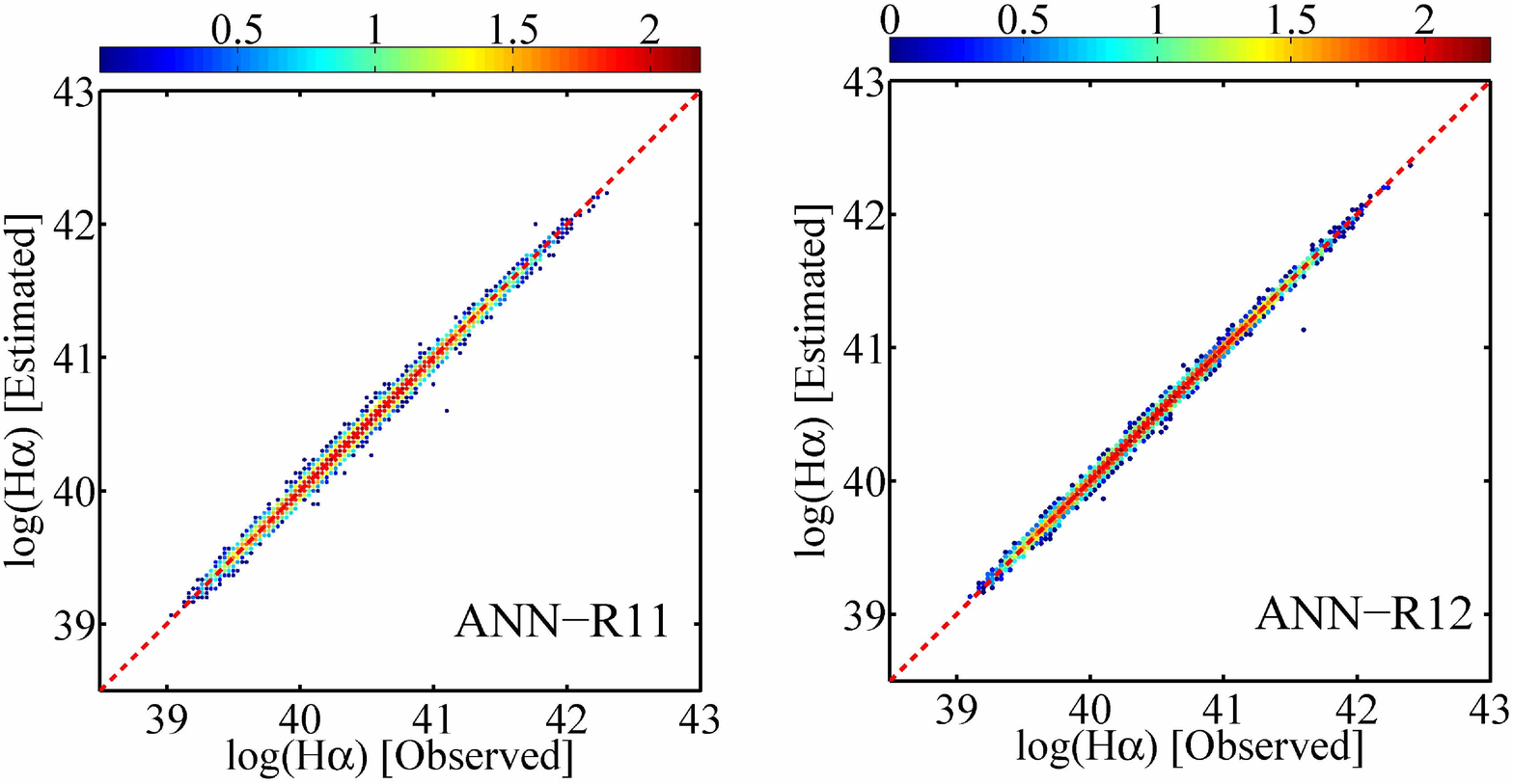}
\caption{The ANN predicted H$\alpha$ luminosities compared to the observed values
for the S/N$>$5 case for star-forming galaxies.  The input data
associated with each run are listed in
Table \ref{sf_runs}.  All the luminosities are in units of erg/s.}
\label{ann_sf_ha}
\end{figure}

\begin{figure}
\centering
\includegraphics[width=8.8cm,height=4.8cm,angle=0]{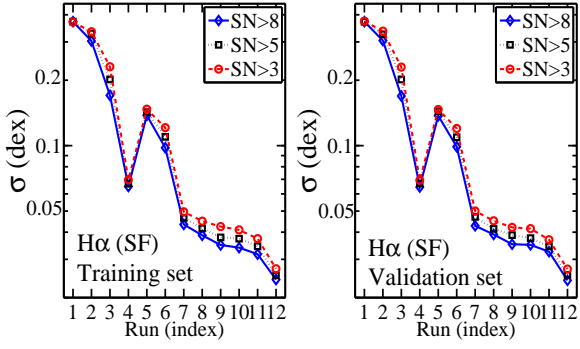}
\caption{The standard deviation of the difference between estimated and
observed H$\alpha$ for the training (left) and validation (right) sets for
star-forming galaxies. The
two sets show same behavior, indicating that the network is robust.}
\label{sd_sf_ha}
\end{figure}

\begin{figure}
\centering
\includegraphics[width=7.4cm,height=3.4cm,angle=0]{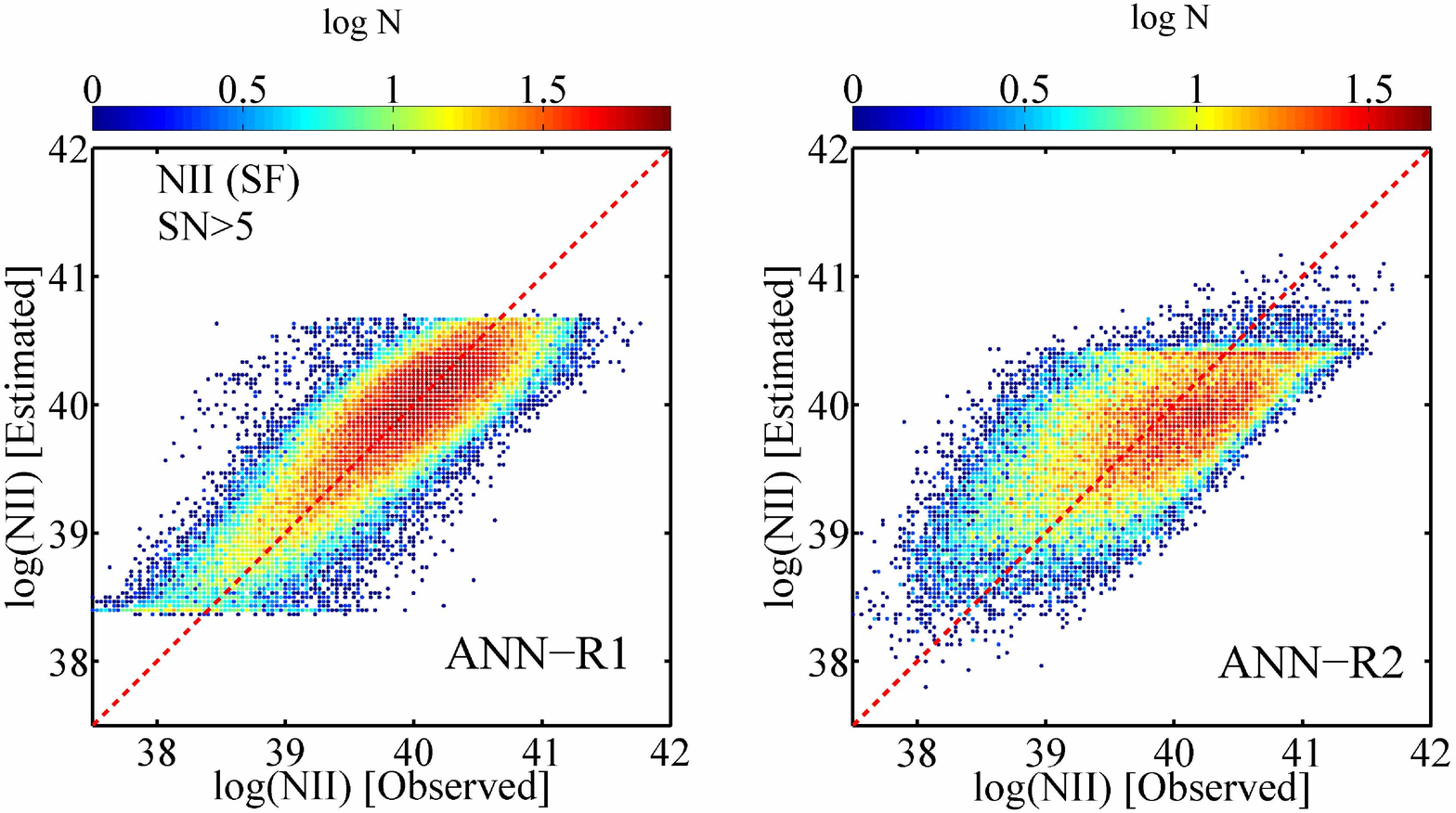}
\includegraphics[width=7.4cm,height=3.4cm,angle=0]{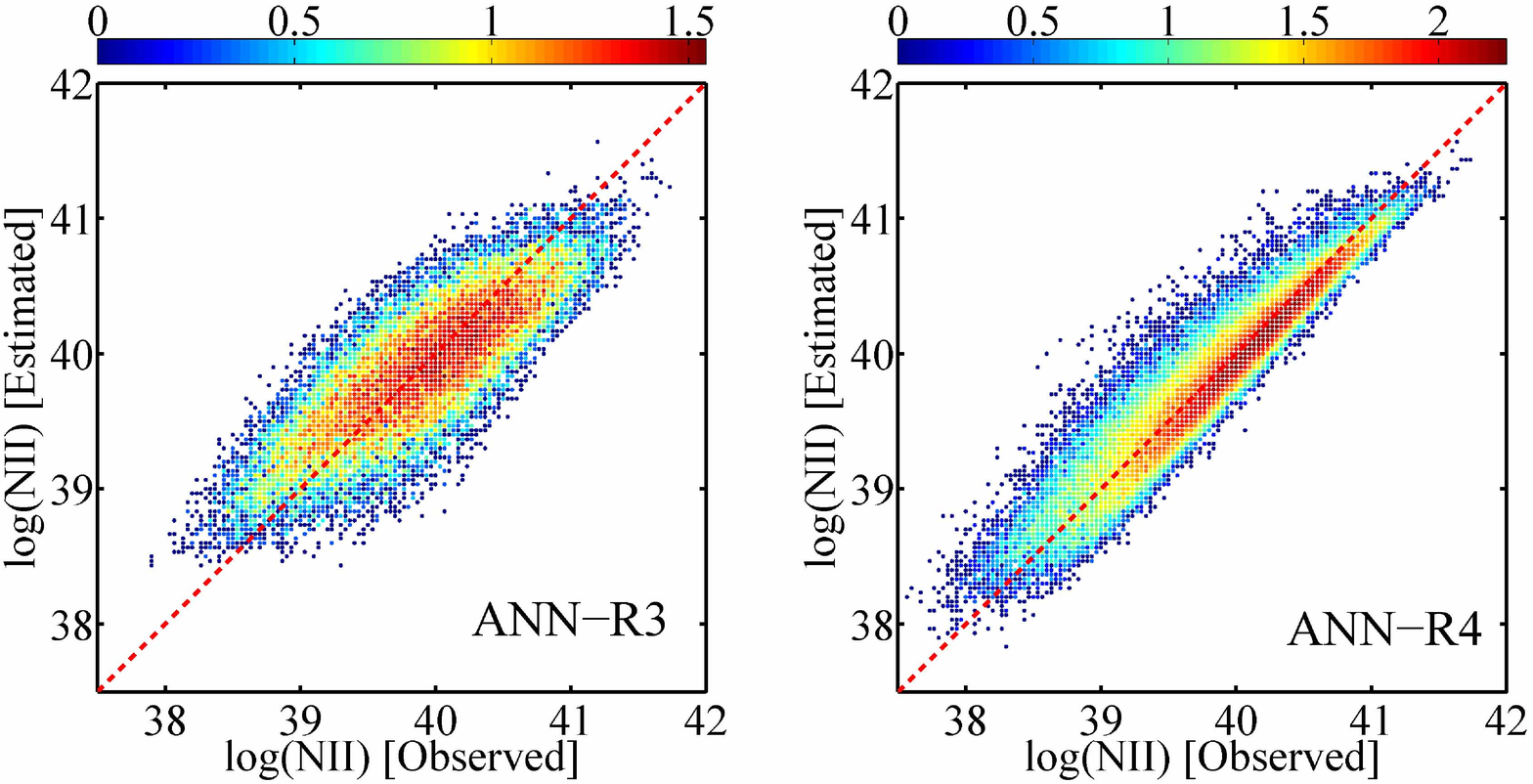}
\includegraphics[width=7.4cm,height=3.4cm,angle=0]{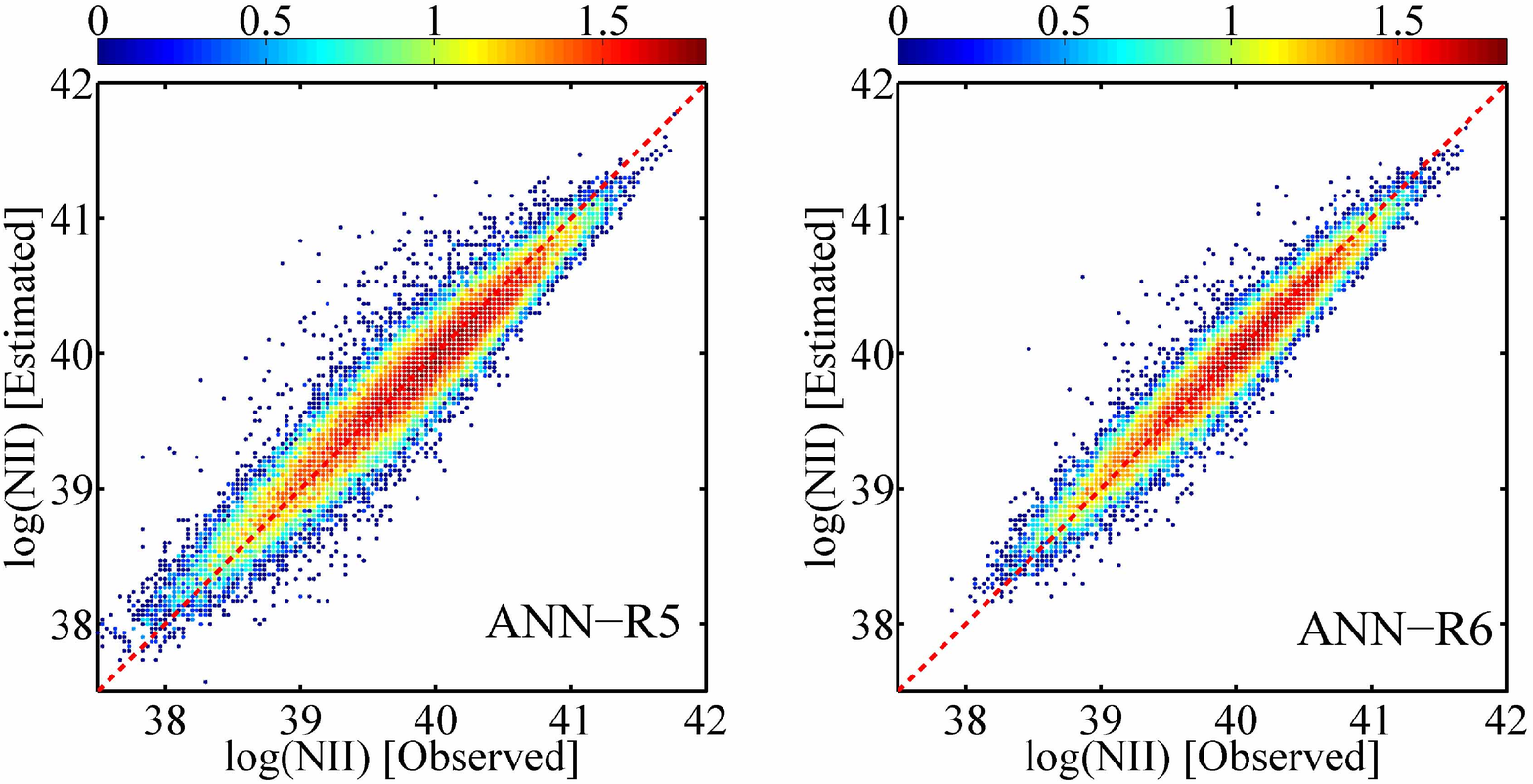}
\includegraphics[width=7.4cm,height=3.4cm,angle=0]{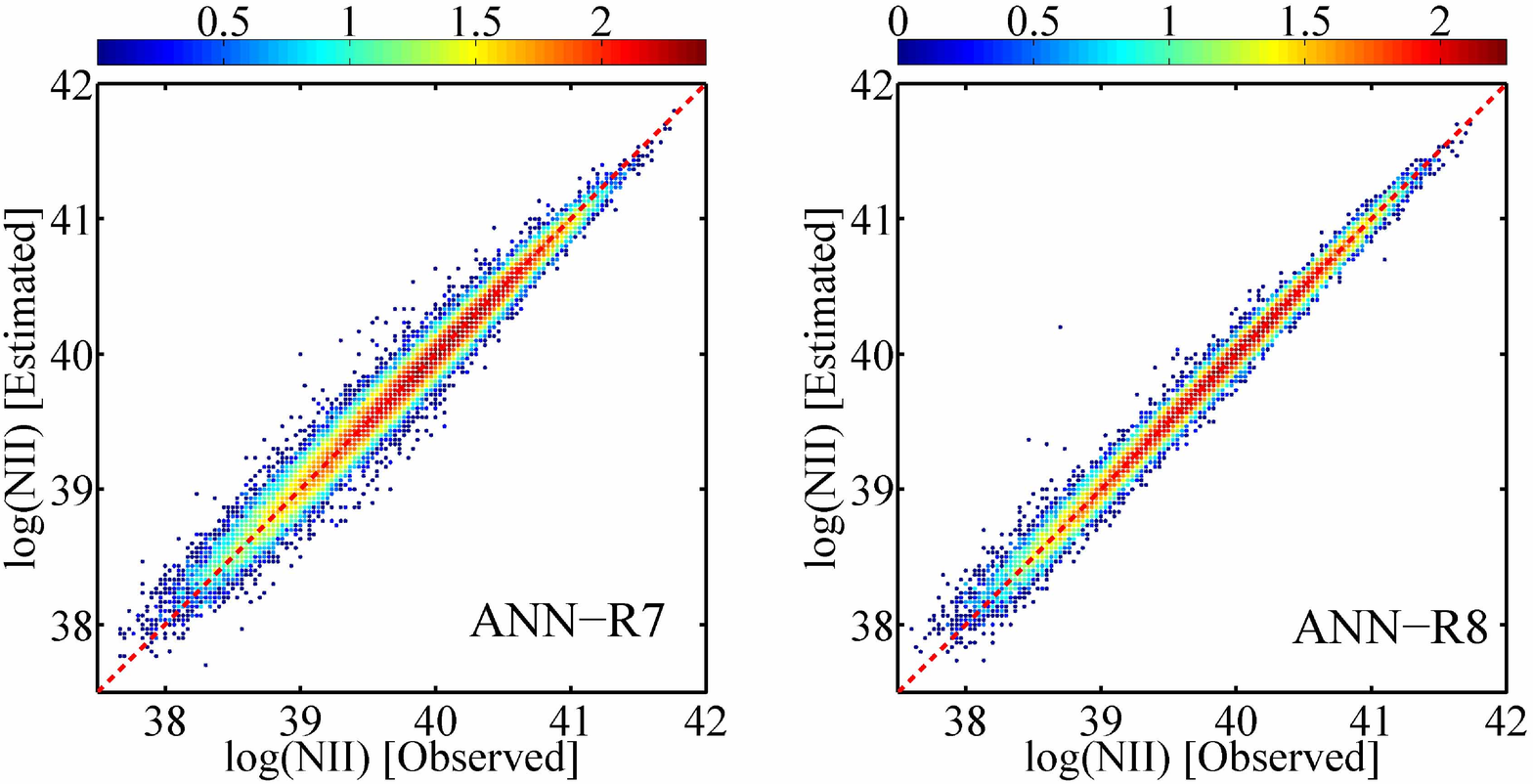}
\includegraphics[width=7.4cm,height=3.4cm,angle=0]{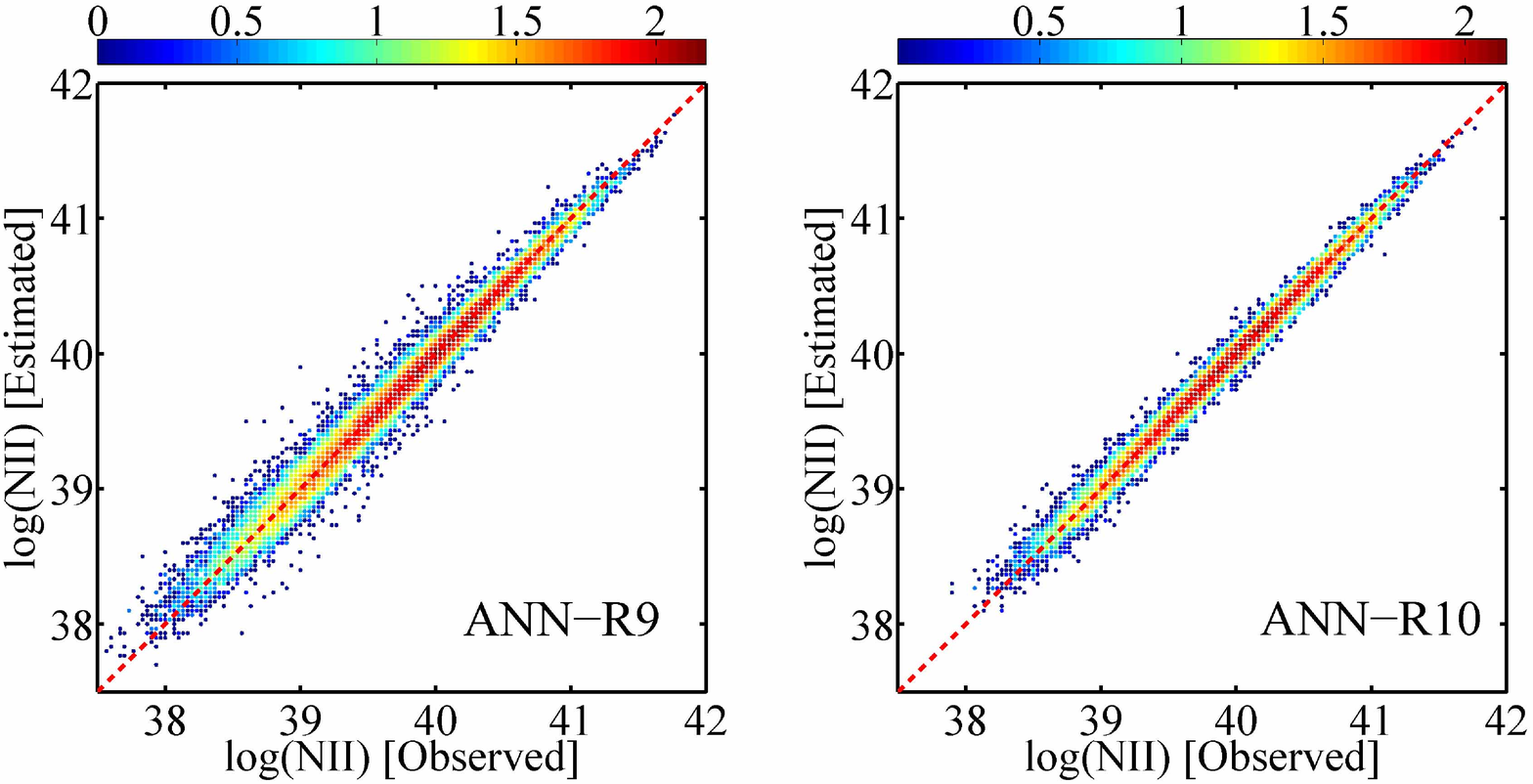}
\includegraphics[width=7.4cm,height=3.4cm,angle=0]{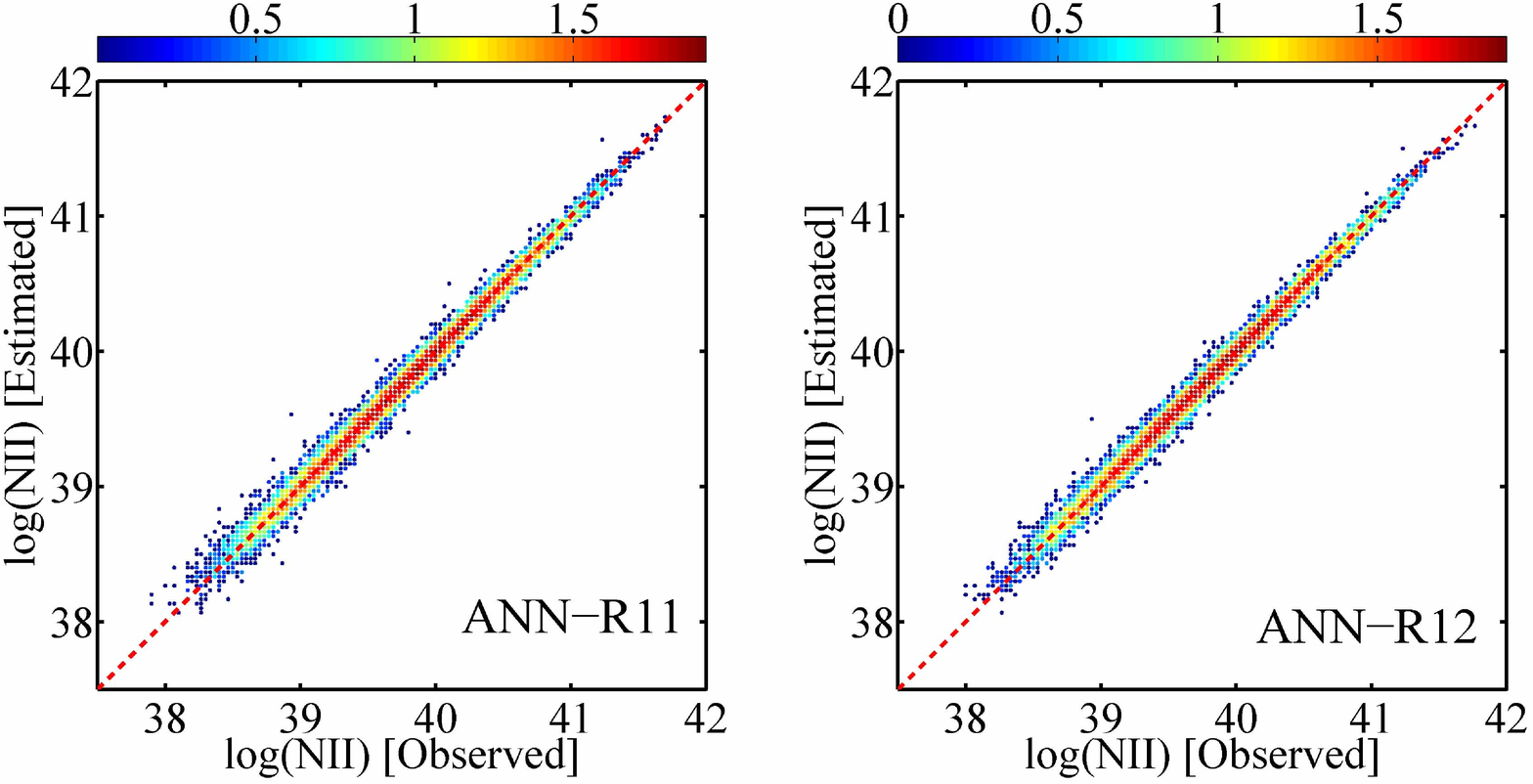}
\caption{The ANN predicted [NII] $\lambda$ 6584 luminosities compared to the
observed values for the S/N$>$5 case for star-forming galaxies.
The input data associated with each run are listed in
Table \ref{sf_runs}. All the luminosities are in units of erg/s.}
\label{ann_sf_n2}
\end{figure}

\begin{figure}
\centering
\includegraphics[width=8.8cm,height=4.8cm,angle=0]{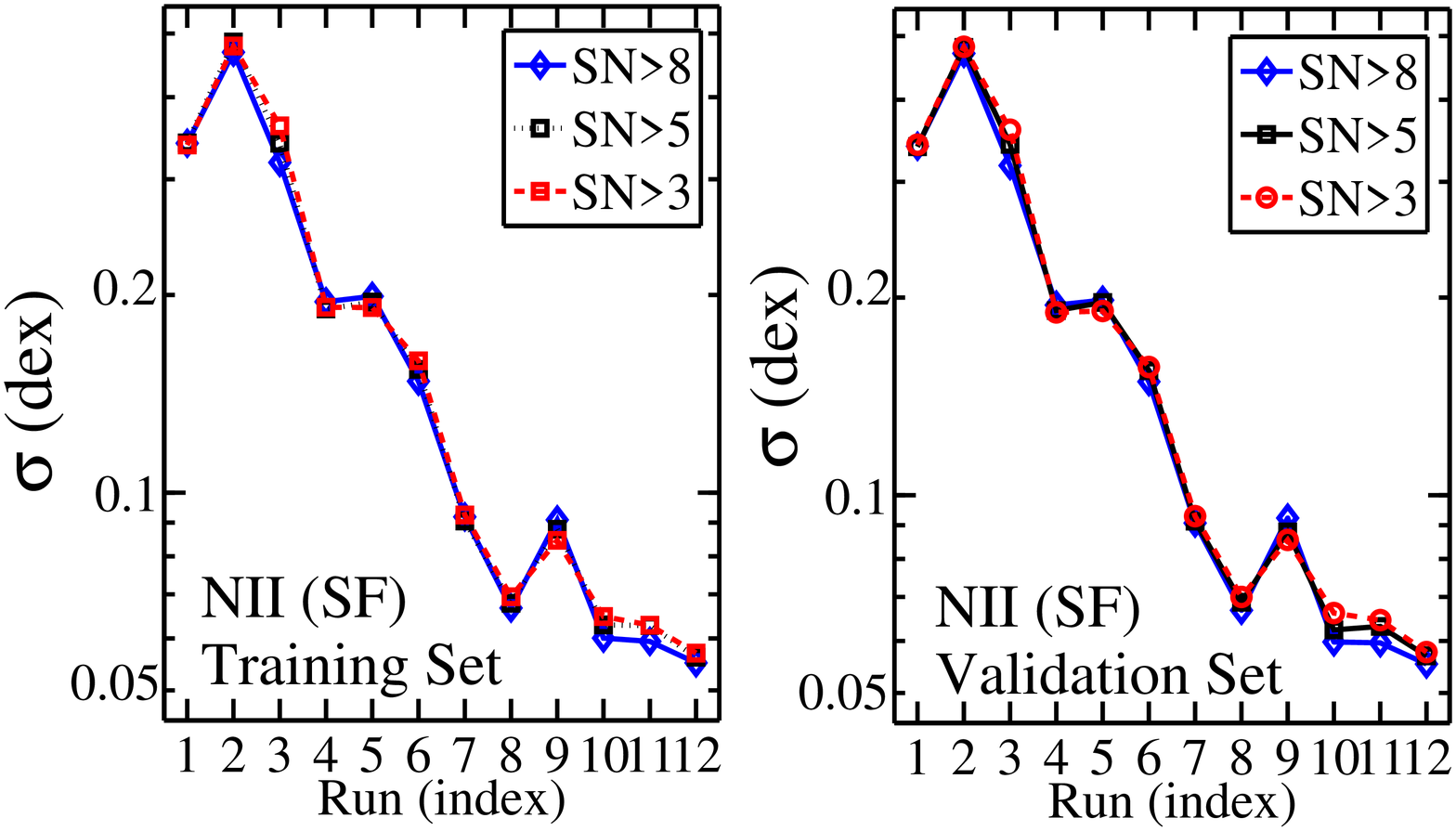}
\caption{The standard deviation of the difference between estimated and
observed [NII]$\lambda$6584 for the training (left) and validation (right)
sets for star-forming galaxies. The two sets show same behavior,
indicating that the network is robust.}
\label{sd_sf_n2}
\end{figure}

Figures \ref{ann_sf_ha} and \ref{ann_sf_n2} show the comparison of
predicted versus measured H$\alpha$ and [NII] line luminosities,
respectively.  Only the S/N$>$5 runs are shown for brevity. In Figures
\ref{sd_sf_ha} (H$\alpha$) and \ref{sd_sf_n2} ([NII]) a summary of the
standard deviations is shown for each of the runs for all 3 S/N
thresholds.  Let us first consider the ANN's performance in predicting
the flux of the H$\alpha$ line (Figures \ref{ann_sf_ha} and
\ref{sd_sf_ha}).  The first item of note is that the training and the
validation sets are identically behaved (Figure \ref{sd_sf_ha}) showing
that the training of the network is robust and repeatable.
As can be seen, from Figures \ref{ann_sf_ha} and \ref{sd_sf_ha},
the physical parameter that is the single most
important ingredient in the accurate prediction of the H$\alpha$ luminosity
is, unsurprisingly, H$\beta$.  All of the runs that include the H$\beta$
line have
standard deviations $<$ 0.07 dex.  Interestingly, the scatter varies
little between the different S/N thresholds with less than 0.01 dex
difference in the scatters between the S/N$>3$ and S/N$>8$ versions of
a given run.
Run R8 is perhaps one of the most useful combinations from a practical
perspective, including two strong emission lines ([OIII] $\lambda$
5007 and H$\beta$) that are relatively close in wavelength, plus fibre
stellar mass.
For this run, the ANN is able to predict the H$\alpha$ luminosity with a
standard deviation of only 0.04 dex, even when the S/N threshold is as
low as 3.

We turn next to the ANN predictions of [NII] $\lambda$6584 (Figures
\ref{ann_sf_n2} and \ref{sd_sf_n2}).  Again, the training and validation
sets are identical, demonstrating that the ANN's results are robust
and not dependent on the particular galaxies used in the training
set.  The standard deviations tend to be higher in general for
the predictions of [NII] than for H$\alpha$.
Interestingly, there is almost no dependence of the
accuracy of the [NII] line prediction on the S/N of the training
set.  Although the dependence of the ANN's performance is less
dramatically linked to the use of H$\beta$ for
predicting [NII] as it was for H$\alpha$, the H$\beta$ still appears to
be the most important of the 3 strong emission lines used here ([OII],
[OIII] and H$\beta$).  Of the runs using a single emission line (R2 -- R4),
the latter, which uses H$\beta$ has the lowest scatter.  Of the runs that
combine mass and a single emission line (R5 -- R7), again it is
the latter which minimizes the scatter.  As we saw was the case for the
prediction of H$\alpha$, run R8, which is attractive for its practical
combination of fibre stellar
mass, H$\beta$ and [OIII]$\lambda$ 5007, has one of the
smallest scatters ($\sim 0.07$ dex).  Indeed, for both H$\alpha$ and [NII]
the accuracy of the ANN's prediction is improved little by adding
[OII].

\subsection{On the use of fibre vs. total stellar mass.}

\begin{figure}
\centering
\includegraphics[width=8.8cm,height=4.8cm,angle=0]{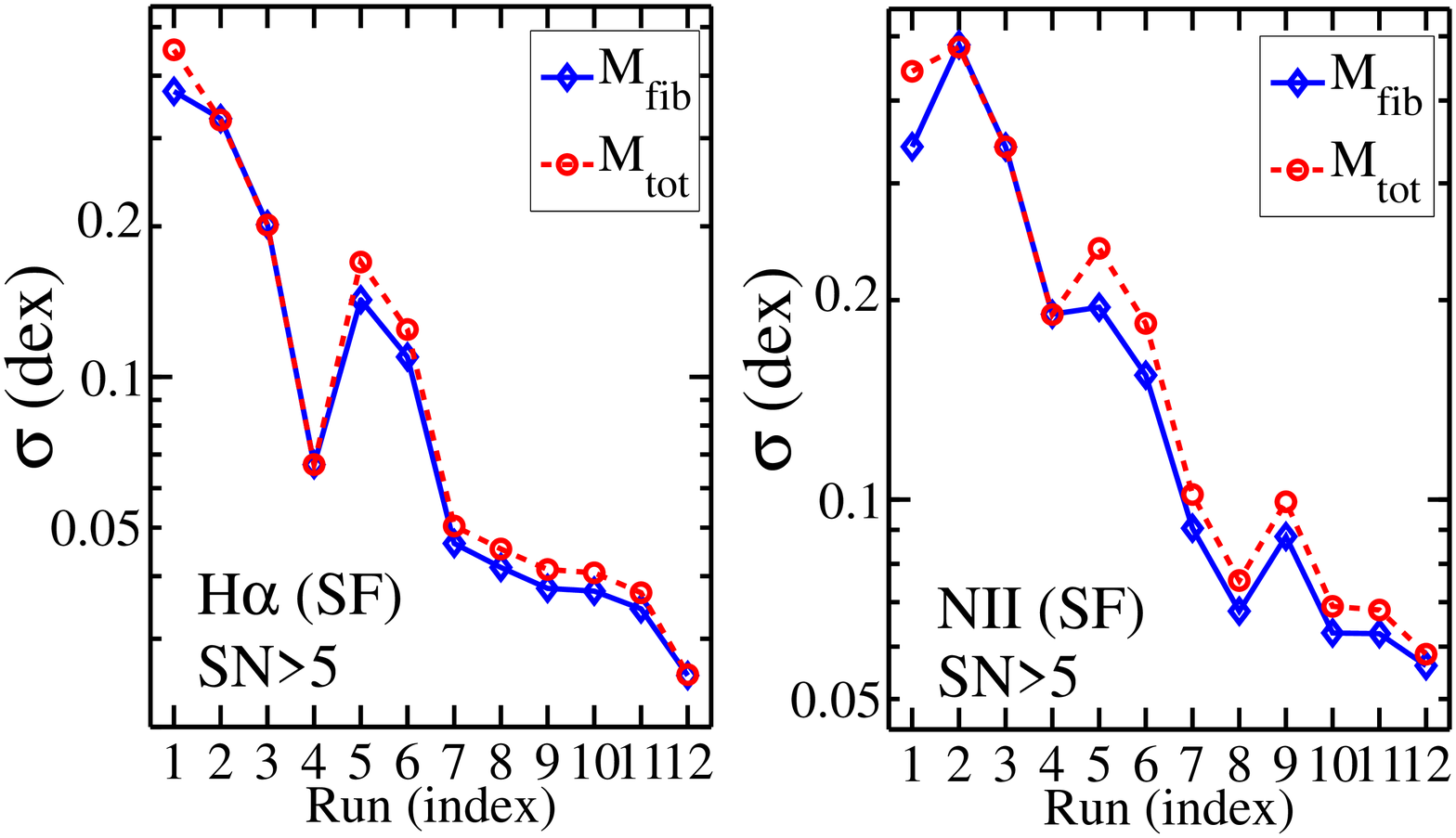}
\caption{A comparison between different runs (1-12) with total and fibre
mass for S/N$>5$ for star-forming galaxies.
Using the total stellar mass leads to a slightly
enhanced scatter, although the difference is probably negligible for
most applications.}
\label{mass_test}
\end{figure}

Up until now, we have used the fibre stellar mass under the assumption that
predicting the spectral line strengths within a given aperture will be
most sensitive to the mass enclosed in that same area.  However,
whilst total galaxy stellar masses might be readily available from broad-band
photometry, determining aperture masses requires an additional analysis
step.  Therefore, in the spirit of user-friendliness, we investigate
the impact of substituting total stellar mass instead of fibre
mass in the runs listed in Table \ref{sf_runs}.  In Figure \ref{mass_test}
we compare the standard deviations of the 12 runs for the S/N $>$ 5
case with this mass substitution for both the H$\alpha$ and [NII] predictions
(left and right panels respectively).  Since some of the runs (R2, R3 and
R4) do not use mass as an input parameter the substitution of total
stellar mass for fibre stellar mass has no impact.  For the other
runs, we see that, as expected, the use of total stellar mass (red points)
yields a slightly higher scatter, but the difference is typically $< 0.01$ dex
in the standard deviations.  We conclude that total stellar mass can
be used as a convenient alternative for aperture stellar mass, at least
over the range of covering fractions probed by the SDSS data (typically
0.1 -- 0.4).  Nonetheless, we
investigate explicitly in Section \ref{test_size_sec} the dependence of
our calibrations on covering fraction.

\section{The prediction of [NII] and H$\alpha$ luminosities for the general case of emission line galaxies}\label{mixed_sec}

In the previous section, we demonstrated that an ANN can be
successfully used to predict the H$\alpha$ and [NII] $\lambda$ 6584
line luminosities in star forming galaxies.  In many cases, the accuracy of
the luminosity prediction is within 0.1 dex.  However, in practice, one can
not always assume that the ionization of the interstellar medium is
dominated by stars.  Ionization by alternative sources, such as
planetary nebulae, shocks or AGN will yield different line strengths.
This fact forms the basis of diagnostic diagrams such as the BPT.

We therefore investigate whether the ANN is able to predict line
luminosities for emission line galaxies whose dominant source of ionization
is unknown.  In practice, the two most likely sources to dominate in
integrated spectra such as those from the SDSS are stars and AGN.
In order to define a general sample of emission line galaxies, we
therefore drop the first requirement in
Section \ref{sf_sec}, that galaxies must be classified as
star-forming.  The full sample of galaxies is referred to as the `mixed'
sample, to reflect the combination of star-forming, AGN and other minority
classes, such as LINERs.  We repeat all of the ANN runs listed in Table
\ref{sf_runs}, with the three S/N thresholds, and adopting the same
strategy of splitting each sample into training and validation sets.
Table \ref{run_sizes_mix} lists the number of galaxies in the mixed
sample that is used in each of the different runs for the three
S/N thresholds tested in the previous section.

\begin{table}
\begin{center}
\caption{The  number of galaxies in each run with different S/N thresholds for the mixed galaxies.}
\begin{tabular}{c|c|c|c}
\hline\hline
Runs & N$_{gal}$ (S/N$>3$)& N$_{gal}$ (S/N$>5$)& N$_{gal}$ (S/N$>8$)\\
\hline
R1  & 264792& 264792&264792\\
R2  & 189738 &123768&79780\\
R3  & 122333 &67686&33030\\
R4  & 212757 & 174502&126286\\
R5  & 189738 & 123768&79780\\
R6  & 122333 &67686&33030\\
R7  & 212757 &174502& 126286\\
R8  & 166022 &107014&65725\\
R9  & 120507 &67306&32963\\
R10 &116334  &65633&32528\\
R11 &67708  & 37242&19405\\
R12 &67034  & 36858&19168\\
\hline
\end{tabular}
\label{run_sizes_mix}
\end{center}
\end{table}

\begin{figure}
\centering
\includegraphics[width=8.8cm,height=4.8cm,angle=0]{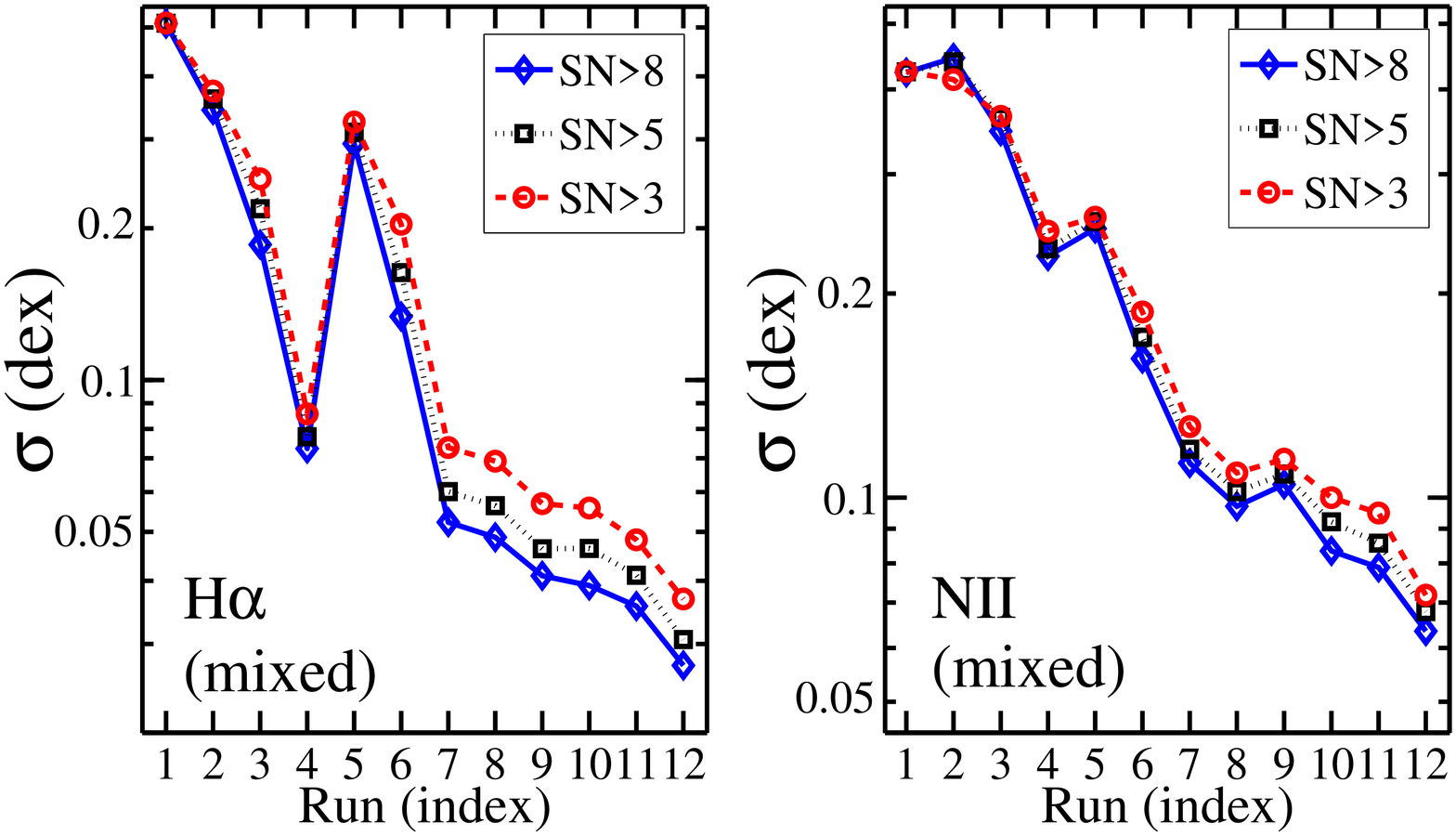}
\caption{A comparison between different runs (1-12) with the ANN
methods for the S/N$>5$ mixed galaxy sample. The left and the right panels are related to H$\alpha$ and [NII], respectively.}
\label{fig-std-ha-nii-mix}
\end{figure}

The scatter in the ANN predicted luminosities compared to the observed value
for the mixed sample is shown in Figure \ref{fig-std-ha-nii-mix},
which includes the output from all three S/N cuts. Here, again, the
validation and training sets show the same behavior so we only show
the results from validation set.  As can be seen, the difference in
scatters for
different S/N cuts is more significant than that of the star-forming sample,
for both H$\alpha$ (Figure \ref{sd_sf_ha}) and [NII] (Figure \ref{sd_sf_n2}).
The difference is more noticeable for H$\alpha$. This shows that the mixed
sample (adding the AGN galaxies to the SF sample) can generate  more
complicated patterns in the dataset and these patterns have more
dependency on the choice of S/N. From  Figure \ref{fig-std-ha-nii-mix},
however, it seems that adding more information can give a more coherent result.

\begin{figure}
\centering
\includegraphics[width=8.8cm,height=4.8cm,angle=0]{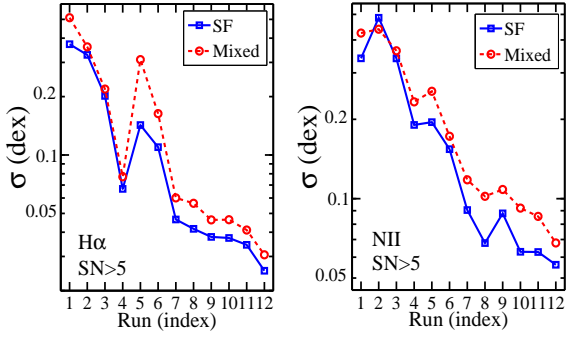}
\caption{A comparison between different runs (1-12) with the ANN for the
results obtained for SF (blue points) and mixed (red points)
samples. The left and the right panels
are related to H$\alpha$ and [NII], respectively.}
\label{fig-std-ha-nii-sf-mix}
\end{figure}

We compare again the use of fibre and total stellar masses.  As was previously
found for the star-forming galaxies, here also  total stellar mass performs
essentially as well as the fibre stellar mass.  A direct comparison
of the ANN's performance between using only star-forming galaxies
and including other classes is shown in Figure \ref{fig-std-ha-nii-sf-mix}.
As expected, the scatter is higher when all galaxy classes are included,
but the ANN can still readily yield scatters $<$ 0.1 dex for both
H$\alpha$ and [NII] for several of the runs. Again,  the maximum
information used as input data (run 12) yields the lowest scatter.

\section{Derivation of basis function coefficients}\label{fx_sec}

\begin{figure}
\centering
\includegraphics[width=8.8cm,height=4.8cm,angle=0]{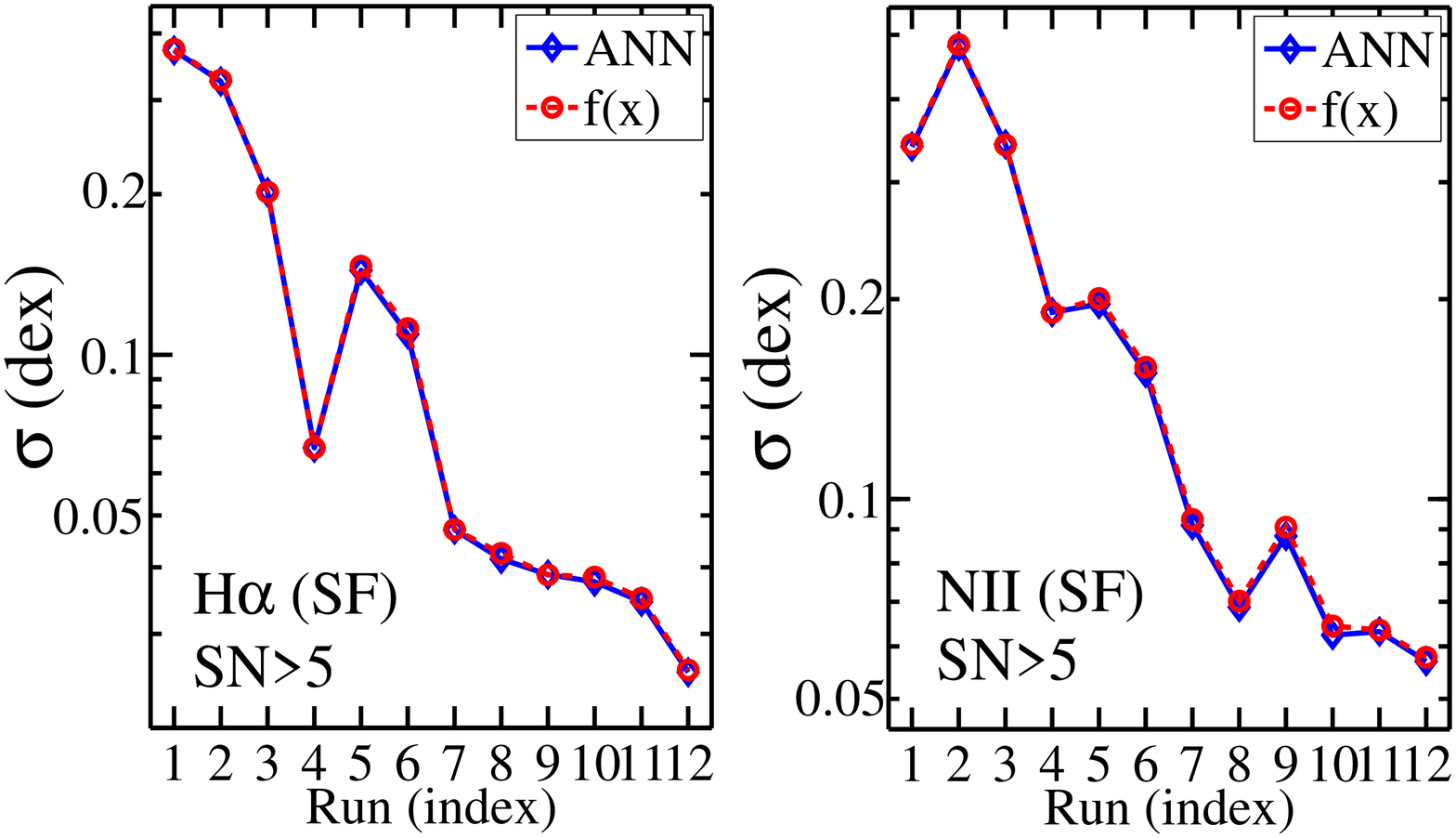}
\caption{A comparison between different runs (1-12) with ANN and f(x)
methods for S/N$>5$ in the star-forming sample. The left and the right
panels are related to H$\alpha$ and [NII], respectively.}
\label{fig-ann-reg-h-n}
\end{figure}

We now investigate how effectively the selected polynomial basis function
(i.e., Eq. \ref{eq-basis-function}) can predict the results of the ANN.
In Figure \ref{fig-ann-reg-h-n} we show a comparison between different
runs (1-12) with the ANN and f(x) methods for S/N$>5$ in the
star-forming sample.
The left and the right panels are related to H$\alpha$ and [NII], respectively.
There is an excellent match between the performance of the two methods,
indicating that no accuracy is lost in the use of the polynomial.

\begin{figure}
\centering
\includegraphics[width=8.8cm,height=4.8cm,angle=0]{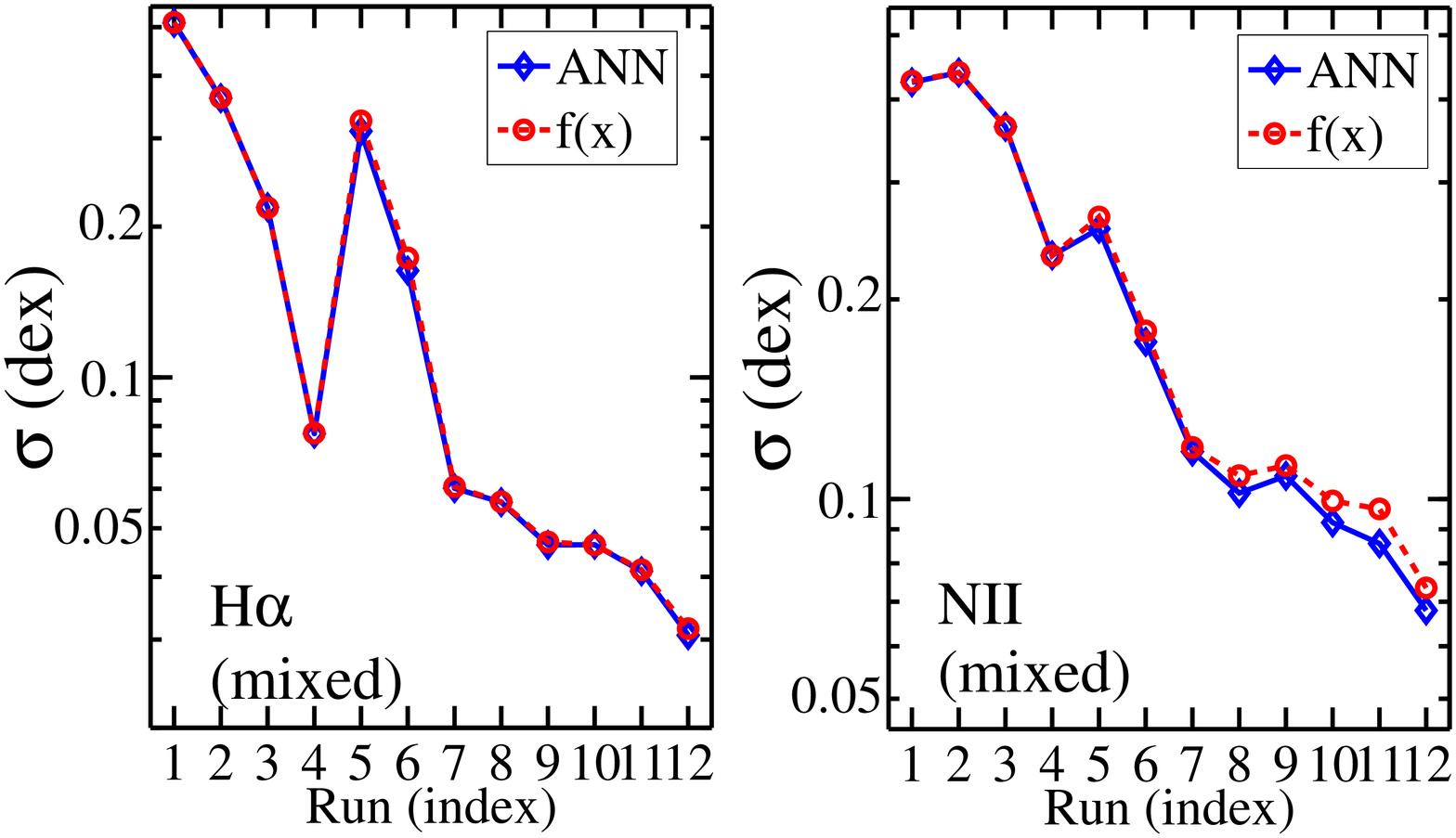}
\caption{A comparison between different runs (1-12) with ANN and f(x)
methods for S/N$>5$. The left and the right panels are
related to H$\alpha$ and [NII], respectively.  In general, the basis
function can accurately reproduce the ANN results.}
\label{fig-std-ha-nii-mix-fx-3}
\end{figure}

We repeat this exercise for the general case of emission line galaxies,
which represents a mix of star-forming and AGN classes, with the results
shown in Figure \ref{fig-std-ha-nii-mix-fx-3}.
The left panel, for estimating H$\alpha$, shows that the polynomial
once again excellently reproduces the results of the ANN. However, the
scatter in the [NII] line luminosity predictions is slightly higher
when the polynomial is used, compared to the ANN, but the effect is
very small, $< 0.01$ dex.

In Appendix \ref{coeff_sec} we present all coefficients for estimating
the H$\alpha$
and [NII] luminosities of star-forming and `mixed' galaxies.

\section{Testing the ANN calibration}\label{test_sec}

In this section we test the robustness of the ANN calibration in three
ways.  First, we investigate the importance of the selected parameters used in ANN and  the size and sampling
of the training set.  Second, we examine whether the scatter in predicted line luminosities
depends on the physical parameters (such as mass and metallicity) of the
sample.  Finally, we test whether the calibrations derived for the SDSS
data work well when applied to an independent dataset.

\subsection{Size and sampling of the training set}\label{test_size_sec}

 A first consideration in the ANN application is the number of neurons
used.  For example, using a large number of the neurons in a network can
increase its power, but it can also result in over-fitting. To prevent
this, the networks' parameters can be tuned in such a way that the
training and validation sets show the same scatter.  In the work
presented here, we find that 8-10 neurons works well with
our dataset. We also check several optimization algorithms and find
that the Levenberg-Marquardt and Bayesian regulation back propagation
provide the same results as one other, and are superior to other
optimization algorithms such as a gradient descent back propagation.

The ANN will only work well if the training set
is a good representation of the data for which the
prediction is desired. The scatter in the training and validation sets
also depends on the size of the training set.     As mentioned previously,
we repeat each run 20 times, with different initialization.
For higher numbers of galaxies selected for training ANN the outputs
of each run cover the same range. In other words, here, there is a stability.
When we reduce the number of galaxies for training  the output of  the runs
are  in significantly different ranges. We find that a training sample
of  $>3000$ galaxies can predict an acceptable
result in which the behavior of the training and validation sets are the same.

\subsection{The effect of the physical parameters on estimated parameters}

\begin{figure}
\centering
\includegraphics[width=8.2cm,height=4.1cm,angle=0]{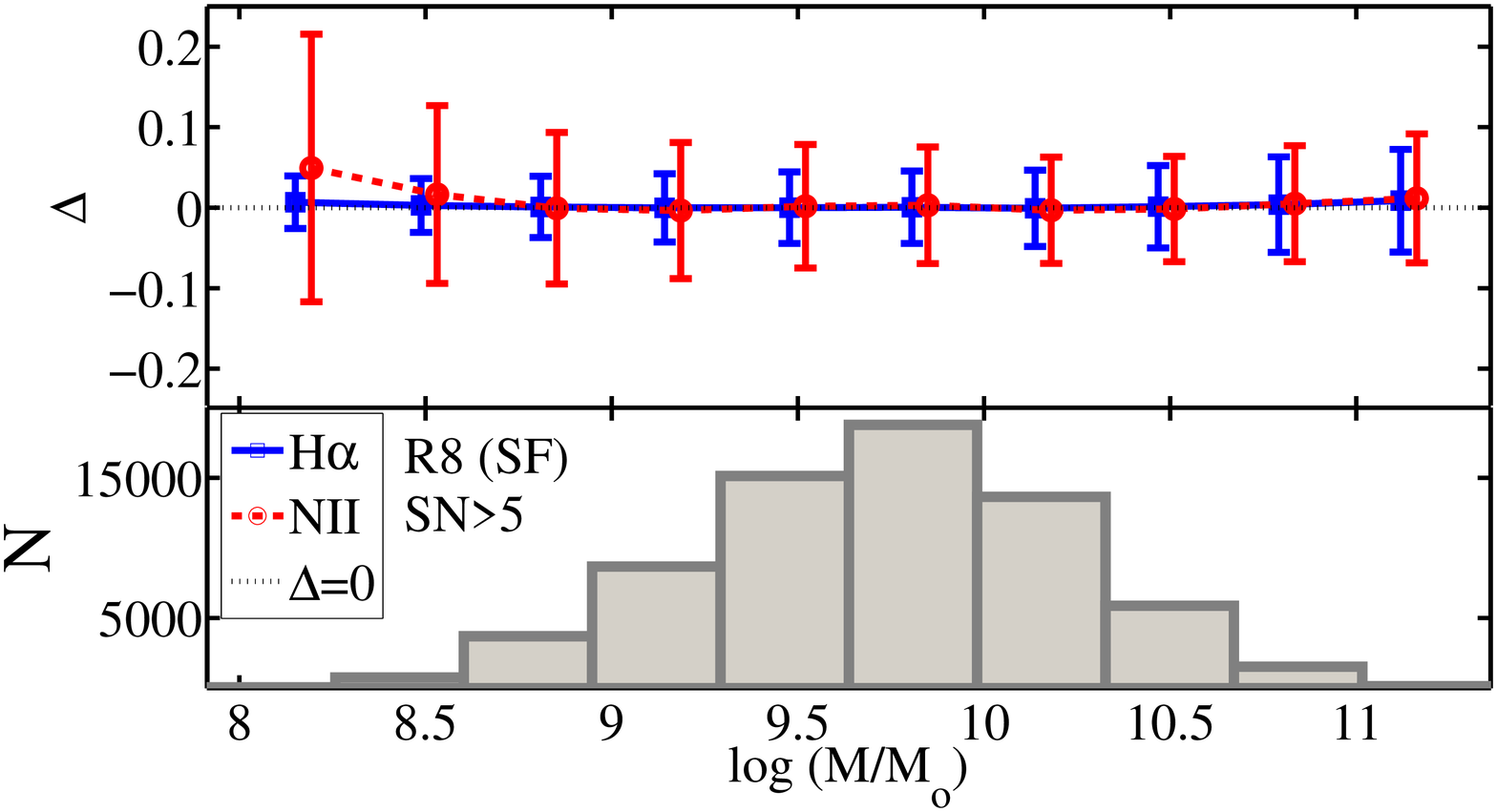}
\includegraphics[width=8.2cm,height=4.1cm,angle=0]{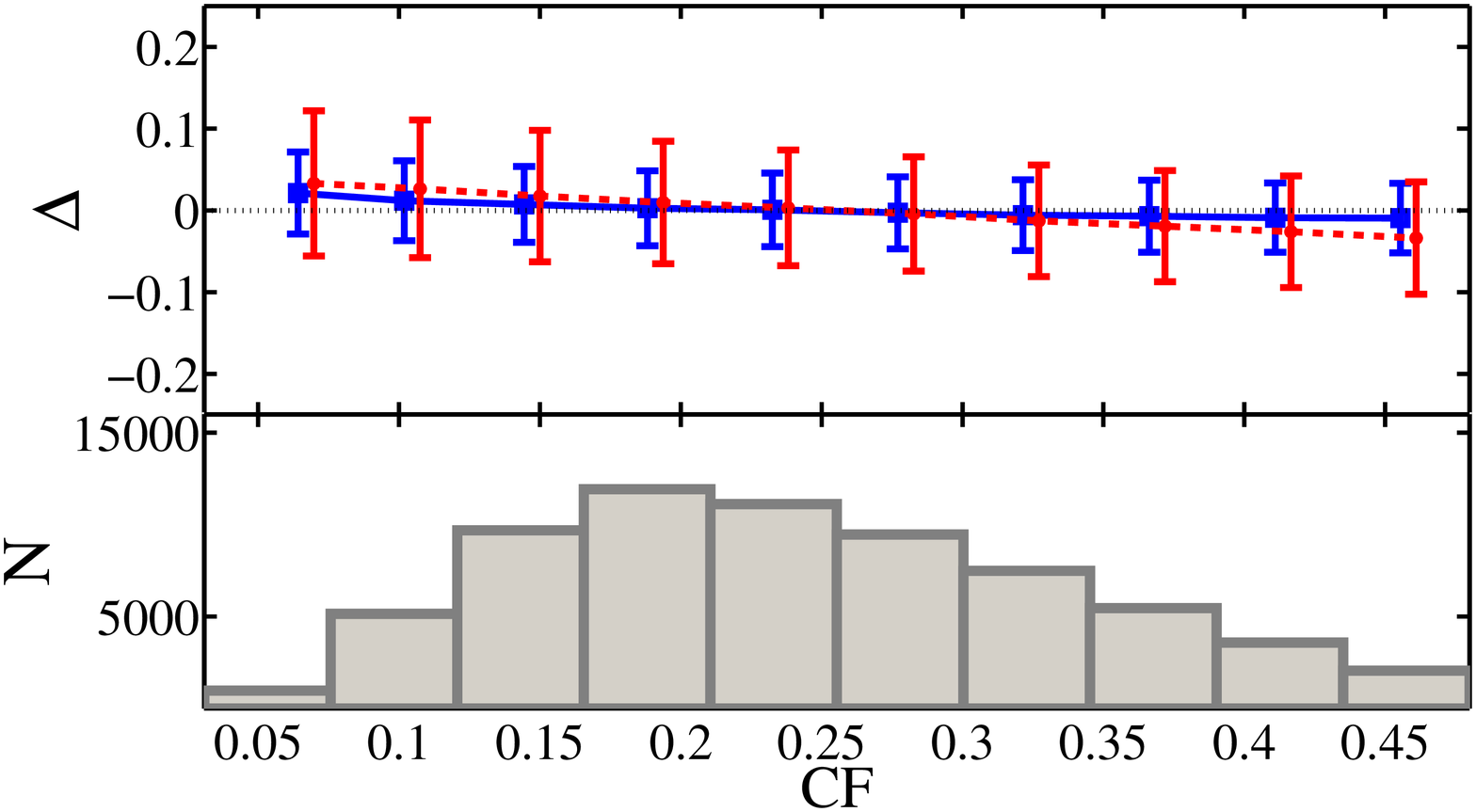}
\includegraphics[width=8.2cm,height=4.1cm,angle=0]{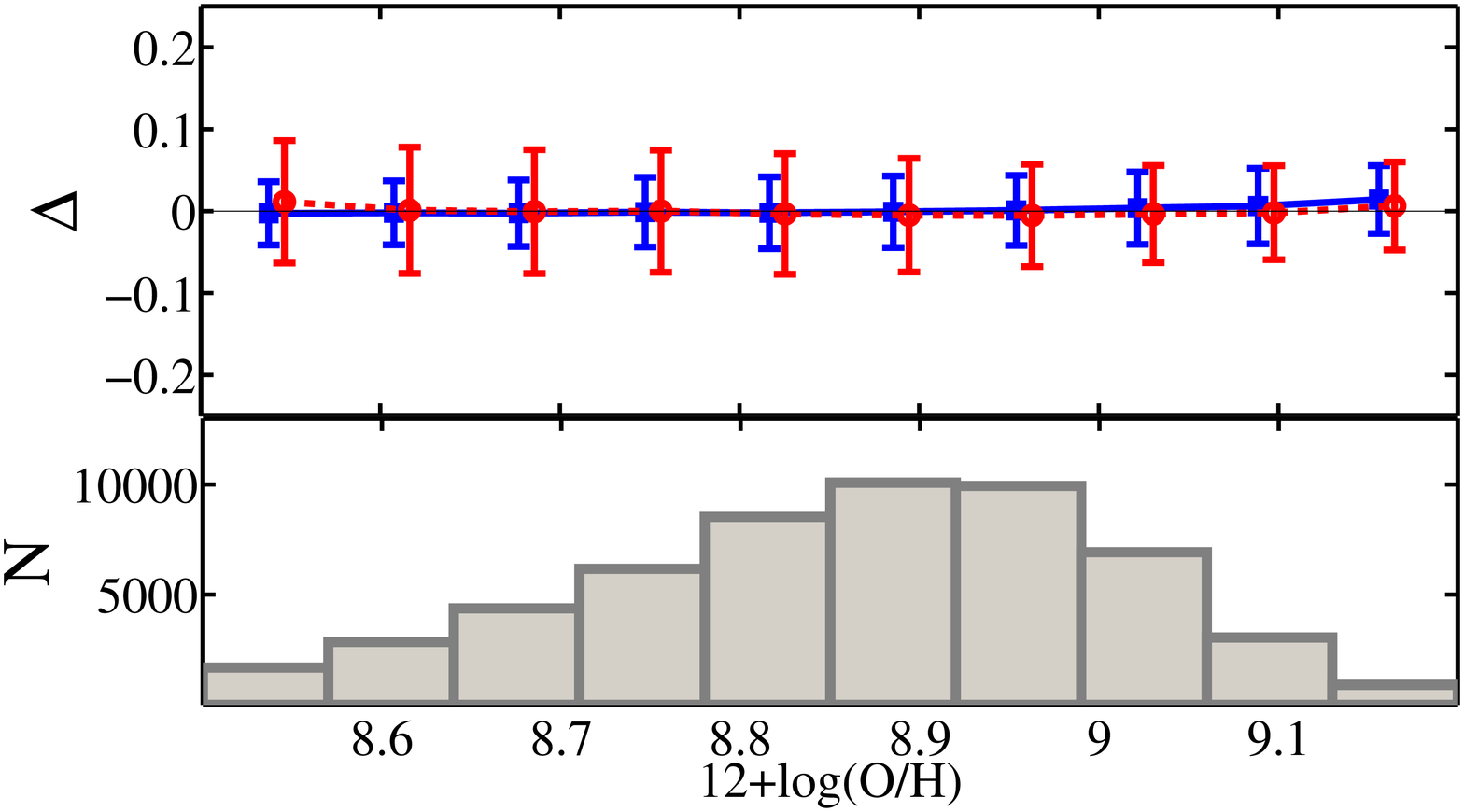}
\includegraphics[width=8.2cm,height=4.1cm,angle=0]{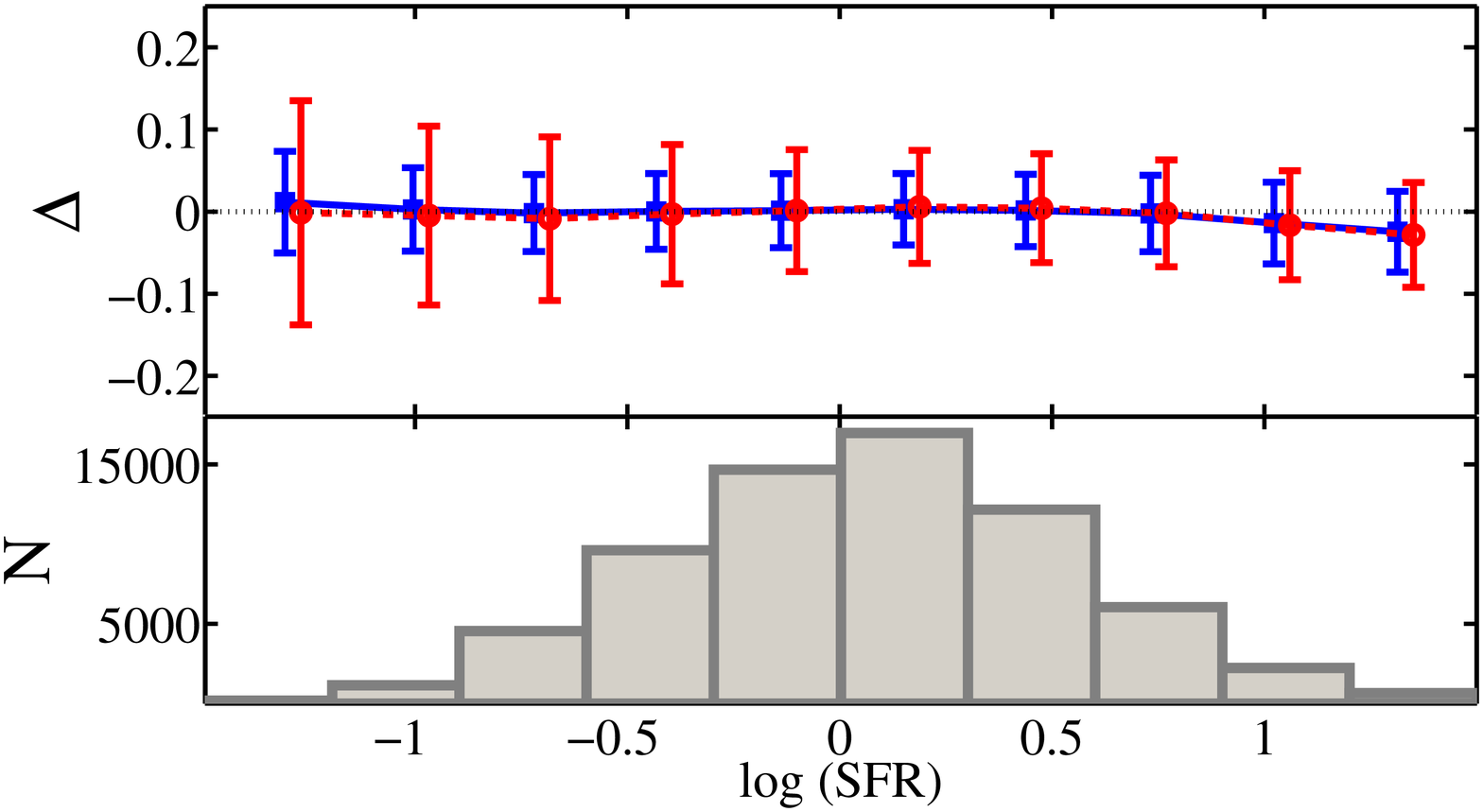}
\caption{The difference in the estimated and observed parameters H$\alpha$
and [NII] vs. total stellar mass, covering fraction (CF),
metallicity and SFR,  for star forming galaxies}
\label{fig-delta-sf}
\end{figure}

\begin{figure}
\centering
\includegraphics[width=8.2cm,height=4.1cm,angle=0]{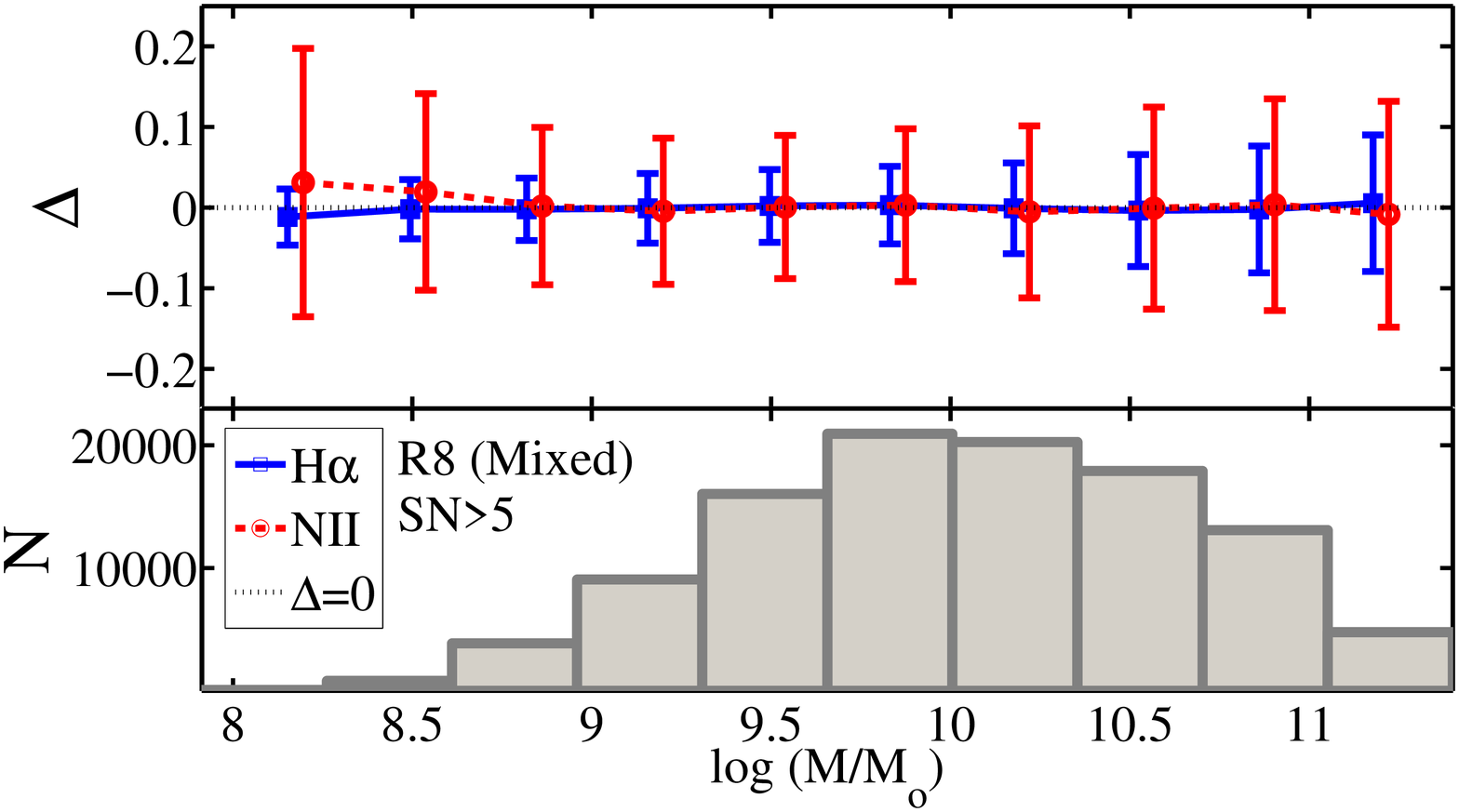}
\includegraphics[width=8.2cm,height=4.1cm,angle=0]{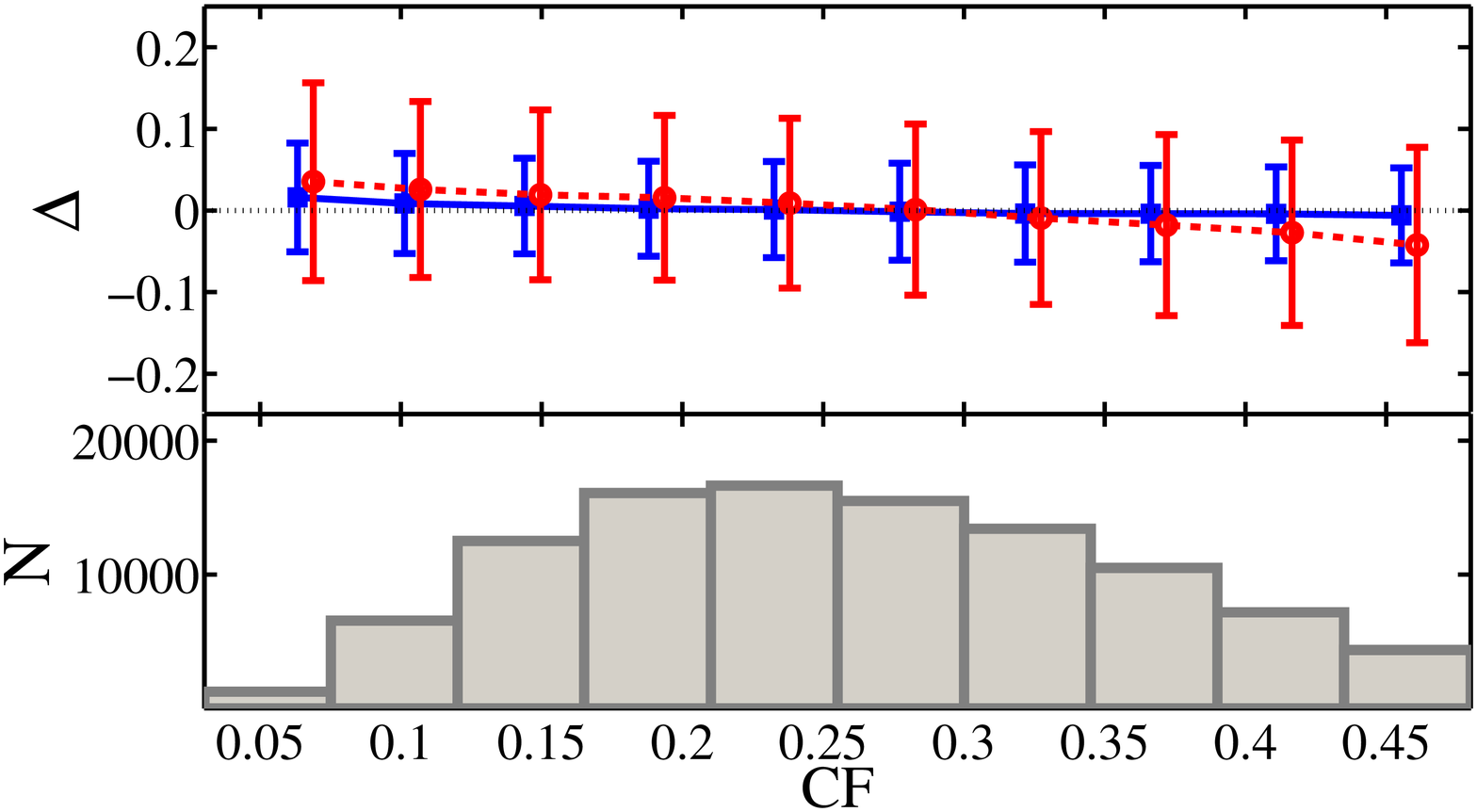}
\caption{The difference in the estimated and observed parameters H$\alpha$
and [NII] vs. total stellar mass and covering fraction (CF), for the
mixed galaxies.}
\label{fig-delta-mix}
\end{figure}

In Figures \ref{fig-delta-sf} and \ref{fig-delta-mix} we test the performance
of the ANN as a function of various physical and observational
parameters, for star-forming and mixed galaxies, respectively.
In each figure we calculate the offset of the line luminosity predictions
(of either H$\alpha$ or [NII]) relative to the observed luminosities as

\begin{equation}
\Delta = \rm{log (L_{pred}/L_{obs})}.
\end{equation}

The blue lines in Figures \ref{fig-delta-sf} and \ref{fig-delta-mix}
show the offsets for H$\alpha$ and the red lines show offsets for
[NII].  In all panels of these two figures, we adopt the results
of run R8 with a S/N$>$5 for demonstration purposes, although our
conclusions do not depend on the choice of run.

For star-forming galaxies, we investigate
in Figure \ref{fig-delta-sf}, the difference in the predicted
and observed line luminosities as a function of total stellar mass,
$r$-band covering fraction (CF), metallicity and SFR.  The $r$-band
covering fraction is simply the fraction of light in the SDSS
$r$-band that is included by the fibre, and is computed from
the combination of total and fibre magnitudes.

From Figure \ref{fig-delta-sf} it can be seen that
there is no systematic offset of $\Delta$ from zero as a function of total
stellar mass or SFR for either H$\alpha$ or [NII].  The results are also
equally robust to changes in the covering fraction, indicating that
the calibrations provided in this paper may be applied to galaxies
with a range of masses and obtained with a variety of apertures.
Nonetheless, there is an extremely mild trend (although
statistically insignificant) of $\Delta$ [NII]
with covering fraction, such that the ANN slightly over-predicts the
luminosity at large covering fractions, and slightly under-predicts the
luminosity at the smallest apertures.  The third panel of  Figure \ref{fig-delta-sf}
shows that this is unlikely to be due abundance gradients, since
$\Delta$ again shows no strong dependence on O/H.

Similar trends are seen for the mixed sample of star-forming plus AGN
galaxies in Figure  \ref{fig-delta-mix}: there is no trend of $\Delta$
H$\alpha$ with total stellar mass, although the scatter is larger.
The larger scatter is to be expected from Figure \ref{fig-std-ha-nii-sf-mix}
where we have shown that the mixed class has a higher scatter in H$\alpha$
and [NII] for most of the runs, compared with the star-forming only
sample.  The same mild, though statistically insignificant, trend is
seen with covering fraction.  No comparison is done for metallicity or
SFR in
the mixed sample,
since the presence of an AGN negates the application of strong line
metallicity and SFR calibrations.

We conclude that the calibrations presented in this paper show
no systematic deviations as a function of stellar mass, SFR, covering
fraction or metallicity, and should hence be applicable to
a wide range of observed data.

\subsection{Application to an independent dataset}

In order to test the assertion of the previous subsection, we now
demonstrate the applicability of the calibrations we have
derived in this work to another dataset.  Specifically, we will
use the publicly available data of the Galaxy
and Mass Assembly (GAMA) survey\footnote{www.gama-survey.org}.
 The GAMA survey is a joint European-Australasian project based
around a spectroscopic campaign using the Anglo-Australian Telescope
(AAT). The GAMA target catalogue is based on selection from the
SDSS (\citealt{Baldry-10,Robotham-10}), but provides independent
spectroscopy and derived data products (e.g. \citealt{Driver-11,Hill-11,
Taylor-11}).   We compile emission line fluxes, redshifts and
fibre stellar masses from the publicly available GAMA data release 2.
Total stellar masses are derived using the GAMA-provided flux scales
appropriate to our cosmology \citep{Taylor-11}.  Balmer emission
line fluxes have already been corrected for underlying stellar absorption.
We impose a redshift cut $z > 0.003$ in order to avoid stellar confusion.

\begin{figure}
\centering
\includegraphics[width=8.6cm,height=4.2cm,angle=0]{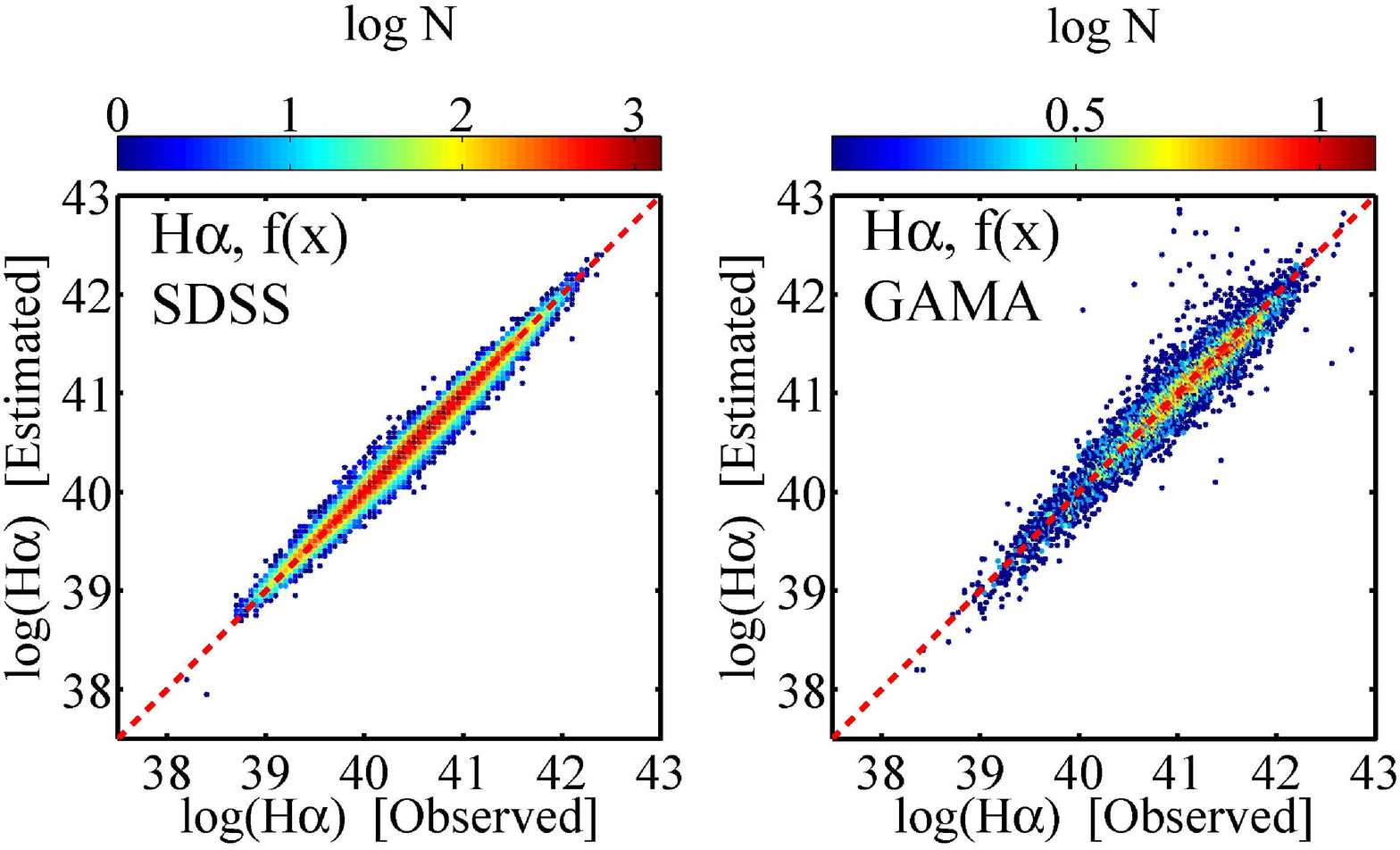}
\includegraphics[width=8.6cm,height=4.2cm,angle=0]{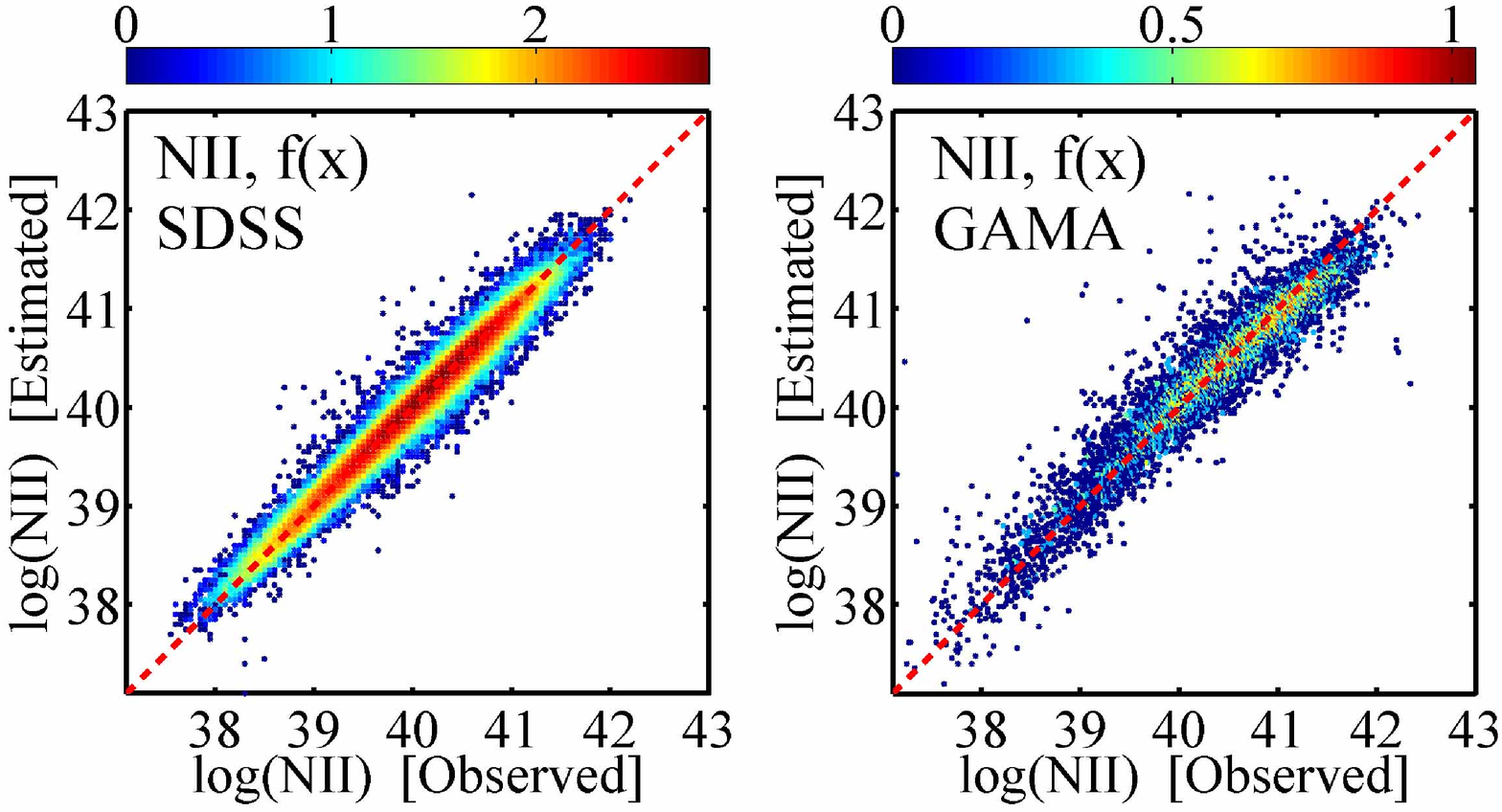}
\caption{A comparison between data taken from SDSS (left panels) and
GAMA (right panels)  using the  f(x) method for run R8.
The top and the bottom panels are related to H$\alpha$ and [NII],
respectively. All the luminosities are in units of erg/s. }
\label{fig-gama-hn}
\end{figure}

In order to test the application of the basis function coefficients
presented in Appendix A, we select a sample of mixed
emission line galaxies from GAMA that are consistent with the
requirements of run R8.  That is, we select the $\sim$ 8000 GAMA
galaxies with well determined stellar masses and with emission
lines [OIII] and H$\beta$ having S/N$>$3.  The predicted line
luminosities of H$\alpha$ (upper panel) and [NII] (lower panel) obtained
 from Eq. \ref{eq-basis-function}, f(x), are shown in the right
hand panels of Figure \ref{fig-gama-hn}.
For a comparison, we plot the results obtained
with the SDSS data set in the left panels. The two datasets show a
similar behavior, although the GAMA data set shows a slightly larger scatter
($\sim0.05$ dex) compared to the SDSS. A small fraction of GAMA sample
($\sim1\%$) have very large offsets. In the SDSS data set, this kind of
offset is not seen in the validation and  training sets for either
the ANN method nor  for the f(x) method. It is likely that the
high scatter is due (at least in part) to the different methods
and accuracies of calibration, data processing and stellar mass
determination.  For example, the GAMA data show
a higher mean value for the two input emission lines and an average
0.25 dex difference in the stellar masses. Despite these differences,
and the differences that are likely inherent to other datasets,
Figure \ref{fig-gama-hn}
demonstrates that the methods presented here yield generally very
good line luminosity predictions.  However, we close this section
with one final caveat.  The ANN-based calibrations that we have presented
in this paper were obtained using a low redshift training set (the SDSS).
At high redshifts (above 1.0 -- 1.5), it seems likely that the pattern of line
ratios on which the ANN relies will change  (\citealt{Liu-08,Brinchmann-08,Kewley-13b}), and the calibrations
presented here are likely to not be applicable.

\section{Example applications of line luminosity predictions}\label{app_sec}

Having tested the robustness of the ANN method, in this final section,
we provide two example applications of the line luminosity predictions.

\subsection{Constructing the BPT diagram}

\begin{figure}
\centering
\includegraphics[width=8.6cm,height=4.2cm,angle=0]{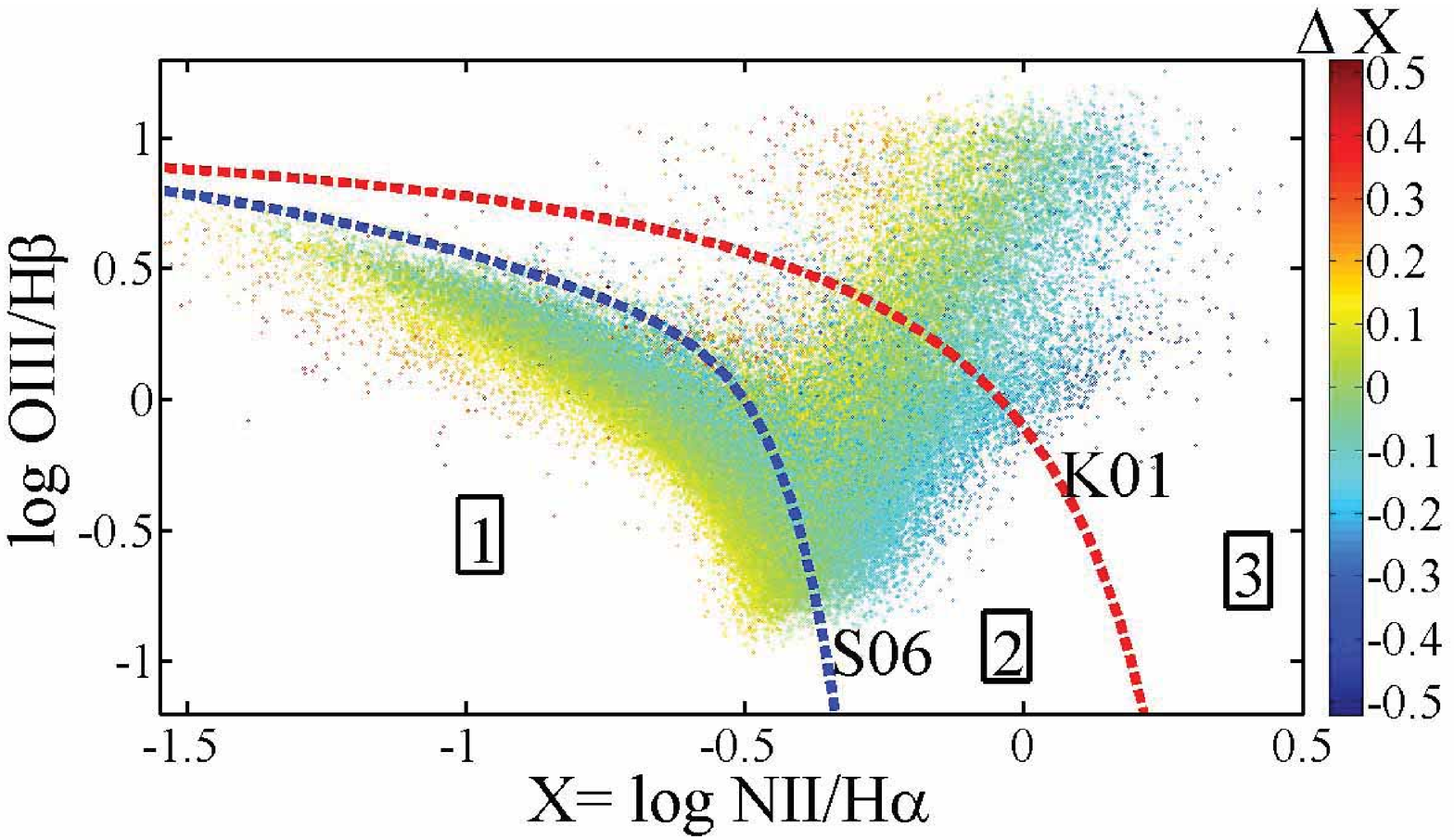}
\includegraphics[width=8.6cm,height=4.2cm,angle=0]{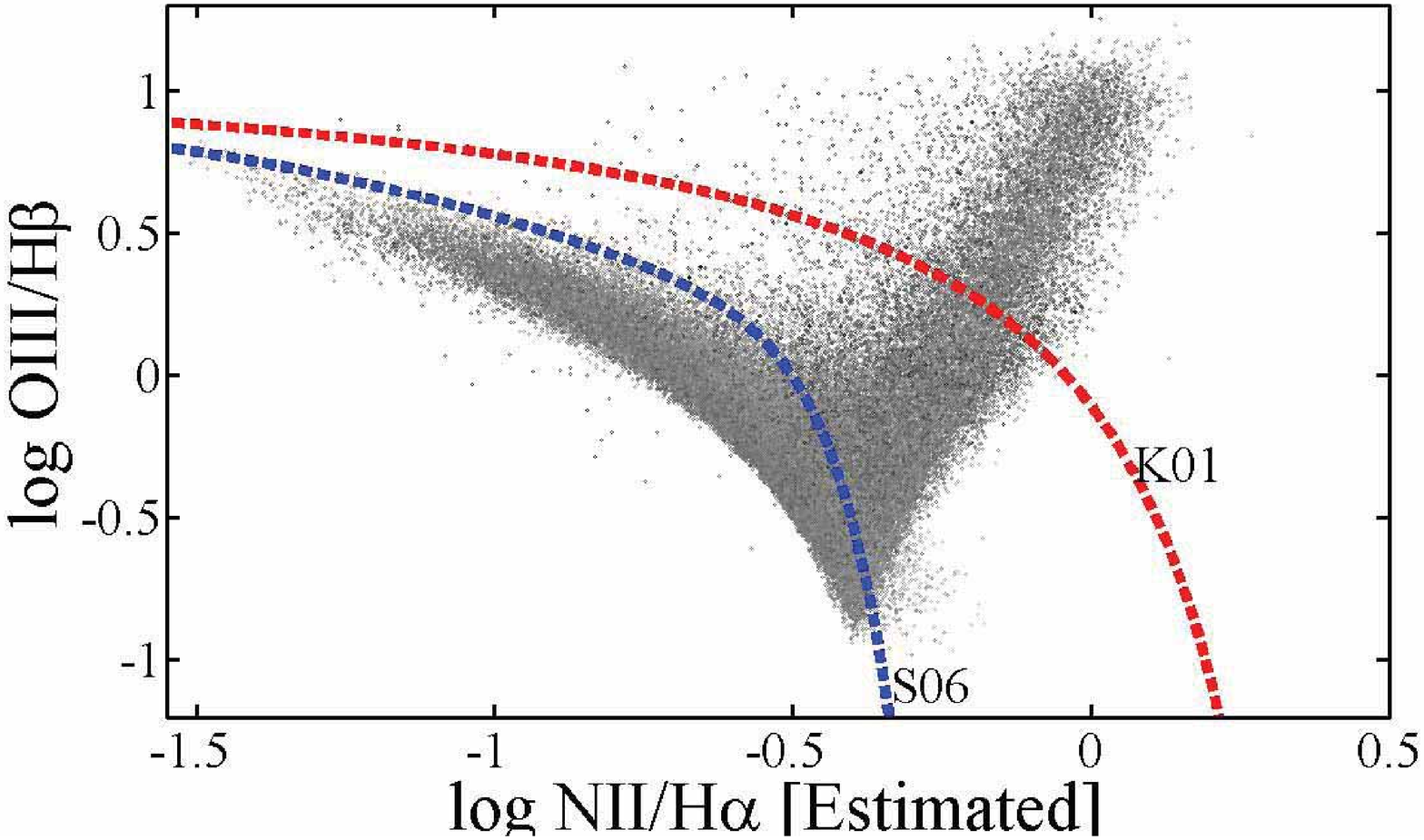}
\includegraphics[width=8.2cm,height=4.2cm,angle=0]{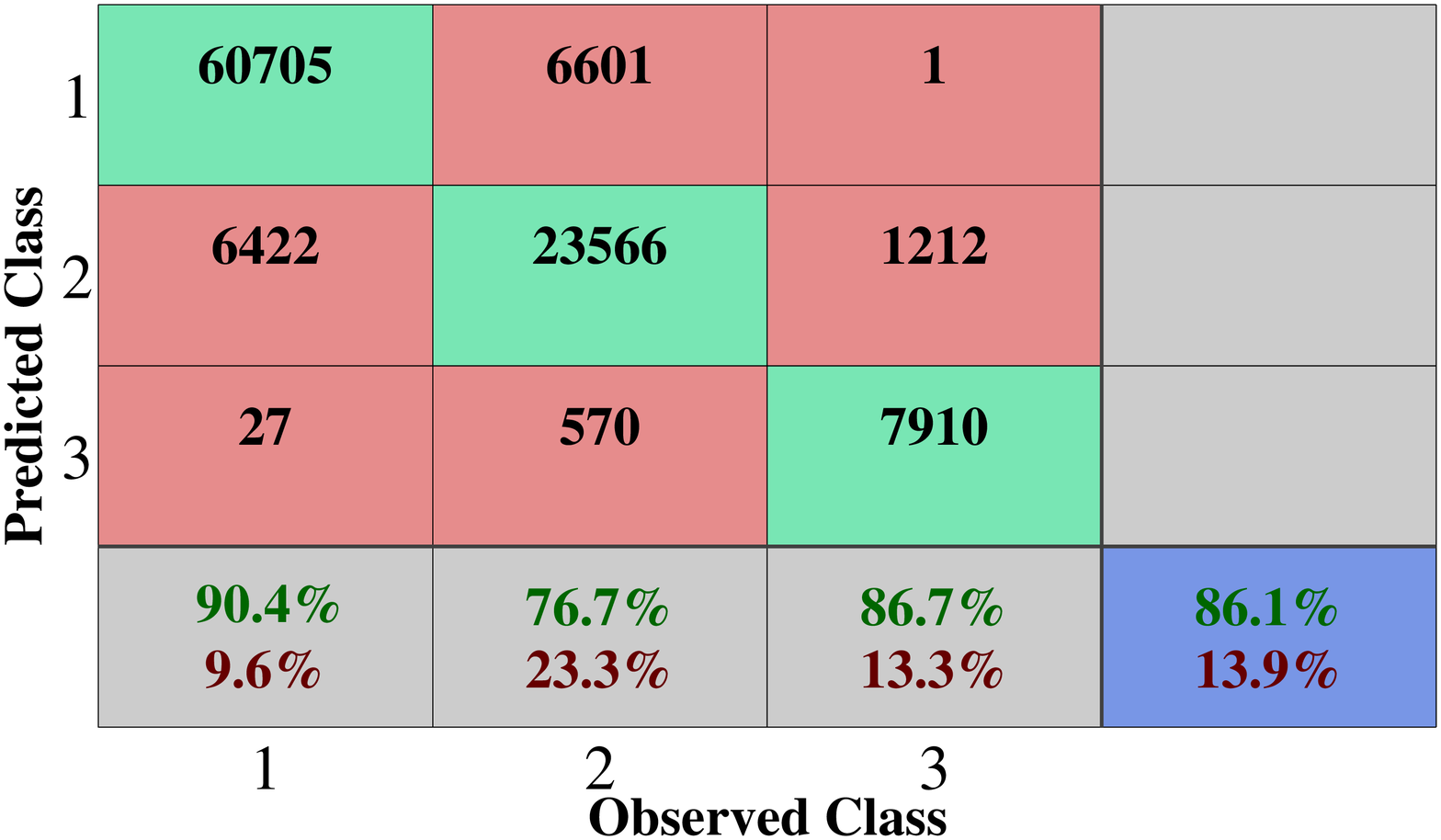}
\caption{Top plot: The BPT diagram associated with Run 8. The horizontal axis
shows the observed value of log ([NII]/H${\alpha}$) as X.  The points are
colour-coded as $\Delta$X = log ([NII]/H$_{\alpha})_{(pred)}$ - log ([NII]/H${\alpha})_{(obs)}$.
The dotted red and blue lines  indicate the demarcation lines given by K01, and S06.
The middle panel shows the BPT diagram when we use the predicted value of [NII]/H${\alpha}$.
The difference between the two BPT diagrams is represented in the bottom panel as
a confusion matrix.  86 per cent of SDSS galaxies are correctly classified.
See text for more explanation.}
\label{fig-BPT}
\end{figure}

The BPT diagram is an important  tool for distinguishing AGN and SF galaxies
from each other. To demonstrate the application of our method (which we
have shown works well for both star-forming and AGN dominated galaxies)
in this regard, we compare the [NII] and H${\alpha}$ of the observed
(original) data with those that are predicted by our method  and construct
 two BPT diagrams.  The top panel of Figure \ref{fig-BPT} shows a BPT diagram
that is constructed for Run 8 for the mixed sample. The horizontal axis
shows the observed value of log (NII/H${\alpha}$) as X.  The points
are colour-coded by $\Delta$X = log (NII/H${\alpha})_{(pred)}$ - log
(NII/H${\alpha})_{(obs)}$. Almost all [NII]/H$\alpha$ line ratios are predicted
within $\Delta X<0.2$ dex and $\sigma\sim0.1$ dex. The dotted red and
blue lines indicate the demarcation lines given by \cite{K-01} (K01), and
\cite{Stasinska-06} (S06), respectively.  We thus label  three classes in the
plot: Class 1=SF, Class2=composite, Class3=AGN, for use in the confusion
matrix (described
below). The middle panel shows the BPT diagram obtained
when we use the predicted values of NII/H$\alpha$.

The bottom panel of Figure \ref{fig-BPT} shows
the confusion matrix, which quantifies how well the predicted [NII] and
H${\alpha}$ luminosities translate to correct AGN classifications.
The numbers in the squares
of the confusion matrix indicate the number of galaxies thus classified.
Rows represent the predicted class (1, 2 or 3) and columns represent
the actual observed class. For example, if  column 1 is added up
(60705+6422+27), the total number of galaxies in class1 (star-forming)
is obtained.  The green squares show the numbers of galaxies whose
predicted class is the same as the observed class. The 6 red boxes give the
numbers of galaxies incorrectly classified.  For example,  only one galaxy
is mis-classified from class3 to class1, whereas 27 galaxies go from class1
to class3 and so on.  In the bottom row,
the green and red values give the percentages of galaxies correctly or
incorrectly (respectively) classified for a given class.  For example,
90.4 per cent of SF galaxies are correctly classified; 9.6$\%$ of the
SF galaxies go to other classes. The blue block shows that 86.1$\%$
of galaxies in all classes are correctly placed, demonstrating the
successful application of our method for AGN classifications.

\subsection{Metallicity calibrations and conversions}

There are many metallicity calibrations that have been developed from
different methodologies and datasets. \cite{KE08} investigate the SDSS
mass-metallicity relation with 10 different metallicity calibrations,
encompassing those based on both theoretical and empirical methods.
Several of these methods are variants on the R$_{23}$ method, requiring
[OII] $\lambda$ 3727, [OIII] $\lambda$ 4959, [OIII$] \lambda$ 5007
and H$\beta$ (e.g., \citealt{KK-04}, KK04).  R$_{23}$ diagnostics have
the advantage of requiring only 3 emission lines\footnote{In practice,
only [OIII] $\lambda$5007, the stronger member of the doublet, is
needed, as the total [OIII] flux can be obtained by scaling.} spanning a relatively
narrow wavelength range.  The well-known principal disadvantage of
R$_{23}$ calibrations is the need to break the degeneracy between
upper and lower branches.  [NII] is frequently used to achieve this
(e.g. Kewley \& Ellison 2008), so one application of the line predictions
presented in this work is to break this degeneracy.

A second metallicity application of the predictions presented herein
involves conversions between metallicity indicators.  As demonstrated
by Kewley \& Ellison (2008), the factor of five spread in metallicity
between calibrations may hinder the combination of different datasets.
Kewley \& Ellison therefore derive polynomial fits between metallicities
derived from different calibrations, which can be used to convert between
diagnostics.  However, an alternative to converting via the Kewley \&
Ellison polynomials, is to predict the additional line fluxes needed to
calculate the metallicity in the desired diagnostic.

\begin{figure}
\centering
\includegraphics[width=8.6cm,height=4.2cm,angle=0]{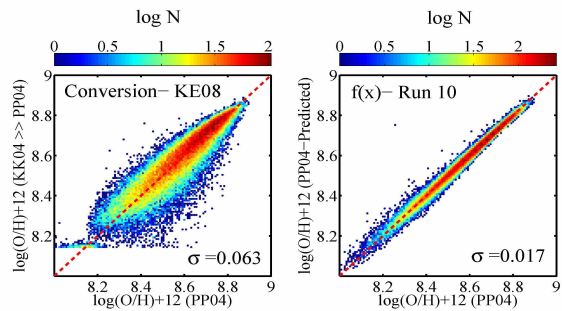}
\caption{In both panels, the measured metallicity from the PP04
diagnostic is shown on the x-axis.  This is compared, in
the left panel, with the PP04 metallicity derived from the
KK04 R$_{23}$ diagnostic plus the conversion given in Kewley
\& Ellison (2008). The right panel shows the case in which the
PP04 metallicity is obtained using Run 10 and the polynomial-based
prediction of [NII] and H$\alpha$.  The ANN-based approach yields a
much tighter relation with the directly measured metallicities.}
\label{fig-metal}
\end{figure}

As an example, we consider the case where we would like to convert
an R$_{23}$ based calibration to one which uses [NII] and H$\alpha$,
(e.g., \citealt{PP-04},  PP04).  We use Run 10 in this demonstration,
since it combines the emission lines used in the basic R$_{23}$ diagnostic.
We will consider the PP04 metallicity as our `target' metallicity, but
imagine the case where only [OII] $\lambda$ 3727, [OIII] $\lambda$ 4959,
[OIII$] \lambda$ 5007 and H$\beta$ are observed.  We begin by adopting
the predictive approach described in this paper, namely we predict the
[NII] and H$\alpha$ luminosities using the polynomial functions described
in Section \ref{fx_sec} for Run 10, and can hence calculate PP04
metallicities.  It is of course possible to calculate real PP04 metallicities
for our sample, since the SDSS data have measured values of [NII]
and H$\alpha$.  We can hence compare the PP04 metallicities measured in
the data, with that based on predicted [NII] and H$\alpha$ line
luminosities.  The right panel of Figure \ref{fig-metal} compares the
predicted and measured PP04 metallicities.  Clearly, the agreement is
excellent, with small scatter (0.017 dex) and very few outliers.

Alternatively, we can repeat our experiment that begins with the
observed [OII] $\lambda$ 3727, [OIII] $\lambda$ 4959,
[OIII$] \lambda$ 5007 and H$\beta$, and first calculate an R$_{23}$
metallicity.  We adopt the KK04 calibration for this test.  We
can then use the conversion in Kewley \& Ellison (2008) to determine
the equivalent PP04 metallicity. The result is shown in the
left panel of Figure \ref{fig-metal}. The horizontal axis is the same as the
right panel, namely the PP04 metallicity that can be determined directly
from the data. The vertical axis is now the metallicity of PP04 that has
been converted from KK04. Although the overall correlation is good,
the scatter is larger (0.063 dex) and there are some cases (albeit a minority)
where the conversion fails (see the region of converted metallicities
of log O/H + 12 $\sim$ 8.2).  Whilst the conversions of Kewley \& Ellison
(2008) are still advantageous in cases where line fluxes are not provided
(i.e. only metallicities are quoted), the more complex relations
encapsulated by the ANN-based calibrations offer an alternative route to
robust metallicity conversions.

\section{Conclusions}

Artificial neural networks (ANNs) can be used to predict spectral line
luminosities that are useful for a variety of applications.  After an overview of formalism
of the ANN, we present a simple case study for the prediction of the
Balmer decrement (H$\alpha$/H$\beta$) which is compared to the
analytic prediction of Groves et al. (2012).  Since the ANN predicts
the expected observed (rather than intrinsic) line luminosity, our
prediction of H$\alpha$ can be combined with the actual observed
H$\beta$ in order to determine the galaxy's internal extinction.
We show that using
the same input data, we can decrease the scatter in the predictions
of the decrement from 0.058 (Groves et al. 2012) to 0.038 dex
(ANN).

We then present an application of artificial neural networks
to the prediction of H$\alpha$ and [NII] $\lambda$ 6584 luminosities in
galaxy spectra.  The ANN is trained on large samples (many tens of thousands)
of galaxies from the SDSS, and tested for biases in training
sample size, galaxy properties and aperture covering fraction.
Twelve runs with different input parameters are tested, representing
varying combinations of stellar masses and strong emission lines.
Many of the runs can predict line luminosities with scatters $<$ 0.1 dex,
regardless of whether the galaxy is star-forming or has an AGN and
whether total or fibre stellar mass is used.  These results indicate
that the ANN is a robust tool for predicting the line luminosity of H$\alpha$
and [NII] in low redshift galaxies.  Indeed, testing on an independent
dataset (GAMA) yields scatter in the line luminosities only
$\sim$ 0.05 dex larger than the SDSS data set,
likely due to different calibration/reduction procedures.

In order
to demonstrate potential applications of our method, we investigate
the measurement of gas-phase metallicities and AGN classification.
The ANN yields a considerable improvement in the prediction of
gas-phase metallicities from diagnostics that require the [NII] and
H$\alpha$ lines, compared to the empirically determined conversions of
Kewley \& Ellison (2008) and an AGN classification is successful in 86
per cent of SDSS galaxies.

In order to make the ANN results useful to the general community,
we have derived polynomial basis function coefficients in a
condensed matrix form that accurately reproduces the ANN results.

\section*{Acknowledgments}

SLE acknowledges the receipt of NSERC Discovery grants which
funded this research.  We are grateful to Trevor Mendel for providing
the SDSS magnitude corrections used in this work and to Maritza
Lara-Lopez and Joe Liske for advice on GAMA data retrieval.

Funding for the SDSS and SDSS-II has been provided by the Alfred
P. Sloan Foundation, the Participating Institutions, the National
Science Foundation, the U.S. Department of Energy, the National
Aeronautics and Space Administration, the Japanese Monbukagakusho, the
Max Planck Society, and the Higher Education Funding Council for
England. The SDSS Web Site is http://www.sdss.org/.

The SDSS is managed by the Astrophysical Research Consortium for the
Participating Institutions. The Participating Institutions are the
American Museum of Natural History, Astrophysical Institute Potsdam,
University of Basel, University of Cambridge, Case Western Reserve
University, University of Chicago, Drexel University, Fermilab, the
Institute for Advanced Study, the Japan Participation Group, Johns
Hopkins University, the Joint Institute for Nuclear Astrophysics, the
Kavli Institute for Particle Astrophysics and Cosmology, the Korean
Scientist Group, the Chinese Academy of Sciences (LAMOST), Los Alamos
National Laboratory, the Max-Planck-Institute for Astronomy (MPIA),
the Max-Planck-Institute for Astrophysics (MPA), New Mexico State
University, Ohio State University, University of Pittsburgh,
University of Portsmouth, Princeton University, the United States
Naval Observatory, and the University of Washington.

GAMA is funded by the STFC (UK), the ARC (Australia), the AAO, and the
participating institutions.

\clearpage
\appendix
\section{The coefficients of the basis function}\label{coeff_sec}

Here, we present the coefficients of Eq. \ref{eq-basis-function} for
star-forming and mixed galaxies for all of the runs of Table
\ref{sf_runs}. In each case we have: $f(C,X)=X^{T} C X + X^{T} C^{'} + C^{''}$.
We present the result for the S/N$>5$ cut, although as shown in this paper,
the dependence on S/N is not large.  We first present the coefficients
for star-forming galaxies in Sections \ref{sf_app_ha} and \ref{sf_app_n2}
for H$\alpha$ and [NII], respectively.  The coefficients for H$\alpha$ and [NII] in the
mixed galaxies are in Sections \ref{mix_app_ha} and \ref{mix_app_n2}.   In all cases, the masses
are total values and are in
units of solar masses and luminosities are in units of erg/s.

\subsection{Star forming galaxies, H$\alpha$}\label{sf_app_ha}

\subsubsection{Run 1}
\[ X= \left( \begin{array}{c}
\rm{log}(M_{tot}/M_\odot)\\
\end{array} \right)\]
\[ C = \left( \begin{array}{rrr}
-0.03296\\
\end{array} \right)\]
\[ C^{'} = \left( \begin{array}{r}
1.24672\\
\end{array} \right)\]
$C^{''}$= 31.25272\\

$\sigma$=0.460 dex

\subsubsection{Run 2}

\[ X= \left( \begin{array}{c}
\rm{log}~L(OIII) \lambda 5007\\
\end{array} \right)\]
\[ C = \left( \begin{array}{rrr}
-0.15360\\
\end{array} \right)\]
\[ C^{'} = \left( \begin{array}{r}
13.10619\\
\end{array} \right)\]
$C^{''}$= -237.75141\\

$\sigma= $0.328 dex

\subsubsection{Run 3}
\[ X= \left( \begin{array}{c}
\rm{log}~L(OII) \lambda 3727\\
\end{array} \right)\]
\[ C = \left( \begin{array}{rrr}
-0.09694\\
\end{array} \right)\]
\[ C^{'} = \left( \begin{array}{r}
8.87324\\
\end{array} \right)\]
$C^{''}$= -159.23380\\

$\sigma=$0.202 dex

\subsubsection{Run 4}
\[ X= \left( \begin{array}{c}
\rm{log}~ L(H\beta)\\
\end{array} \right)\]
\[ C = \left( \begin{array}{rrr}
-0.01622\\
\end{array} \right)\]
\[ C^{'} = \left( \begin{array}{r}
2.34114\\
\end{array} \right)\]
$C^{''}$= -27.09286\\

$\sigma=$0.067 dex

\subsubsection{Run 5}
\[ X= \left( \begin{array}{c}
\rm{log}(M_{tot}/M_\odot)\\
\rm{log}~L(OIII) \lambda 5007\\
\end{array} \right)\]
\[ C = \left( \begin{array}{rrr}
-0.02273&-0.02675\\
-0.02675&-0.05434\\
\end{array} \right)\]
\[ C^{'} = \left( \begin{array}{r}
3.14945\\
5.59911\\
\end{array} \right)\]
$C^{''}= $-103.99888\\

$\sigma=$0.174 dex

\subsubsection{Run 6}
\[ X= \left( \begin{array}{c}
\rm{log}(M_{tot}/M_\odot)\\
\rm{log}~L(OII) \lambda 3727\\
\end{array} \right)\]
\[ C = \left( \begin{array}{rrr}
0.15201&-0.07548\\
-0.07548&0.00162\\
\end{array} \right)\]
\[ C^{'} = \left( \begin{array}{r}
3.44755\\
2.25416\\
\end{array} \right)\]
$C^{''}= $-41.39747\\

$\sigma=$0.129 dex

\subsubsection{Run 7}
\[ X= \left( \begin{array}{c}
\rm{log}(M_{tot}/M_\odot)\\
\rm{log}~ L(H\beta)\\
\end{array} \right)\]
\[ C = \left( \begin{array}{rrr}
0.01016&-0.00472\\
-0.00472&-0.00085\\
\end{array} \right)\]
\[ C^{'} = \left( \begin{array}{r}
0.26847\\
1.15591\\
\end{array} \right)\]
$C^{''}= $-4.20487\\

$\sigma$=0.051 dex

\subsubsection{Run 8}
\[ X= \left( \begin{array}{c}
\rm{log}(M_{tot}/M_\odot)\\
\rm{log}~L(OIII) \lambda 5007\\
\rm{log}~L(H\beta)\\
\end{array} \right)\]
\[ C = \left( \begin{array}{rrr}
0.03111&-0.01046&-0.01025\\
-0.01046&-0.00021&-0.00688\\
-0.01025&-0.00688&0.02283\\
\end{array} \right)\]
\[ C^{'} = \left( \begin{array}{r}
1.12673\\
0.72863\\
-0.04120\\
\end{array} \right)\]
$C^{''}= $1.09006\\

$\sigma=$0.046 dex

\subsubsection{Run 9}
\[ X= \left( \begin{array}{c}
\rm{log}(M_{tot}/M_\odot)\\
\rm{log}~L(OII) \lambda 3727\\
\rm{log}~ L(H\beta)\\
\end{array} \right)\]
\[ C = \left( \begin{array}{rrr}
0.02220&-0.00141&-0.00747\\
-0.00141&-0.08002&0.08326\\
-0.00747&0.08326&-0.08025\\
\end{array} \right)\]
\[ C^{'} = \left( \begin{array}{r}
0.34978\\
-0.36266\\
1.02108\\
\end{array} \right)\]
$C^{''}= $5.64838\\

$\sigma$=0.042 dex

\subsubsection{Run 10}
\[ X= \left( \begin{array}{c}
\rm{log}(M_{tot}/M_\odot)\\
\rm{log}~ L(OII) \lambda 3727\\
\rm{log}~ L(OIII) \lambda 5007\\
\rm{log}~ L(H\beta)\\
\end{array} \right)\]
\[ C = \left( \begin{array}{rrrr}
0.02388&-0.00471&-0.00838&-0.00002\\
-0.00471&0.02498&-0.01381&0.00751\\
-0.00838&-0.01381&-0.06316&0.07245\\
-0.00002&0.00751&0.07245&-0.08338\\
\end{array} \right)\]
\[ C^{'} = \left( \begin{array}{r}
0.65118\\
-1.56593\\
0.51798\\
1.42717\\
\end{array} \right)\]
$C^{''}$=9.82980\\

$\sigma$=0.041 dex

\subsubsection{Run 11}
\[ X= \left( \begin{array}{c}
\rm{log}(M_{tot}/M_\odot)\\
\rm{log}~ L(OII) \lambda 3727\\
\rm{log}~ L(OII) \lambda 3729\\
\rm{log}~ L(OIII) \lambda 4959\\
\rm{log}~ L(OIII) \lambda 5007\\
\rm{log}~ L(H\beta)\\
\end{array} \right)\]
\[ C = \left( \begin{array}{rrrrrr}
0.00849&-0.01119&-0.02542&-0.00758&-0.00338&0.04340\\
-0.01119&0.24333&-0.28838&0.00784&0.03451&-0.00215\\
-0.02542&-0.28838&0.18608&0.16751&-0.15217&0.08324\\
-0.00758&0.00784&0.16751&-0.18046&0.07107&-0.05579\\
-0.00338&0.03451&-0.15217&0.07107&-0.08688&0.13757\\
0.04340&-0.00215&0.08324&-0.05579&0.13757&-0.17301\\
\end{array} \right)\]
\[ C^{'} = \left( \begin{array}{r}
0.22781\\
0.55526\\
0.75596\\
-0.92821\\
-0.09195\\
1.13597\\
\end{array} \right)\]
$C^{''}$=-9.22439\\

$\sigma$=0.037 dex
\clearpage

\subsubsection{Run 12}

\[ X= \left( \begin{array}{c}
\rm{log}(M_{tot}/M_\odot)\\
\rm{log}~ L(OII) \lambda 3727 \\
\rm{log}~ L(OII) \lambda 3729\\
\rm{log}~ L(OIII) \lambda 4959\\
\rm{log}~ L(OIII) \lambda 5007\\
\rm{log}~ L(H\beta)\\
\rm{log}~ L(SII) \lambda 6717\\
\rm{log}~ L(SII) \lambda 6731\\
\end{array} \right)\]
\[ C = \left( \begin{array}{rrrrrrrr}
0.00968&-0.02743&-0.01413&0.00426&0.00049&0.03736&0.05072&-0.05871\\
-0.02743&0.07812&-0.02066&0.00922&-0.08175&0.23562&-0.21780&0.00315\\
-0.01413&-0.02066&0.02049&0.01218&-0.05831&0.08275&0.25872&-0.30051\\
0.00426&0.00922&0.01218&-0.3118&0.31590&-0.11115&0.18424&-0.09006\\
0.00049&-0.08175&-0.05831&0.31590&-0.15913&-0.08423&-0.10801&0.16496\\
0.03736&0.23562&0.08275&-0.11115&-0.08423&0.35655&-0.49776&-0.00398\\
0.05072&-0.21780&0.25872&0.18424&-0.10801&-0.49776&-0.44227&0.80051\\
-0.05871&0.00315&-0.30051&-0.09006&0.16496&-0.00398&0.80051&-0.52487\\
\end{array} \right)\]
\[ C^{'} = \left( \begin{array}{r}
0.40251\\
-0.11074\\
0.41208\\
-1.06788\\
1.33069\\
1.47554\\
1.53207\\
-2.62153\\
\end{array} \right)\]
$C^{''}$=-1.41523\\

$\sigma$=0.026 dex
\clearpage

\subsection{Star forming galaxies, NII}\label{sf_app_n2}

\subsubsection{Run 1}
\[ X= \left( \begin{array}{c}
\rm{log}(M_{tot}/M_\odot)\\
\end{array} \right)\]
\[ C = \left( \begin{array}{rrr}
-0.16813\\
\end{array} \right)\]
\[ C^{'} = \left( \begin{array}{r}
4.13524\\
\end{array} \right)\]
$C^{''}$= 15.37728\\

$\sigma$=0.452 dex

\subsubsection{Run 2}
\[ X= \left( \begin{array}{c}
\rm{log}~L(OIII) \lambda 5007\\
\end{array} \right)\]
\[ C = \left( \begin{array}{rrr}
-0.22813\\
\end{array} \right)\]
\[ C^{'} = \left( \begin{array}{r}
19.00561\\
\end{array} \right)\]
$C^{''}$= -355.07588\\

$\sigma= $0.485 dex

\subsubsection{Run 3}
\[ X= \left( \begin{array}{c}
\rm{log}~L(OII) \lambda 3727\\
\end{array} \right)\]
\[ C = \left( \begin{array}{rrr}
-0.20077\\
\end{array} \right)\]
\[ C^{'} = \left( \begin{array}{r}
17.26172\\
\end{array} \right)\]
$C^{''}$= -329.24057\\

$\sigma=$0.341 dex

\subsubsection{Run 4}
\[ X= \left( \begin{array}{c}
\rm{log}~ L(H\beta)\\
\end{array} \right)\]
\[ C = \left( \begin{array}{rrr}
-0.08098\\
\end{array} \right)\]
\[ C^{'} = \left( \begin{array}{r}
7.59421\\
\end{array} \right)\]
$C^{''}$= -134.11928\\

$\sigma=$0.191 dex

\subsubsection{Run 5}
\[ X= \left( \begin{array}{c}
\rm{log}(M_{tot}/M_\odot)\\
\rm{log}~L(OIII) \lambda 5007\\
\end{array} \right)\]
\[ C = \left( \begin{array}{rrr}
-0.15073&0.00102\\
0.00102&-0.09215\\
\end{array} \right)\]
\[ C^{'} = \left( \begin{array}{r}
3.71279\\
7.98303\\
\end{array} \right)\]
$C^{''}= $-154.44567\\

$\sigma=$0.244 dex

\subsubsection{Run 6}
\[ X= \left( \begin{array}{c}
\rm{log}(M_{tot}/M_\odot)\\
\rm{log}~L(OII) \lambda 3727\\
\end{array} \right)\]
\[ C = \left( \begin{array}{rrr}
-0.00206&-0.02532\\
-0.02532&-0.06718\\
\end{array} \right)\]
\[ C^{'} = \left( \begin{array}{r}
2.73641\\
6.69499\\
\end{array} \right)\]
$C^{''}= $-127.06397\\

$\sigma=$0.189 dex

\subsubsection{Run 7}
\[ X= \left( \begin{array}{c}
\rm{log}(M_{tot}/M_\odot)\\
\rm{log}~ L(H\beta)\\
\end{array} \right)\]
\[ C = \left( \begin{array}{rrr}
-0.15371&0.03475\\
0.03475&-0.03697\\
\end{array} \right)\]
\[ C^{'} = \left( \begin{array}{r}
0.60523\\
3.20966\\
\end{array} \right)\]
$C^{''}= $-47.62000\\

$\sigma$=0.103 dex

\subsubsection{Run 8}
\[ X= \left( \begin{array}{c}
\rm{log}(M_{tot}/M_\odot)\\
\rm{log}~L(OIII) \lambda 5007\\
\rm{log}~L(H\beta)\\
\end{array} \right)\]
\[ C = \left( \begin{array}{rrr}
0.05580&0.04715&-0.09749\\
0.04715&-0.38793&0.37129\\
-0.09749&0.37129&-0.33060\\
\end{array} \right)\]
\[ C^{'} = \left( \begin{array}{r}
3.08934\\
-0.16444\\
0.23358\\
\end{array} \right)\]
$C^{''}= $2.50062\\

$\sigma=$0.077 dex

\subsubsection{Run 9}
\[ X= \left( \begin{array}{c}
\rm{log}(M_{tot}/M_\odot)\\
\rm{log}~L(OII) \lambda 3727\\
\rm{log}~ L(H\beta)\\
\end{array} \right)\]
\[ C = \left( \begin{array}{rrr}
-0.16622&0.03788&0.01900\\
0.03788&-0.50049&0.43634\\
0.01900&0.43634&-0.43066\\
\end{array} \right)\]
\[ C^{'} = \left( \begin{array}{r}
-1.02627\\
3.84885\\
0.63659\\
\end{array} \right)\]
$C^{''}= $-64.41316\\

$\sigma$=0.103 dex

\subsubsection{Run 10}
\[ X= \left( \begin{array}{c}
\rm{log}(M_{tot}/M_\odot)\\
\rm{log}~ L(OII) \lambda 3727\\
\rm{log}~ L(OIII) \lambda 5007\\
\rm{log}~ L(H\beta)\\
\end{array} \right)\]
\[ C = \left( \begin{array}{rrrr}
0.04888&-0.03780&0.06129&-0.05993\\
-0.03780&-0.02406&0.06520&-0.01936\\
0.06129&0.06520&-0.53989&0.47404\\
-0.05993&-0.01936&0.47404&-0.45100\\
\end{array} \right)\]
\[ C^{'} = \left( \begin{array}{r}
2.09873\\
-1.22501\\
-1.67510\\
2.62808\\
\end{array} \right)\]
$C^{''}$=14.47485\\

$\sigma$=0.070 dex

\subsubsection{Run 11}
\[ X= \left( \begin{array}{c}
\rm{log}(M_{tot}/M_\odot)\\
\rm{log}~ L(OII) \lambda 3727\\
\rm{log}~ L(OII) \lambda 3729\\
\rm{log}~ L(OIII) \lambda 4959\\
\rm{log}~ L(OIII) \lambda 5007\\
\rm{log}~ L(H\beta)\\
\end{array} \right)\]
\[ C = \left( \begin{array}{rrrrrr}
0.01892&-0.05874&-0.02687&0.04600&0.01915&0.00109\\
-0.05874&0.32541&-0.49221&-0.04912&0.24652&-0.03436\\
-0.02687&-0.49221&0.40693&0.21087&-0.24713&0.09164\\
0.04600&-0.04912&0.21087&-0.74986&0.52075&0.03981\\
0.01915&0.24652&-0.24713&0.52075&-0.99480&0.52839\\
0.00109&-0.03436&0.09164&0.03981&0.52839&-0.62532\\
\end{array} \right)\]
\[ C^{'} = \left( \begin{array}{r}
1.34917\\
1.36630\\
2.70244\\
0.47628\\
-4.52718\\
1.91962\\
\end{array} \right)\]
$C^{''}$=-25.86109\\

$\sigma$=0.069 dex
\clearpage
\subsubsection{Run 12}
\[ X= \left( \begin{array}{c}
\rm{log}(M_{tot}/M_\odot)\\
\rm{log}~ L(OII) \lambda 3727 \\
\rm{log}~ L(OII) \lambda 3729\\
\rm{log}~ L(OIII) \lambda 4959\\
\rm{log}~ L(OIII) \lambda 5007\\
\rm{log}~ L(H\beta)\\
\rm{log}~ L(SII) \lambda 6717\\
\rm{log}~ L(SII) \lambda 6731\\
\end{array} \right)\]
\[ C = \left( \begin{array}{rrrrrrrr}
0.01942&-0.04737&-0.03570&0.07703&-0.00425&-0.05106&0.12038&-0.07015\\
-0.04737&0.26655&-0.11395&-0.12365&0.04978&0.34980&-0.74976&0.32137\\
-0.03570&-0.11395&0.18781&-0.00216&-0.17901&0.10025&0.94379&-0.93487\\
0.07703&-0.12365&-0.00216&-0.8032&0.79926&0.09203&0.16379&-0.17860\\
-0.00425&0.04978&-0.17901&0.79926&-0.98218&-0.02844&-0.06786&0.45552\\
-0.05106&0.34980&0.10025&0.09203&-0.02844&-0.05603&-0.57067&0.17117\\
0.12038&-0.74976&0.94379&0.16379&-0.06786&-0.57067&-1.15587&1.28448\\
-0.07015&0.32137&-0.93487&-0.17860&0.45552&0.17117&1.28448&-1.03579\\
\end{array} \right)\]
\[ C^{'} = \left( \begin{array}{r}
0.64958\\
0.62648\\
-0.14575\\
1.89409\\
-2.97822\\
-2.39165\\
10.37734\\
-5.03625\\
\end{array} \right)\]
$C^{''}$=-30.71568\\

$\sigma$=0.060 dex
\clearpage

\subsection{Mixed galaxies, H$\alpha$}\label{mix_app_ha}

\subsubsection{Run 1}
\[ X= \left( \begin{array}{c}
\rm{log}(M_{tot}/M_\odot)\\
\end{array} \right)\]
\[ C = \left( \begin{array}{rrr}
-0.10888\\
\end{array} \right)\]
\[ C^{'} = \left( \begin{array}{r}
2.51728\\
\end{array} \right)\]
$C^{''}$= 26.00809\\

$\sigma$=0.545 dex

\subsubsection{Run 2}
\[ X= \left( \begin{array}{c}
\rm{log}~L(OIII) \lambda 5007\\
\end{array} \right)\]
\[ C = \left( \begin{array}{rrr}
-0.17822\\
\end{array} \right)\]
\[ C^{'} = \left( \begin{array}{r}
15.00137\\
\end{array} \right)\]
$C^{''}$= -274.22246\\

$\sigma= $0.363 dex

\subsubsection{Run 3}
\[ X= \left( \begin{array}{c}
\rm{log}~L(OII) \lambda 3727\\
\end{array} \right)\]
\[ C = \left( \begin{array}{rrr}
-0.11801\\
\end{array} \right)\]
\[ C^{'} = \left( \begin{array}{r}
10.55414\\
\end{array} \right)\]
$C^{''}$= -192.74271\\

$\sigma=$0.218 dex

\subsubsection{Run 4}
\[ X= \left( \begin{array}{c}
\rm{log}~ L(H\beta)\\
\end{array} \right)\]
\[ C = \left( \begin{array}{rrr}
-0.02079\\
\end{array} \right)\]
\[ C^{'} = \left( \begin{array}{r}
2.70384\\
\end{array} \right)\]
$C^{''}$= -34.26546\\

$\sigma=$0.077 dex

\subsubsection{Run 5}
\[ X= \left( \begin{array}{c}
\rm{log}(M_{tot}/M_\odot)\\
\rm{log}~L(OIII) \lambda 5007\\
\end{array} \right)\]
\[ C = \left( \begin{array}{rrr}
-0.31149&0.01205\\
0.01205&-0.13286\\
\end{array} \right)\]
\[ C^{'} = \left( \begin{array}{r}
5.42786\\
11.12470\\
\end{array} \right)\]
$C^{''}= $-224.41751\\

$\sigma=$0.322 dex

\subsubsection{Run 6}
\[ X= \left( \begin{array}{c}
\rm{log}(M_{tot}/M_\odot)\\
\rm{log}~L(OII) \lambda 3727\\
\end{array} \right)\]
\[ C = \left( \begin{array}{rrr}
-0.09149&-0.02375\\
-0.02375&-0.02670\\
\end{array} \right)\]
\[ C^{'} = \left( \begin{array}{r}
3.94498\\
3.55238\\
\end{array} \right)\]
$C^{''}= $-70.02284\\

$\sigma=$0.181 dex

\subsubsection{Run 7}
\[ X= \left( \begin{array}{c}
\rm{log}(M_{tot}/M_\odot)\\
\rm{log}~ L(H\beta)\\
\end{array} \right)\]
\[ C = \left( \begin{array}{rrr}
-0.00935&-0.00141\\
-0.00141&-0.00076\\
\end{array} \right)\]
\[ C^{'} = \left( \begin{array}{r}
0.38873\\
1.08283\\
\end{array} \right)\]
$C^{''}= $-3.31720\\

$\sigma$=0.063 dex

\subsubsection{Run 8}
\[ X= \left( \begin{array}{c}
\rm{log}(M_{tot}/M_\odot)\\
\rm{log}~L(OIII) \lambda 5007\\
\rm{log}~L(H\beta)\\
\end{array} \right)\]
\[ C = \left( \begin{array}{rrr}
-0.00618&-0.02453&0.02424\\
-0.02453&0.02895&-0.01964\\
0.02424&-0.01964&0.00894\\
\end{array} \right)\]
\[ C^{'} = \left( \begin{array}{r}
0.23530\\
-0.27664\\
1.38326\\
\end{array} \right)\]
$C^{''}= $-2.96210\\

$\sigma=$0.060 dex

\subsubsection{Run 9}
\[ X= \left( \begin{array}{c}
\rm{log}(M_{tot}/M_\odot)\\
\rm{log}~L(OII) \lambda 3727\\
\rm{log}~ L(H\beta)\\
\end{array} \right)\]
\[ C = \left( \begin{array}{rrr}
0.00453&-0.01054&0.00639\\
-0.01054&-0.01360&0.02811\\
0.00639&0.02811&-0.03522\\
\end{array} \right)\]
\[ C^{'} = \left( \begin{array}{r}
0.31418\\
-1.07845\\
1.55476\\
\end{array} \right)\]
$C^{''}= $9.43929\\

$\sigma$=0.050 dex
\subsubsection{Run 10}
\[ X= \left( \begin{array}{c}
\rm{log}(M_{tot}/M_\odot)\\
\rm{log}~ L(OII) \lambda 3727\\
\rm{log}~ L(OIII) \lambda 5007\\
\rm{log}~ L(H\beta)\\
\end{array} \right)\]
\[ C = \left( \begin{array}{rrrr}
0.00202&0.00452&-0.01148&0.00660\\
0.00452&-0.03223&0.02274&0.03060\\
-0.01148&0.02274&-0.02739&0.00303\\
0.00660&0.03060&0.00303&-0.04711\\
\end{array} \right)\]
\[ C^{'} = \left( \begin{array}{r}
0.06516\\
-1.91380\\
0.36326\\
2.05773\\
\end{array} \right)\]
$C^{''}$=10.09943\\

$\sigma$=0.049 dex

\subsubsection{Run 11}
\[ X= \left( \begin{array}{c}
\rm{log}(M_{tot}/M_\odot)\\
\rm{log}~ L(OII) \lambda 3727\\
\rm{log}~ L(OII) \lambda 3729\\
\rm{log}~ L(OIII) \lambda 4959\\
\rm{log}~ L(OIII) \lambda 5007\\
\rm{log}~ L(H\beta)\\
\end{array} \right)\]
\[ C = \left( \begin{array}{rrrrrr}
0.00307&-0.01363&0.01718&0.00397&-0.01054&0.00329\\
-0.01363&0.22767&-0.25863&0.01849&0.01397&0.00043\\
0.01718&-0.25863&0.10010&0.15041&-0.09680&0.10263\\
0.00397&0.01849&0.15041&-0.21704&0.14711&-0.10852\\
-0.01054&0.01397&-0.09680&0.14711&-0.12904&0.07748\\
0.00329&0.00043&0.10263&-0.10852&0.07748&-0.07576\\
\end{array} \right)\]
\[ C^{'} = \left( \begin{array}{r}
-0.01535\\
0.09101\\
-0.15203\\
0.38543\\
-0.57733\\
1.29101\\
\end{array} \right)\]
$C^{''}$=0.05811\\

$\sigma$=0.044 dex

\clearpage
\subsubsection{Run 12}
\[ X= \left( \begin{array}{c}
\rm{log}(M_{tot}/M_\odot)\\
\rm{log}~ L(OII) \lambda 3727 \\
\rm{log}~ L(OII) \lambda 3729\\
\rm{log}~ L(OIII) \lambda 4959\\
\rm{log}~ L(OIII) \lambda 5007\\
\rm{log}~ L(H\beta)\\
\rm{log}~ L(SII) \lambda 6717\\
\rm{log}~ L(SII) \lambda 6731\\
\end{array} \right)\]
\[ C = \left( \begin{array}{rrrrrrrr}
0.00483&-0.00687&0.00722&0.00858&-0.02505&0.07127&-0.01173&-0.05020\\
-0.00687&0.07641&-0.09193&0.06505&-0.04749&0.18673&-0.13848&-0.05103\\
0.00722&-0.09193&-0.03008&0.05943&0.00125&0.02944&0.09235&-0.06900\\
0.00858&0.06505&0.05943&-0.3187&0.24274&-0.03248&0.15751&-0.16355\\
-0.02505&-0.04749&0.00125&0.24274&-0.20150&0.08419&-0.21983&0.13379\\
0.07127&0.18673&0.02944&-0.03248&0.08419&-0.15983&-0.25379&0.11570\\
-0.01173&-0.13848&0.09235&0.15751&-0.21983&-0.25379&0.08350&0.27905\\
-0.05020&-0.05103&-0.06900&-0.16355&0.13379&0.11570&0.27905&-0.20249\\
\end{array} \right)\]
\[ C^{'} = \left( \begin{array}{r}
0.41177\\
0.08808\\
0.42989\\
-1.36273\\
1.35978\\
1.75639\\
0.77219\\
-2.37421\\
\end{array} \right)\]
$C^{''}$=4.03677\\

$\sigma$=0.031 dex
\clearpage

\subsection{Mixed galaxies, NII}\label{mix_app_n2}

\subsubsection{Run 1}
\[ X= \left( \begin{array}{c}
\rm{log}(M_{tot}/M_\odot)\\
\end{array} \right)\]
\[ C = \left( \begin{array}{rrr}
-0.13954\\
\end{array} \right)\]
\[ C^{'} = \left( \begin{array}{r}
3.44881\\
\end{array} \right)\]
$C^{''}$= 19.28663\\

$\sigma$=0.486 dex

\subsubsection{Run 2}
\[ X= \left( \begin{array}{c}
\rm{log}~L(OIII) \lambda 5007\\
\end{array} \right)\]
\[ C = \left( \begin{array}{rrr}
-0.17807\\
\end{array} \right)\]
\[ C^{'} = \left( \begin{array}{r}
15.10890\\
\end{array} \right)\]
$C^{''}$= -279.18357\\

$\sigma= $0.441 dex

\subsubsection{Run 3}
\[ X= \left( \begin{array}{c}
\rm{log}~L(OII) \lambda 3727\\
\end{array} \right)\]
\[ C = \left( \begin{array}{rrr}
-0.18259\\
\end{array} \right)\]
\[ C^{'} = \left( \begin{array}{r}
15.88026\\
\end{array} \right)\]
$C^{''}$= -302.97463\\

$\sigma=$0.363 dex

\subsubsection{Run 4}
\[ X= \left( \begin{array}{c}
\rm{log}~ L(H\beta)\\
\end{array} \right)\]
\[ C = \left( \begin{array}{rrr}
-0.08789\\
\end{array} \right)\]
\[ C^{'} = \left( \begin{array}{r}
8.16018\\
\end{array} \right)\]
$C^{''}$= -145.59304\\

$\sigma=$0.232 dex

\subsubsection{Run 5}
\[ X= \left( \begin{array}{c}
\rm{log}(M_{tot}/M_\odot)\\
\rm{log}~L(OIII) \lambda 5007\\
\end{array} \right)\]
\[ C = \left( \begin{array}{rrr}
-0.33638&0.03556\\
0.03556&-0.12638\\
\end{array} \right)\]
\[ C^{'} = \left( \begin{array}{r}
4.42711\\
10.06759\\
\end{array} \right)\]
$C^{''}= $-199.28986\\

$\sigma=$0.287 dex

\subsubsection{Run 6}
\[ X= \left( \begin{array}{c}
\rm{log}(M_{tot}/M_\odot)\\
\rm{log}~L(OII) \lambda 3727\\
\end{array} \right)\]
\[ C = \left( \begin{array}{rrr}
-0.13895&0.00448\\
0.00448&-0.06185\\
\end{array} \right)\]
\[ C^{'} = \left( \begin{array}{r}
2.98432\\
5.71208\\
\end{array} \right)\]
$C^{''}= $-108.88141\\

$\sigma=$0.204 dex

\subsubsection{Run 7}
\[ X= \left( \begin{array}{c}
\rm{log}(M_{tot}/M_\odot)\\
\rm{log}~ L(H\beta)\\
\end{array} \right)\]
\[ C = \left( \begin{array}{rrr}
-0.07553&-0.00295\\
-0.00295&0.00430\\
\end{array} \right)\]
\[ C^{'} = \left( \begin{array}{r}
2.13710\\
0.64146\\
\end{array} \right)\]
$C^{''}= $-3.90142\\

$\sigma$=0.130 dex

\subsubsection{Run 8}
\[ X= \left( \begin{array}{c}
\rm{log}(M_{tot}/M_\odot)\\
\rm{log}~L(OIII) \lambda 5007\\
\rm{log}~L(H\beta)\\
\end{array} \right)\]
\[ C = \left( \begin{array}{rrr}
0.02838&0.16541&-0.21083\\
0.16541&0.08089&-0.15075\\
-0.21083&-0.15075&0.23639\\
\end{array} \right)\]
\[ C^{'} = \left( \begin{array}{r}
3.45256\\
2.15154\\
-1.58771\\
\end{array} \right)\]
$C^{''}= $-8.74722\\

$\sigma=$0.120 dex

\subsubsection{Run 9}
\[ X= \left( \begin{array}{c}
\rm{log}(M_{tot}/M_\odot)\\
\rm{log}~L(OII) \lambda 3727\\
\rm{log}~ L(H\beta)\\
\end{array} \right)\]
\[ C = \left( \begin{array}{rrr}
-0.02146&0.26979&-0.25424\\
0.26979&0.07413&-0.17380\\
-0.25424&-0.17380&0.24658\\
\end{array} \right)\]
\[ C^{'} = \left( \begin{array}{r}
-0.43272\\
2.31412\\
0.42635\\
\end{array} \right)\]
$C^{''}= $-32.43116\\

$\sigma$=0.123 dex

\subsubsection{Run 10}
\[ X= \left( \begin{array}{c}
\rm{log}(M_{tot}/M_\odot)\\
\rm{log}~ L(OII) \lambda 3727\\
\rm{log}~ L(OIII) \lambda 5007\\
\rm{log}~ L(H\beta)\\
\end{array} \right)\]
\[ C = \left( \begin{array}{rrrr}
0.06991&0.02309&0.19932&-0.27420\\
0.02309&0.27687&-0.07584&-0.18603\\
0.19932&-0.07584&0.00979&-0.02231\\
-0.27420&-0.18603&-0.02231&0.31119\\
\end{array} \right)\]
\[ C^{'} = \left( \begin{array}{r}
3.10939\\
-1.90252\\
2.97736\\
-1.49383\\
\end{array} \right)\]
$C^{''}$=13.00394\\

$\sigma$=0.111 dex

\subsubsection{Run 11}
\[ X= \left( \begin{array}{c}
\rm{log}(M_{tot}/M_\odot)\\
\rm{log}~ L(OII) \lambda 3727\\
\rm{log}~ L(OII) \lambda 3729\\
\rm{log}~ L(OIII) \lambda 4959\\
\rm{log}~ L(OIII) \lambda 5007\\
\rm{log}~ L(H\beta)\\
\end{array} \right)\]
\[ C = \left( \begin{array}{rrrrrr}
0.06104&-0.03561&0.07511&0.17917&0.05983&-0.31646\\
-0.03561&0.68302&-0.51306&0.06339&0.04573&-0.32448\\
0.07511&-0.51306&0.45190&0.39281&-0.42422&0.08948\\
0.17917&0.06339&0.39281&-0.63545&0.36572&-0.39956\\
0.05983&0.04573&-0.42422&0.36572&-0.17392&0.29882\\
-0.31646&-0.32448&0.08948&-0.39956&0.29882&0.48174\\
\end{array} \right)\]
\[ C^{'} = \left( \begin{array}{r}
2.27815\\
4.45369\\
-1.35835\\
12.75392\\
-9.79639\\
-4.29598\\
\end{array} \right)\]
$C^{''}$=-23.38822\\

$\sigma$=0.109 dex

\clearpage
\subsubsection{Run 12}
\[ X= \left( \begin{array}{c}
\rm{log}(M_{tot}/M_\odot)\\
\rm{log}~ L(OII) \lambda 3727 \\
\rm{log}~ L(OII) \lambda 3729\\
\rm{log}~ L(OIII) \lambda 4959\\
\rm{log}~ L(OIII) \lambda 5007\\
\rm{log}~ L(H\beta)\\
\rm{log}~ L(SII) \lambda 6717\\
\rm{log}~ L(SII) \lambda 6731\\
\end{array} \right)\]
\[ C = \left( \begin{array}{rrrrrrrr}
0.02457&-0.06768&0.03279&0.14291&-0.05730&-0.12458&0.21699&-0.15910\\
-0.06768&0.41139&-0.11531&0.13233&-0.26493&0.02881&-0.78737&0.59629\\
0.03279&-0.11531&-0.12419&0.10406&-0.12712&0.07877&0.98205&-0.80974\\
0.14291&0.13233&0.10406&-0.6098&0.49268&-0.15328&0.18834&-0.29071\\
-0.05730&-0.26493&-0.12712&0.49268&-0.38996&-0.06633&0.06724&0.39530\\
-0.12458&0.02881&0.07877&-0.15328&-0.06633&0.86524&-0.53906&-0.11910\\
0.21699&-0.78737&0.98205&0.18834&0.06724&-0.53906&0.20432&-0.34769\\
-0.15910&0.59629&-0.80974&-0.29071&0.39530&-0.11910&-0.34769&0.73627\\
\end{array} \right)\]
\[ C^{'} = \left( \begin{array}{r}
1.02135\\
1.33751\\
-0.23101\\
7.39726\\
-6.82928\\
-4.62200\\
14.21418\\
-8.72142\\
\end{array} \right)\]
$C^{''}$=-35.41896\\

$\sigma$=0.078 dex

\end{document}